\documentclass[aps,pra,showpacs,amssymb,nofootinbib,superscriptaddress,twocolumn,longbibliography]{revtex4-1}

\usepackage{comment}
\usepackage[usenames, dvipsnames]{color}
\usepackage{tikz}
\usetikzlibrary{shapes,backgrounds,fit,decorations.pathreplacing}
\usepackage[T1]{fontenc}
\usepackage[utf8]{inputenc}
\usepackage{bbm, bm, amsbsy, amsthm, amsfonts, amsmath, dsfont, graphicx, epsfig, epstopdf, dsfont, multibib, color,relsize,times}
\usepackage{multirow}

\usepackage[colorlinks,breaklinks]{hyperref}
\makeatletter
\newcommand\org@hypertarget{}
\let\org@hypertarget\hypertarget
\renewcommand\hypertarget[2]{%
  \Hy@raisedlink{\org@hypertarget{#1}{}}#2%
  }
\makeatother
\usepackage[figure,table]{hypcap}
\usepackage{MnSymbol}
\usepackage{enumerate}
\usepackage{float}
\hypersetup{
	bookmarksnumbered,
	pdfstartview={FitH},
	citecolor={darkgreen},
	linkcolor={darkred},
	urlcolor={darkblue},
	pdfpagemode={UseOutlines}}
\definecolor{darkgreen}{RGB}{50,190,50}
\definecolor{darkblue}{RGB}{0,0,190}
\definecolor{darkred}{RGB}{238,0,0}
\definecolor{quantum}{RGB}{83,37,127}
\definecolor{quantumlight}{RGB}{169,146,191}
\usepackage{soul}

\newcommand{\ket}[1]{\ensuremath{\left|\right.\!{#1}\!\left.\right\rangle}}

\newcommand{\bra}[1]{\ensuremath{\left\langle\right.\!{#1}\!\left.\right|}}

\newcommand{\ketbra}[2]{\ensuremath{|{\hspace*{0.75pt}#1\hspace*{0.75pt}}\rangle\!\langle{\hspace*{0.75pt}#2\hspace*{0.75pt}}|}}

\newcommand{\nr}{\ensuremath{\hspace*{0.5pt}}}

\newcommand{\subtiny}[3]{\ensuremath{_{\hspace{#1 pt}\protect\raisebox{#2 pt}{\tiny{$ #3$}}}}}
\newcommand{\suptiny}[3]{\ensuremath{^{\hspace{#1 pt}\protect\raisebox{#2 pt}{\tiny{$ #3$}}}}}

\newcommand{\expval}[1]{\ensuremath{\left\langle\right.\hspace*{-1pt} #1 \hspace*{-1pt}\left.\right\rangle}}

\newcommand{\tr}{\textnormal{Tr}}
\newcommand{\floor}[1]{\left\lfloor #1 \right\rfloor}
\newcommand{\ceil}[1]{\left\lceil #1 \right\rceil}
\newcommand{\ignore}[1]{}

\newcommand{\djj}{d\kern-0.4em\char"16\kern-0.1em}

\renewcommand{\thesection}{\arabic{section}}
\renewcommand{\thesubsection}{\arabic{section}.\Alph{subsection}}
\makeatletter
\renewcommand{\p@subsection}{}
\renewcommand{\p@subsubsection}{}
\makeatother

\usepackage[customcolors]{hf-tikz}
\tikzset{style green/.style={
    set fill color=green!50!lime!60,
    set border color=white,
  },
  style cyan/.style={
    set fill color=cyan!90!blue!60,
    set border color=white,
  },
  style orange/.style={
    set fill color=orange!80!red!60,
    set border color=white,
  },
  style hordash/.style={
    set fill color=white,
    set border color=black,
  },
  hor/.style={
    above left offset={-0.09,0.25},
    below right offset={0.09,-0.05},
    #1
  },
  ver/.style={
    above left offset={-0.09,0.35},
    below right offset={0.09,-0.1},
    #1
  }
}

\usepackage[framemethod=tikz]{mdframed}
\definecolor{mycolor}{rgb}{0.122, 0.435, 0.698}
\newmdenv[innerlinewidth=0.5pt, roundcorner=4pt,linecolor=mycolor,innerleftmargin=6pt,
innerrightmargin=6pt,innertopmargin=6pt,innerbottommargin=6pt]{mybox}
\usepackage{paracol}
\usepackage{tcolorbox}
\tcbuselibrary{theorems}

\newtcbtheorem{Definitions}{Definition}%
{colback=mycolor!5,colframe=mycolor,fonttitle=\bfseries,
left=4pt,
right=4pt,
top=5pt,
bottom=5pt}{defi}

\newtcbtheorem{Lemmas}{Lemma}%
{colback=green!5,colframe=green!50!blue,fonttitle=\bfseries,
left=4pt,
right=4pt,
top=5pt,
bottom=5pt
}{lemma}

\newtcolorbox[blend into=figures]{boxdefi}[3][]
{ float*=ht,width=\textwidth,lower separated=false, center upper,
title={#2},label= def:#3,#1}

\begin{document}

\title{Trade-offs between precision and fluctuations in charging finite-dimensional quantum batteries}
\author{Pharnam Bakhshinezhad}
\email{pharnam.bakhshinezhad@tuwien.ac.at}
\thanks{P.B. previously published as Faraj Bakhshinezhad.}
\affiliation{Atominstitut, Technische Universit{\"a}t Wien, Stadionallee 2, 1020 Vienna, Austria}
\affiliation{Department of Physics and Nanolund, Lund University, Box 118, 221 00 Lund, Sweden}
\affiliation{Institute for Quantum Optics and Quantum Information - IQOQI Vienna, Austrian Academy of Sciences, Boltzmanngasse 3, 1090 Vienna, Austria}

\author{Beniamin R. Jablonski}
\affiliation{Institute for Quantum Optics and Quantum Information - IQOQI Vienna, Austrian Academy of Sciences, Boltzmanngasse 3, 1090 Vienna, Austria}
\author{Felix C. Binder}
\email{quantum@felix-binder.net}
\affiliation{School of Physics, Trinity College Dublin, Dublin 2, Ireland}
\author{Nicolai Friis}
\email{nicolai.friis@tuwien.ac.at}
\affiliation{Atominstitut, Technische Universit{\"a}t Wien, Stadionallee 2, 1020 Vienna, Austria}
\affiliation{Institute for Quantum Optics and Quantum Information - IQOQI Vienna, Austrian Academy of Sciences, Boltzmanngasse 3, 1090 Vienna, Austria}

\begin{abstract}
Within quantum thermodynamics, many tasks are modelled by processes that require work sources represented by out-of-equilibrium quantum systems, often dubbed quantum batteries, in which work can be deposited or from which work can be extracted. Here we consider quantum batteries modelled as finite-dimensional quantum systems initially in thermal equilibrium that are charged via cyclic Hamiltonian processes. We present optimal or near-optimal protocols for $N$ identical two-level systems and individual $d$-level systems with equally spaced energy gaps in terms of the charging precision and work fluctuations during the charging process. We analyze the trade-off between these figures of merit as well as the performance of local and global operations.
\end{abstract}

    \maketitle


\section{Introduction}\label{sec:introduction}

The second quantum revolution~\cite{DowlingMilburn2003} has brought about unprecedented access to technologies at the nanoscale, which are currently operating in what has been dubbed the noisy intermediate-scale quantum (NISQ) regime (cf.~\cite{Preskill2018}).
These advances go hand in hand with the desire to further improve the control over quantum systems and to better understand their potential and limitations for storing and processing information. At the same time, residual heat and noise are ever present adversaries in this endeavour, and moving systems away from thermal equilibrium with their surroundings requires sufficient control as well as the investment of time and energy. A framework that aims to address fundamental questions regarding the dynamics, interactions, energetics, and control of quantum systems in the presence of heat baths presents itself in the form of quantum thermodynamics~\cite{GooldHuberRieraDelRioSkrzypczyk2016, VinjanampathyAnders2016}. Indeed, from a fundamental thermodynamic perspective, pure states can only be prepared approximately since Nernst's unattainability principle~\cite{Nernst1906,FreitasGallegoMasanesPaz2018,TarantoBakhshinezhadEtAl2023} \textemdash\ the third law of thermodynamics\textemdash\ requires infinite resources to cool any system to its ground state. To accurately assess the resources required for a specific task, it must therefore in principle be assumed that the respective system is initially in a thermal state and that work must be invested to change this.

Quantum thermodynamics offers a broad spectrum of different scenarios to model such state transformations and corresponding work inputs, but two distinct paradigms can be identified as the conceptual polar opposites of this spectrum (cf.~\cite{ClivazSilvaHaackBohrBraskBrunnerHuber2019a,ClivazSilvaHaackBohrBraskBrunnerHuber2019b,TarantoBakhshinezhadEtAl2023}): {W}ork can be supplied to a target system (i) via a heat flow generated by a temperature gradient between two thermal baths, or (ii) via a direct supply from a coherent work source.
The former scenario can be understood as the operation of a heat engine~\cite{ScovilSchulzDuBois1959,KosloffLevy2014,UzdinLevyKosloff2015,LevyGelbwaserKlimovsky2019,Mitchison2019,WoodsNgWehner2019}, where the heat flow supplies work incoherently to a working substance and the dynamics are globally energy-conserving. On the one hand, this paradigm is appealing from a thermodynamic point of view, since the system is overall closed and external control can be minimal in the sense that an external agent operating the machine is only needed to switch on (and off) interactions between the target and the heat baths.
On the other hand, only a restricted class of state transformations is achievable within this paradigm (cf.~\cite{TarantoBakhshinezhadEtAl2023}) and practical laboratory situations in which quantum technologies are employed are not typically operated using heat engines.

An all-encompassing understanding of possible state transformations and their resource costs must therefore include coherent work sources as in (ii).
Although the specific realizations of these work sources are often not included explicitly in modelling state transformations, doing exactly this will ultimately be necessary to truly obtain fine-grained descriptions that will lead to a better understanding of quantum systems beyond thermal equilibrium.
Such descriptions can be envisioned to provide insights, e.g., regarding the effects of finite-time transformations, finite-size reservoirs, and fluctuations of relevant quantifiers.
A starting point for such a more general approach lies in modelling the work sources \textemdash commonly dubbed \emph{quantum batteries}~\cite{AlickiFannes2013,CampaioliPollockVinjanampathy2019} \textemdash on their own, i.e., independently of the systems that they eventually supply work to. In other words, quantum batteries are considered as quantum systems in which work can be temporarily deposited and from which it can subsequently be extracted.

This approach has recently received a lot of attention (see, e.g., \cite{CampaioliGherardiniQuachPoliniAndolina} for a recent review), with main foci on the charging speed or power~\cite{BinderVinjanampathyModiGoold2015, CampaioliPollockBinderCeleriGooldVinjanampathyModi2017,Gyhm2022}, including different models for batteries and charging systems and the interactions between them~\cite{LeLevinsenModiParishPollock2018, FerraroCampisiAndolinaPellegriniPolini2018,
AndolinaFarinaMariPellegriniGiovannettiPolini2018,
AndolinaEtAl2019,
AndolinaKeckMariGiovannettiPolini2019,
FarinaAndolinaMariPoliniGiovannetti2019,
RossiniAndolinaPolini2019,
CrescenteCarregaSassettiFerraro2020a,
RossiniAndolinaRosaCarregaPolini2020, CentroneMancinoPaternostro2021}, such as e.g., collision models~\cite{SeahPerarnauHaackBrunnerNimmrichter2021,Shaghaghi2022, SalviaPerarnauHaackBrunnerNimmrichter2022}. Other approaches to quantum batteries have considered, for instance, the stability of the charged battery~\cite{RosaRossiniAndolinaPoliniCarrega2020,
GherardiniCampaioliFilippoBinder2020}, charging assisted by strong interactions and thermalization~\cite{HovhannisyanBarraImparato2020} or by feedback control~\cite{MitchisonGooldPrior2021}, the roles of coherence~\cite{CaravelliYanGarciaPintosHamma2021} and dissipation~\cite{Barra2019, Alicki2019}, developed methods for describing fluctuations of the stored work~\cite{GarciaPintosHammaDelCampo2020,CusumanoRudnicki2021,Wang2021}, and analyzed fluctuations as a means of certifying high-dimensional entanglement~\cite{ImaiGuehneNimmrichter}.

Here, we follow the approach of Ref.~\cite{FriisHuber2018}, and consider battery charging realized via cyclic Hamiltonian processes. In this case, the system Hamiltonian returns to its original form at the end of each cycle and the battery-system state, initially assumed to be thermal as we have reasoned above, can be modelled to lie within the unitary orbit of the initial state~\cite{AlickiFannes2013}.
This has the advantage that it allows us to consider the charging process independently of the specifics of other potentially involved auxiliary systems (e.g., the charger systems as in~\cite{AndolinaFarinaMariPellegriniGiovannettiPolini2018, FarinaAndolinaMariPoliniGiovannetti2019}, or external classical power sources). We can thus focus on the properties of the charging process and of the charged battery, and study fundamental bounds on the chosen figures of merit.

We further centre our attention on two particular quantities: the \emph{charging precision}, quantified by the variance of the final battery charge, and the \emph{work fluctuations} arising during the charging process.
\begin{figure}[t!]
\centering
\includegraphics[width=0.45\textwidth]{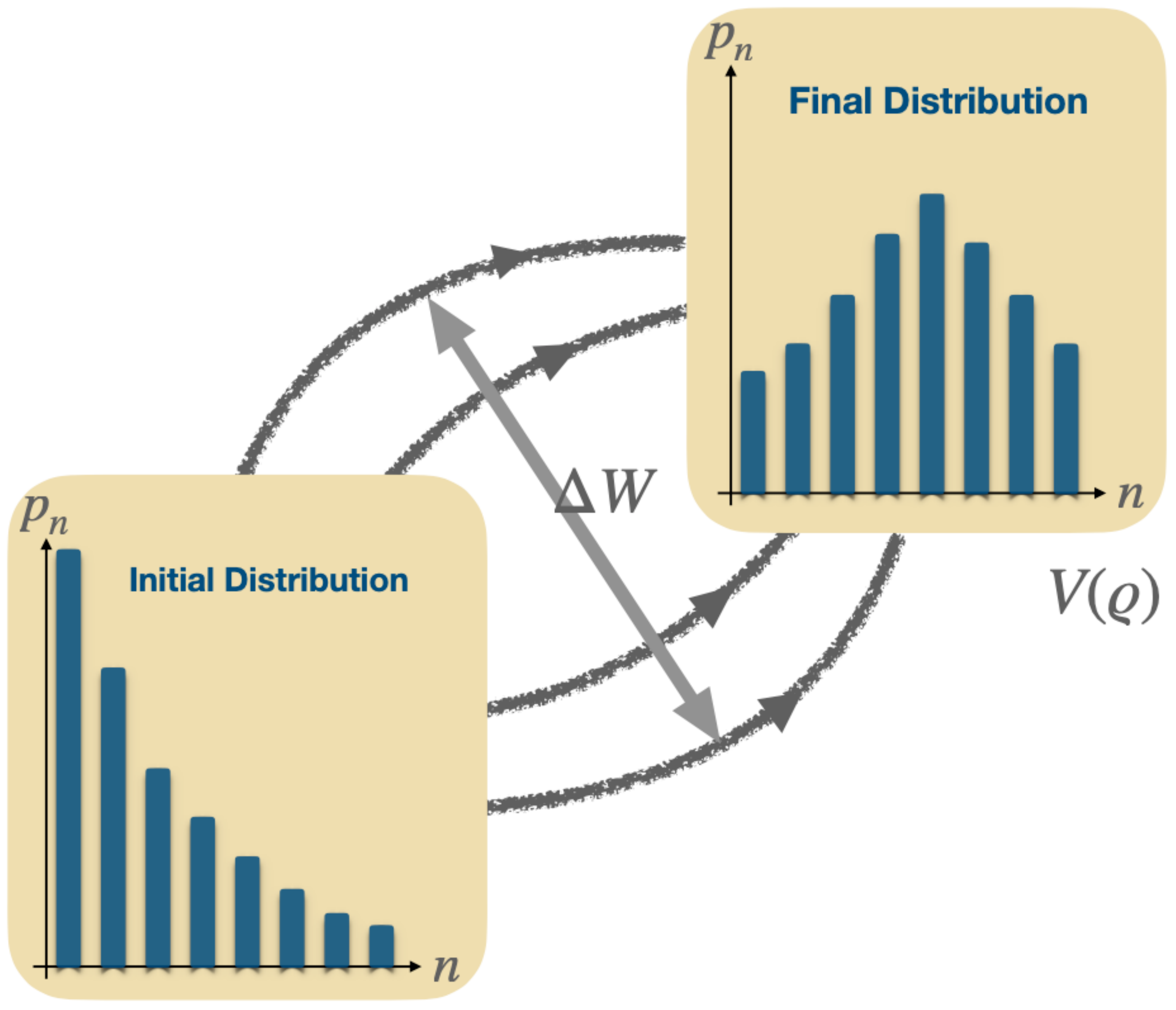}
\vspace*{-2mm}
\caption{\textbf{Charging processes}. Illustration of a unitary charging process for a battery in initial state $\tau$ and final state $\varrho=U\tau U^{\dagger}$, showing the distribution of probability weights in the energy eigenbasis. The charging precision $V(\varrho)$ only depends on the final state $\varrho$, whereas the fluctuations $\Delta W$ depend on the particular evolution from $\tau$ to $\varrho$.}
\label{fig:schematic}
\end{figure}
While the former concerns a property of the final state of the battery, independently of how this state was reached, the latter characterizes the particular charging process, as illustrated in Fig.~\ref{fig:schematic}.
Nevertheless, at fixed final battery charge, both quantities cannot be simultaneously optimized by the same charging procedure~\cite{FriisHuber2018} except for certain special cases\footnote{As noted in Ref.~\cite{FriisHuber2018}, the final energy variance and the work fluctuations coincide when the initial battery state is an energy eigenstate, which further motivates going beyond pure-state batteries~\cite{CrescenteCarregaSassettiFerraro2020b}.}.
In other words, for nonzero-temperature initial states, optimal precision generally implies non-optimal fluctuations, and vice versa.
Consequently, it is of interest to derive optimal protocols for both precision and fluctuations, and to determine trade-offs between these figures of merit.

For quantum batteries realized by quantum harmonic oscillators, such optimal protocols have been derived in Ref.~\cite{FriisHuber2018}, but translating them to finite-dimensional quantum systems frequently considered in pertinent literature (cf.~\cite{BinderVinjanampathyModiGoold2015, CampaioliPollockBinderCeleriGooldVinjanampathyModi2017, FerraroCampisiAndolinaPellegriniPolini2018,
AndolinaFarinaMariPellegriniGiovannettiPolini2018,
AndolinaEtAl2019,
AndolinaKeckMariGiovannettiPolini2019,
FarinaAndolinaMariPoliniGiovannetti2019,
RossiniAndolinaPolini2019,
CrescenteCarregaSassettiFerraro2020a,
RossiniAndolinaRosaCarregaPolini2020,
RosaRossiniAndolinaPoliniCarrega2020,
GherardiniCampaioliFilippoBinder2020,
HovhannisyanBarraImparato2020,
GarciaPintosHammaDelCampo2020,
JuliaFarreSalamonRieraBeraLewenstein2020,
CaravelliCoulterDeWitGarciaPintosHamma2020,
CrescenteCarregaSassettiFerraro2020b,
FerraroAndolinaCampisiPellegriniPolini2019}) has proven to be a formidable task~\cite{JablonskiMSc2019}.
Here, we present advances towards closing this gap: we construct a general protocol that optimizes the charging precision as well as a protocol aiming to minimize the work fluctuations for quantum batteries consisting of $N$ identical two-level systems or of individual $d$-level systems with equally spaced energy levels. We compare the performance of these protocols in terms of both figures of merit, charging precision and work fluctuations, to investigate potential trade-offs between them.
Our results represent a first step towards more \emph{all-encompassing} future analyzes of the performance of work-storage strategies for quantum-thermodynamic systems that take into account more complicated energy-level structures as well as other relevant properties of the charging process (power and fluctuations) and of the battery itself (charging precision and stability of the charge).

This manuscript is structured as follows: in Sec.~\ref{sec:framework}, we provide technical definitions for the systems under study and the relevant figures of merit. In Sec.~\ref{sec:Precision}, we then present the protocol achieving the fundamental precision limit for the considered systems, before turning to the problem of determining a similar protocol for fluctuations in Sec.~\ref{sec:fluctuation}. There, we provide a protocol that is motivated by insights previously gained for harmonic oscillators, evaluate it numerically and argue that it is a good approximation to the optimum. We then explore the trade-off between the two quantities in Sec.~\ref{sec:protocol comparison }. Finally, in Sec.~\ref{sec:non-local}, we study the role of local versus non-local operations in charging $N$-qubit quantum batteries by
comparing the optimal global protocols to the worst-case local protocols, before we present our conclusions in Sec.~\ref{sec:discussion}.


\section{Framework}\label{sec:framework}

In this section, {we first} discuss the basic setup for the type of charging processes {we consider} and provide definitions for the relevant figures of merit, i.e., charging precision and work fluctuations, in Sec.~\ref{sec:precision and fluctuations framework}, before establishing some preliminaries and notation for the particular battery systems we will study in Sec.~\ref{sec:nqubit}, i.e., batteries consisting of $N$ non-interacting identical two-level systems, here simply referred to as $N$-\emph{qubit batteries}, and what we call \emph{qudit batteries}, where the system to be charged consists of a single $d$-dimensional quantum system with equally spaced energy levels.


\subsection{Precision and work fluctuations of charging processes}
\label{sec:precision and fluctuations framework}

We consider a $d$-dimensional quantum system with associated Hamiltonian $H\,=\,\sum_{n=0}^{d-1} E_{n}\,\ketbra{n}{n}$ as a quantum battery, where $E_n$ and $\ket{n}$ represent the $n$th energy eigenvalue and its corresponding energy eigenstate, respectively. Without loss of generality, we assume these eigenvalues to be labelled such that energies are non-decreasing with increasing $n$, i.e., the $E_{n}$ are ordered such that $E_{n'}\geq E_{n}$ for $n'>n\ \forall n, n'$, and we set $E_{0}=0$. We further assume that the battery is initially uncharged, i.e., contains no work that is extractable via cyclic Hamiltonian processes. In such a process, the evolution of the system is given by a time-dependent Hamiltonian $H(t)\,=\, H\,+\,V(t),$ where the cyclic potential $V(t)$ includes all time-dependency of the Hamiltonian and satisfies the condition $V(t_{\mathrm{f}})\,=\,V(0)\,=\,0$, in which $t_{\mathrm{f}}$ represents the duration of the protocol. Such processes can be represented on the system Hilbert space as unitary operations~\cite{AlickiFannes2013}. In general, systems from which no work can be extracted by such operations are in so-called \emph{passive} states~\cite{PuszWoronowicz1978}, that is, states whose average energy cannot be lowered by unitary operations. Any passive state must hence be diagonal with respect to the energy eigenbasis and its eigenvalues must be ordered non-increasingly with increasing energy. The notion of passivity can even be relaxed to restrict the operations used for work extraction, e.g., to Gaussian operations in continuous-variable systems~\cite{BrownFriisHuber2016}. Here, however, we wish to further restrict the class of initial battery states to those that are \textit{completely passive}{: passive} states for which any number of identical copies also remains passive. For a given Hamiltonian, the only completely passive states are thermal states, and we hence consider initial battery states in Gibbs form,
\begin{align}
    \tau(\beta) &=\, \frac{e^{-\beta H}}{\mathcal{Z}(\beta)},
\end{align}
where $\beta=1/T$ is the inverse ambient temperature, $\mathcal{Z}=\tr\bigl[e^{-\beta H}\bigr]$ is the partition function of the canonical ensemble, and we use units where $\hbar\,=\,k_{\textup{B}}=1$ throughout the manuscript. Besides being motivated by Nernst's principle as we have argued in Sec.~\ref{sec:introduction}, this choice of initial state thus ensures that no work is extractable before the charging process.

We are then interested in charging procedures realized by cyclic Hamiltonian processes and thus need to consider the unitary orbit of the initial thermal state, i.e.,
\begin{align}
    \tau(\beta)\rightarrow {\varrho}    &=
    U\, \tau(\beta)\,U^{\dagger}.
\end{align}
Since the initial state is passive, all unitary operations increase (or leave invariant) the average energy. We therefore consider increasing the average energy of the battery by a fixed amount $\Delta E=\tr\bigl[H\left(\varrho-\tau(\beta)\right)\bigr]$. For any given charge $\Delta E$, there exists a set of unitaries transforming a fixed initial state to final states with the same average energy.
This non-uniqueness of the final state and of the unitaries leading to it provides an opportunity for optimization of different quantities of interest, such as {the} charging precision, energy fluctuations, and charging speed, all of which may play important roles during charging processes.

The first quantity that we will focus on here is the \emph{charging precision}{. We describe this quantity} by the energy variance of the final state, given by
\begin{align}
    V(\varrho)=\tr\bigl[H^{2}\, \varrho\bigr]
    -\bigl(\tr\bigl[H\, \varrho\bigr]\bigl)^2,
\end{align}
such that a smaller variance corresponds to higher precision and vice versa. We note that for any given unitary $U$, the final-state variance $V(\varrho)$ is non-negative but should be viewed relative to the variance of the initial state $\tau(\beta)$, and the former may be smaller or larger than the latter. {T}he unitarily achievable values of $V(\varrho)$ for any fixed energy input depend on the temperature of the initial state~\cite{FriisHuber2018}. Moreover, for infinite-dimensional systems, the increase in variance is not bounded from above for any nonzero energy increase $\Delta E$. Here, however, we wish to analyze the fundamental precision limit for fixed values of $\Delta E$ for \textit{finite-dimensional} systems {and all quantities of interest are hence bounded}. In addition, it is worth mentioning that the precision only depends on the initial and the final states, and the type of dynamics relating these states do not play any role in the characterization of the precision. This means that the optimization of the precision reduces to finding the state with the minimum variance for a given energy input.

The second quantity that we analyze here can be viewed as {a} figure of merit for the quality of the charging process: specifically, we are interested in minimizing the \emph{work fluctuations} for a given energy increase. {In general, there are a number of inequivalent ways to define work in the quantum domain}\footnote{For a selection of different approaches, see, e.g., \cite{ReebWolf2014,AlickiHorodeckiMPR2004,BrunnerLindenPopescuSkrzypczyk2012,Aberg2013,GallegoEisertWilming2016,NiedenzuHuberBoukobza2019,BeyerLuomaStrunz2020}, or the discussion in~\cite{GooldHuberRieraDelRioSkrzypczyk2016}.}
{Here, we focus on work quantified by \emph{two-point measurement} (TPM) scheme~\cite{TalknerLutzHaenggi2007}. There, two ideal\footnote{Note that when taking the assumption of initially thermal states seriously, projective measurements are not ideal because measurement apparatuses cannot themselves be prepared in pure states to begin with~\cite{GuryanovaFriisHuber2020}. Consequently, correction terms apply in principal, in particular to work estimation procedures~\cite{DebarbaEtAl2019}{. B}ut since we are here interested in fundamental bounds independently of specifics of the measurement devices, we will not include such corrections.} projective measurements, one prior to and one after the action of the transformation (here represented by $U$), are used to estimate the work performed on the system in terms of differences between the respective measurement outcomes $E_{m}$ and $E_{n}$. The work value assigned to such a pair of outcomes is $W_{m\rightarrow n}=E_{n}-E_{m}$ and the work estimate is then obtained as the average
\begin{align}
\label{eq:def. work}
\expval{W} =\sum_{m,n=0}^{d-1} p_{m \to n}\big(E_n-E_m\big)\,.
\end{align}
Here, $p_{m \to n}$ is the transition probability of the energy eigenstate $\ket{m}$ to $\ket{n}$ starting from the initial state $\tau$,
\begin{align}
p_{m \to n}=\, p_m \,|\bra{n}U\ket{m}|^2,
\label{eq:transition prob.}
\end{align}
and $p_{m}=\bra{m}\tau\ket{m}$ is the probability to find the initial state in the eigenstate $\ket{m}$. For unitary processes such as those we consider here one finds that the average work matches the change in average energy, $\expval{W}=\Delta E$.} The work fluctuations $(\Delta W)^2$ of the charging process {can then be} quantified by the average squared distance from the average energy increase,
\begin{align}
\label{eq:def. fluct.}
(\Delta W)^2=\sum_{m,n=0}^{d-1} p_{m \to n}\big(E_n-E_m-\Delta E\big)^2.
\end{align}
{As such, $(\Delta W)^2$ represents the second statistical moment of the work distribution obtained in the TPM scenario. Thus, if one raises the TPM scheme to a definition of work, the work fluctuations provide a basis for confidence statements about the estimated (average) work input required to charge the battery, whereas the energy variance allows one to quantify one's confidence regarding the estimated charge itself.}

{However, the relation between the work fluctuations and energy variances is generally complicated: in our case, the quantities satisfy}
\begin{align}
\label{eq:fluct en var}
(\Delta W)^2& = V(\varrho)+V(\tau)
-2\left[ \tr[\tilde{H}\,H\tau]- E(\tau)\,E(\varrho)\right],
\end{align}
where $\tilde{H}=U^{\dagger}H\,U$ and $E(\varrho)=\tr[H\varrho]$.

Based on Eq.~(\ref{eq:fluct en var}), one can easily see that if the battery is initially in the ground state\footnote{Or, more generally, any eigenstate of the system Hamiltonian~\cite{FriisHuber2018}.}, i.e., $T=0$, the work fluctuations coincide with the final-state variance, i.e., $(\Delta W)^2=(\Delta \sigma)^2= V(\varrho)$, but in general, these quantities do not coincide and cannot be simultaneously minimized. Detailed comparisons of optimal protocols for precision and fluctuations have so far only been available for harmonic-oscillator batteries. Here, we therefore want to analyze and compare achievable performances for finite-dimensional batteries, whose specific compositions we will describe next.


\subsection{$N$-qudit model: single-qudit and $N$-qubit batteries}
\label{sec:nqubit}

As a specific realization of a quantum battery, let us now describe a multipartite system consisting of $N$ identical non-interacting $d$-dimensional subsystems with equidistant energy levels (here referred to as `qudits'). We will then consider two special cases of such a system more closely: $N$-qubit quantum batteries (arbitrary $N$ but $d=2$) and single-qudit batteries ($N=1$ but arbitrary $d$).

The $N$-qudit system is described by a set of local Hamiltonians $H_{d_i}=\mathds{1}_{d}^{\otimes i-1}\otimes H_{d}\otimes \mathds{1}_{d}^{\otimes N-i}$, where $\mathds{1}_{d}$ denotes the $d$-dimensional identity matrix, and  $H_{d}\,=\,\sum_{n=0}^{d-1} E_{n}\,\ketbra{n}{n}$ is the Hamiltonian of a $d$-dimensional system with equally spaced energy levels, i.e., $E_{n}=n\omega$.

Since the individual battery systems are not interacting with each other, the joint initial thermal state is an uncorrelated product state of the form $\tau_{\textup{tot}}(\beta)=\tau_{d}(\beta)^{\otimes N}$, where
\begin{align}
    \tau_{d} (\beta)=\sum_{n=0}^{d-1}p_n \ketbra{n}{n},    \qquad p_n=\frac{1-e^{-\beta \omega}}{1-e^{-\beta \omega d}}e^{-\beta n \omega}.
\label{eq: thermal state d}
\end{align}
Taking into account the degeneracy of the energy levels of the joint system, the total Hamiltonian can be written as
\begin{align}
\label{eq:Hamiltonian Nqubit}
H_{\textup{tot}}&=\sum_{m=0}^{N(d-1)} \sum_{i_m=1}^{g_{d}(m)} m\nr\omega \,\ketbra{m,i_m}{m,i_m}\nonumber\\
&=\sum_{m=0}^{N(d-1)} \sum_{i_m=1}^{g_{d}(m)} E_{m,i_m} \,\ketbra{m,i_m}{m,i_m},
\end{align}
where $m$ and $i_m$ indicate the $m$th distinct energy level and $i$th level with  energy $m$, respectively, while $g_{d}(m)$ indicates the degeneracy of the $m$th distinct energy eigenvalue, such that $E_{m,i_m}=E_{m,j_m}$ for all $i_{m},j_{m}=1,\ldots,g_{d}(m)$ and $E_{m,i_m}\neq E_{n,j_n}$ for all $i_{m},j_{n}$ as long as $m\neq n$. For instance, for an $N$-qubit system, the degeneracy of energies is given by $g_{\textup{2}}(m)=\tbinom{N}{m}$. In terms of this notation for the eigenstates of the joint Hamiltonian, one can rewrite the initial thermal state as
\begin{align}
\label{eq:initial state Nqubit}
\tau_{\textup{tot}}(\beta)=\sum_{m=0}^{N(d-1)} \sum_{i_m=1}^{g_{d}(m)} \frac{e^{-\beta m\nr \omega}}{\mathcal{Z}_{d}(\beta)^N} \ketbra{m,i_m}{m,i_m},
\end{align}
where the initial probability distribution on the diagonal with respect to the energy eigenbasis is independent of the index $i_m$, i.e., $p_{m,i_m}:= {e^{-\beta m\nr\omega}}/{\mathcal{Z}_{d}(\beta)^N}$ where $p_{m,i_m}$ indicates the initial probability weight of the eigenstate $\ket{m,i_m}$. Using Eqs.~(\ref{eq:Hamiltonian Nqubit}) and~(\ref{eq:initial state Nqubit}), the initial average energy of the total system takes the form
\begin{align}
    E\bigl(\tau_{\textup{tot}} (\beta)\bigr)&=\mathrm{Tr}[\tau_{\textup{tot}}(\beta) \, H_{\textup{tot}}]=N\omega (\tfrac{1}{e^{\beta \omega}-1}-\tfrac{d}{e^{\beta\omega d}-1}) ,
\end{align}
where we have made use of the fact that $\mathrm{Tr}[\tau_{d}(\beta)^{\otimes N} \, H_{d_i}]=\mathrm{Tr}[\tau_{d}(\beta) \, H_{d}].$

Due to the symmetry $g_{d}(n)=g_{d}\bigl(N(d-1)-n\bigr)$ with regards to the sizes of the degenerate subspaces, identifying the maximal average energy within the unitary orbit of the initial state simply corresponds to rearranging the probability weights in such a way that the populations of the eigenstates $\ket{n}$ and $\ket{d-n}$ are exchanged, such that $\tau_{d}(\beta)\mapsto \tau_{d}(-\beta)$. Using dimensionless quantities $\epsilon_0:=E\bigl(\tau (\beta)\bigr)/\omega$ and $\epsilon:=E(\varrho)/\omega$ to describe the initial and final energies, respectively, we then have $\epsilon_0 \leqslant \epsilon \leqslant N(d-1)- \epsilon_0$. Since $\epsilon=\epsilon_0 +\Delta \epsilon$, this implies that the charge $\Delta\epsilon$ of the battery satisfies
\begin{align}
0 \leqslant \Delta\epsilon \leqslant  N(d-1)- 2\epsilon_0.
\label{eq:input energy range}
\end{align}

In the following, we investigate the precision and work fluctuations achievable with unitary charging processes for two different special cases of the system described by this model: single-qudit batteries $(N=1)$ and $N$-qubit batteries ($d=2$).


\section{Fundamental charging precision limits for arbitrary temperatures}\label{sec:Precision}

\begin{figure*}[t!]
(a) \includegraphics[width=0.31\textwidth,trim={0cm 0mm 0cm 0mm}]{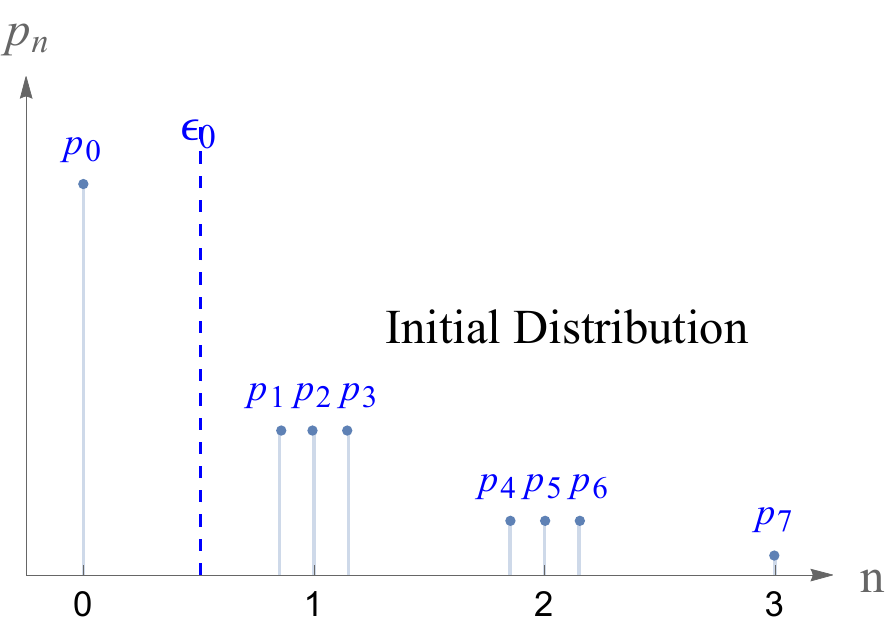}
(b) \includegraphics[width=0.3\textwidth,trim={0cm 0mm 0cm 0mm}]{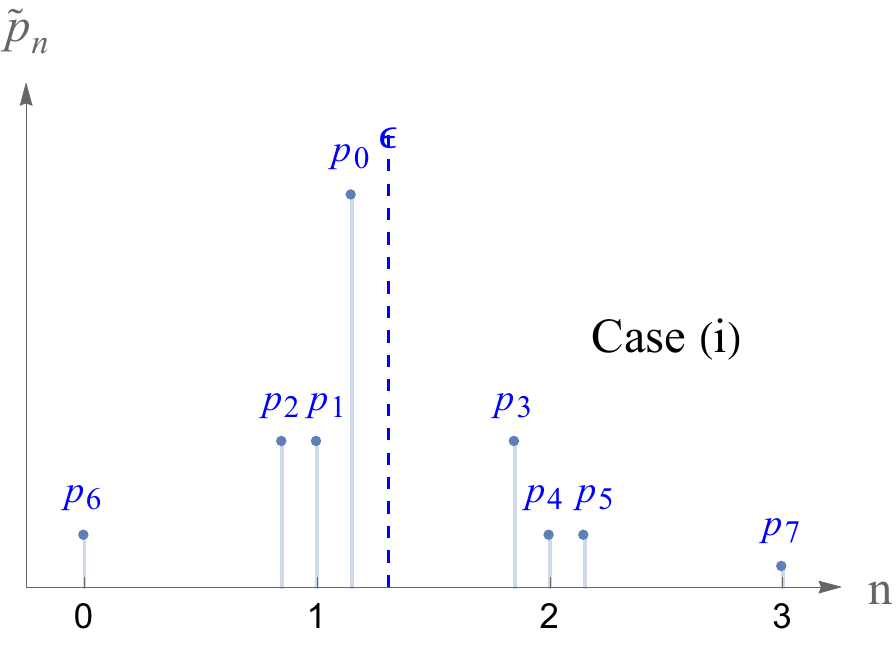}
(c) \includegraphics[width=0.3\textwidth,trim={0cm 0mm 0cm 0mm}]{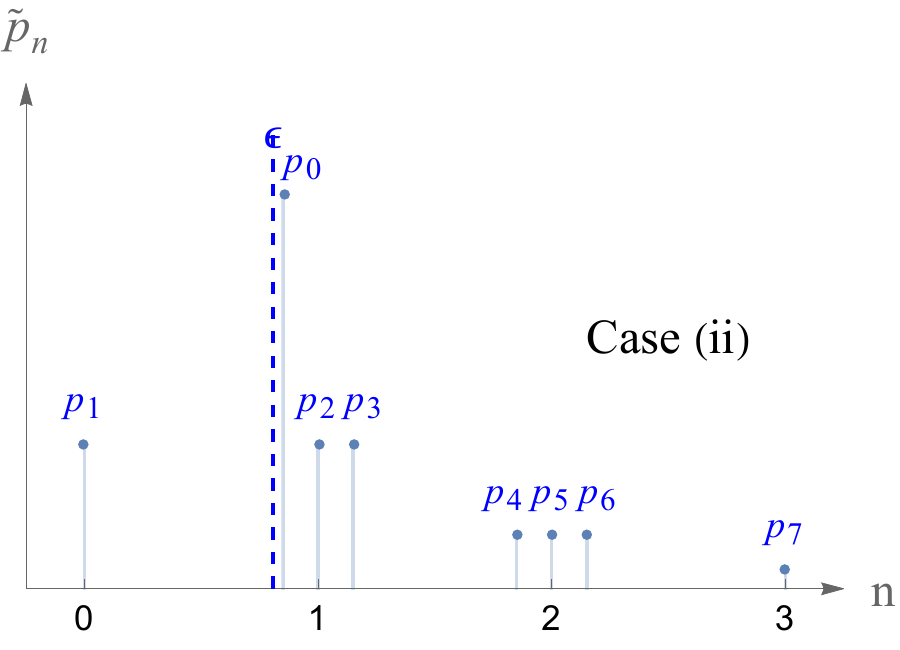}\\
(d) \includegraphics[width=0.3\textwidth,trim={0cm 0mm 0cm 0mm}]{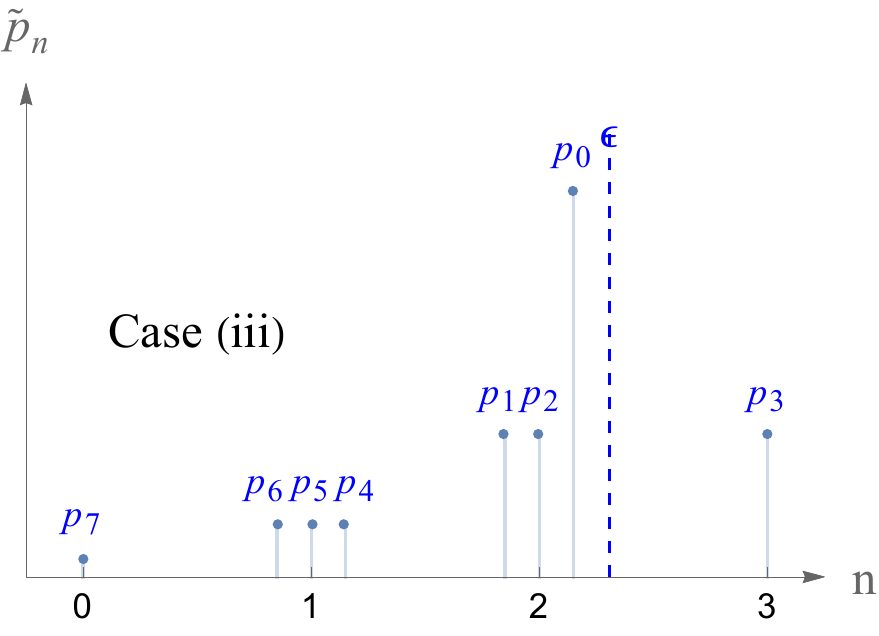}
(e) \includegraphics[width=0.3\textwidth,trim={0cm 0mm 0cm 0mm}]{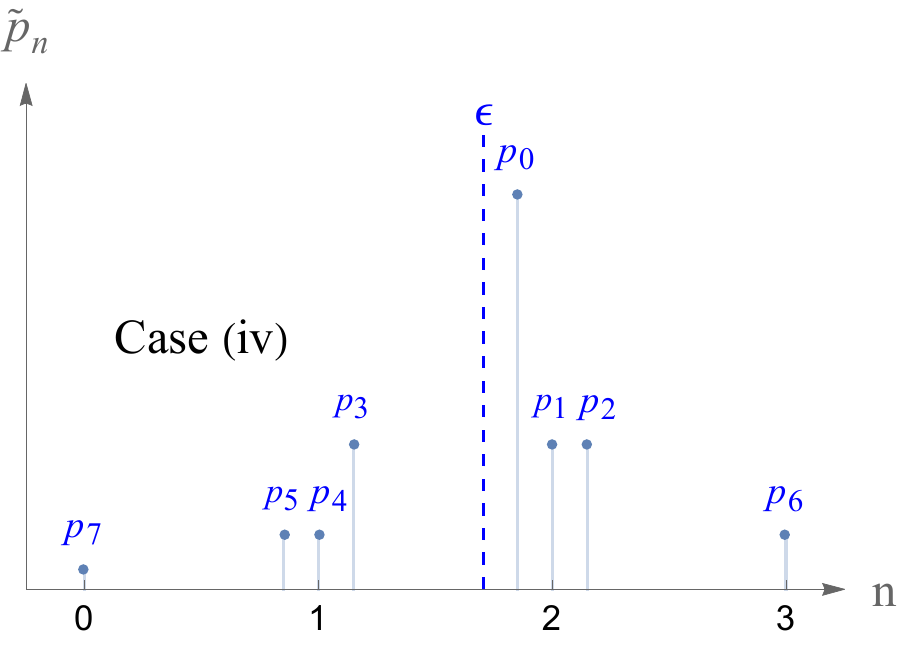}
\caption{\textbf{Step I of the optimal-precision protocol}. The four cases (i)-(iv) of reordering the initial probability weights (a) are illustrated in panels (b)-(e), respectively, by varying the target energy $\epsilon$ (and hence $k$) for fixed $N=3$. The vertical axes show the sizes of the respective probability weights, labelled by $s=0,1,\ldots,7$, while the horizontal axes show the energy levels (degeneracies are indicated by groups of vertical lines) as well as the initial energy $\epsilon_{0}$ in (a) and the different target energies $\epsilon$ in (b)-(e) as dashed lines.}
\label{fig:Step I of the optimal-precision protocol}
\end{figure*}

We are now in the position to determine the optimal protocol minimizing the variance for a specific charge $\Delta\epsilon$. To briefly reiterate, the {(single-qudit or $N$-qubit) system} is {initially} prepared in the {state} $\tau(\beta)$ with energy $ \epsilon_0\,=\,\epsilon\bigl(\tau(\beta)\bigr)${. Then} the energy of the system is unitarily increased to reach the target energy $\epsilon=\epsilon_0 +\Delta \epsilon$. The goal is to choose the unitary operation such that the energy variance $V(\varrho)$ of the final state is minimized for fixed $\Delta\epsilon$ and fixed inverse initial temperature $\beta$. However, direct minimization of $V(\varrho)$ is generally difficult even for fixed initial temperature and input energy, owing to the number of parameters required to describe the involved unitaries~\cite{SpenglerHuberHiesmayr2010}. This problems is exacerbated by the fact that the unitaries achieving the minimum variance are not unique and our desire to specify the result as a function of both $\Delta\epsilon$ and $\beta$. We therefore employ an optimization protocol for $V(\varrho)$ that makes use of an auxiliary quantity $\tilde{V}$, the \emph{average squared distance} (ASD) from the target energy, which we define as
\begin{align}
\begin{split}
  \tilde{V}(\epsilon)&= \sum_{n} \,\sum_{i_n=1}^{g_{d}(n)}\, \tilde{p}_{n,i_n} (E_n-\omega\epsilon)^2\\
    &= \omega^{2} \sum_{n}\, \sum_{i_n=1}^{g_{d}(n)}\,\tilde{p}_{n,i_n} (n-\epsilon)^2
\end{split}
\end{align}
for a given probability distribution $\{\tilde{p}_{n,i_n}\}_{n}$. In general, the quantities $\tilde{V}(\epsilon)$ and $V(\varrho)$ do not coincide, but when the distribution $\{\tilde{p}_{n,i_n}\}_{n,i_n}$ matches the probability distribution of the final state $\varrho$ with respect to the energy eigenbasis, i.e., when $\tilde{p}_{n,i_n}=\bra{n,i_n}\varrho\ket{n,i_n}\ \forall\,n, i_n$, we have $\sum_{n,i_n} \, \tilde{p}_{n,i_n}n=\epsilon$, such that $\tilde{V}(\epsilon)=V(\varrho)$. In this way, the optimization can be carried out in terms of a protocol that aims to minimize the ASD with respect to a fixed target energy in every step, while the average energy changes throughout the protocol and only reaches the target value at the end. The ASD thus allows us to obtain the optimal protocol for the variance in a simple way. This protocol can be divided into two distinct steps:

\begin{enumerate}[I.]
\item{In the first step, illustrated in Fig.~\ref{fig:Step I of the optimal-precision protocol}, the initial probability distribution is rearranged such that the larger probability weights are assigned to energy levels closer to the target energy $\epsilon$. That is, the resulting distribution $\{\tilde{p}\suptiny{1}{0}{(\mathrm{I})}_{n,i_n}\}_{n,i_n}$ satisfies $\tilde{p}\suptiny{1}{0}{(\mathrm{I})}_{m,j_m}\geq \tilde{p}\suptiny{1}{0}{(\mathrm{I})}_{n,i_n}$ for all $m,n$ with $|m-\epsilon|<|n-\epsilon|$, and hence corresponds to the minimum ASD with respect to $\epsilon$ in the unitary orbit of the initial state.}
\item{In the second step, unitary two-level rotations that minimally increase the ASD per unit of energy shift are used to adjust the average energy to match the target energy $\epsilon$.}
\end{enumerate}

\noindent In the following, we first describe these steps in more detail for the $N$-qubit system. It is then straightforward to adapt the $N$-qubit protocol to a single-qudit system of dimension $d$ by considering the former protocol for $N=d-1$ with the additional replacement $g_{2}(m)\mapsto1$ for all $m$.
For ease of notation we will drop the subscript on the degeneracy factor for qubits from now on and use $g(m)$ instead of $g_{2}(m)$.\\

For the sake of notational simplicity, we further define the new variable
\begin{align}
    s(m, i_m):=
    \begin{cases}
        1 & \text{for}\ m=0\\
        \sum_{n=0}^{m-1} g(n)+i_m & \text{for}\ m>0
    \end{cases},
\end{align}
to label the eigenstates of the joint system using only a single index $s=1,2,\ldots,2^{N}$. Since each value of $s$ uniquely identifies a pair of values $\{m(s),i_{m}(s)\}$, we use the notation $p(s):=p_{m(s),i_{m}(s)}$ such that $p(s)\geqslant p(s')$ for all $s \leqslant s'$ with $s,s'\in \{1,2,\hdots , 2^N\}$. This allows us to order the probability weights with respect to the energy eigenstates in non-increasing order using only the parameter~$s$.\\

\noindent\textbf{\label{subsec:part1}Step I of the protocol:} In the first step, we rearrange the initial-state probability weights $p(s)$ to form a new probability distribution $\{\tilde{p}(s)\}_{s}$, such that the largest value, $p(1)$, is associated with the energy level closest to $\epsilon$, the second-largest weight, $p(2)$, is associated with the second-closest level and so on, to reach the minimal value of $\tilde{V}(\epsilon)$ in the unitary orbit of the initial state. In order to do so, we first need to find the closest energy level (labelled $k$) to the desired target energy $\epsilon$, which is given by
\begin{align}
\label{eq:k}
    k=\begin{cases}
\lfloor \epsilon \rfloor   & ~~~~\textnormal{if}~\epsilon- \lfloor \epsilon \rfloor \,<\,\lceil \epsilon \rceil-  \epsilon \\
\\
\lceil \epsilon \rceil   & ~~~\textnormal{if}~\epsilon- \lfloor \epsilon \rfloor \, \geqslant \,\lceil \epsilon \rceil-  \epsilon
    \end{cases},
\end{align}
where $\lfloor \epsilon \rfloor$ and $\lceil \epsilon \rceil$ denote the floor and ceiling functions, i.e., the closest integers to $\epsilon$ that are smaller or larger than $\epsilon$, respectively.

We can then identify four different cases, labelled (i)-(iv) in the following, depending on the signs of the quantities $\epsilon-k$ and $\lfloor \frac{N}{2} \rfloor-k$, where the latter represents a constraint arising from the finite system dimension. That is, the details of how the probability weights are reordered depend upon whether $k$ is closer to the lowest or to the highest energy level. For all four cases, the resulting density matrix after step~I is diagonal with respect to the energy eigenbasis, and the corresponding diagonal probability weights are given by:\\

\noindent\textbf{Case (i):} If $k=\lfloor \epsilon \rfloor $ and $k \leqslant \lfloor \frac{N}{2} \rfloor $,
\begin{align}
\tilde{p}(s)\,=\,
\begin{cases}
 p(\sum_{j=m}^{2k-m} \binom{j}{N}-i_m+1)    &m \,<\, k  \\[0.5mm]
 p(\binom{k}{N}-i_k+1)   & m\,=\,k \\[0.5mm]
       p(\sum_{j=2k-m+1}^{m-1} \binom{j}{N}+i_m)   & k \,<\, m\, \leqslant \,2k+1  \\[0.5mm]
 p(s)   &  m>2k+1
\end{cases}.
\end{align}

\noindent\textbf{Case (ii):} If $k=\lfloor \epsilon \rfloor $ and $k > \lfloor \frac{N}{2} \rfloor $,
\begin{align}
\tilde{p}(s)\,=\,
\begin{cases}
 p(\sum_{j=m}^{N} \binom{j}{N}-i_m+1)    &m\, \leqslant\, 2k -N\\[0.5mm]
 p(\sum_{j=m}^{2k-m}\binom{j}{N}-i_k+1)   &2k-N\, <\, m<k \\[0.5mm]
        p( \binom{k}{N}-i_m+1)   &  m\,=\,k \\[0.5mm]
  p(\sum_{j=2k-m+1}^{m-1} \binom{j}{N}+i_m)   &  m\,>\,k
\end{cases}.
\end{align}

\noindent\textbf{Case (iii):} If $k=\lceil \epsilon \rceil $ and $k \leqslant \lfloor \frac{N}{2} \rfloor $,
\begin{align}
\tilde{p}(s)\,=\,
\begin{cases}
 p(\sum_{j=m}^{2k-m-1} \binom{j}{N}-i_m+1)    &m\, <\, k  \\[0.5mm]
p(i_k)   & m\,=\,k \\[0.5mm]
        p(\sum_{j=2k-m}^{m-1} \binom{j}{N}+i_m)   & k \,<\, m \,\leqslant\, 2k  \\[0.5mm]
 p(s)   &  m\,>\,2k+1
\end{cases}.
\end{align}

\noindent\textbf{Case (iv):} If $k=\lceil \epsilon \rceil $ and $k > \lfloor \frac{N}{2} \rfloor $,
\begin{align}
\tilde{p}(s)\,=\,
\begin{cases}
 p(\sum_{j=m}^{N} \binom{j}{N}-i_m+1)    &m \,\leqslant\, 2k -N-1\\[0.5mm]
 p(\sum_{j=m}^{2k-m-1}\binom{j}{N}-i_k+1)   &2k-N-1 < m<k \\[0.5mm]
        p( i_m)   &  m\,=\,k \\[0.5mm]
  p(\sum_{j=2k-m+1}^{m-1} \binom{j}{N}+i_m)   &  m\,>\,k
\end{cases}.
\end{align}

\begin{figure*}[ht!]
(a)\ \includegraphics[width=0.45\textwidth,trim={0cm 4mm 0cm 0mm}]{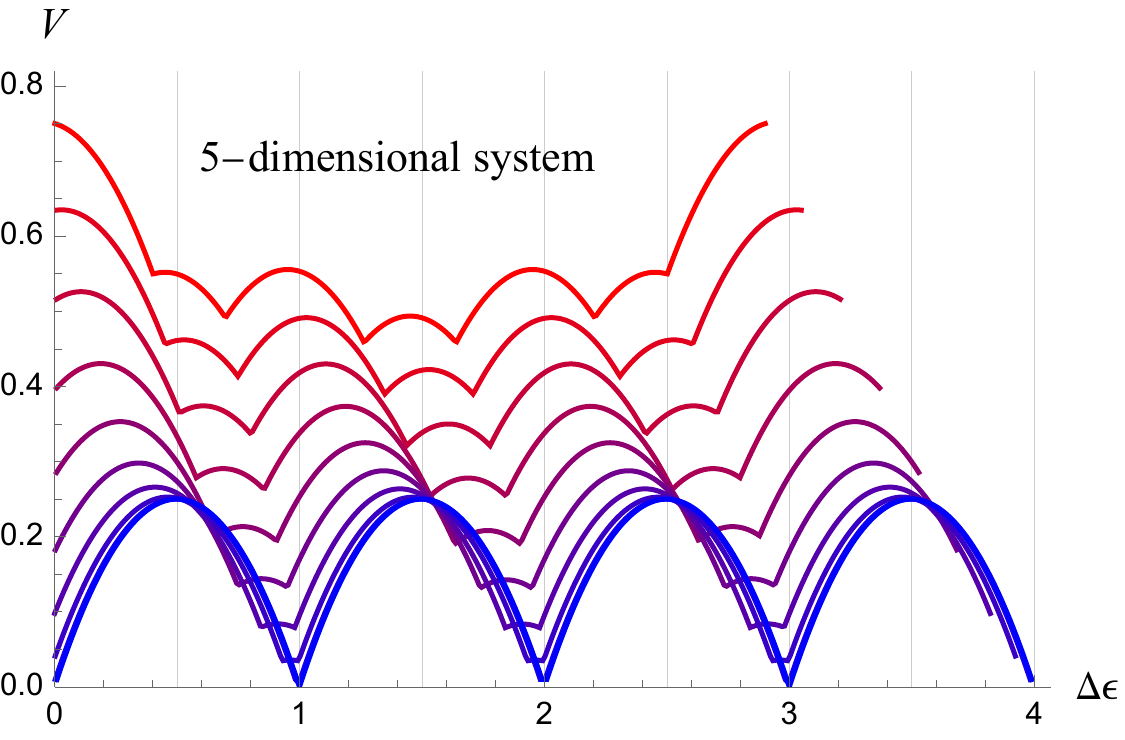}
(b)\includegraphics[width=0.45\textwidth,trim={0cm 4.5mm 0cm 0mm}]{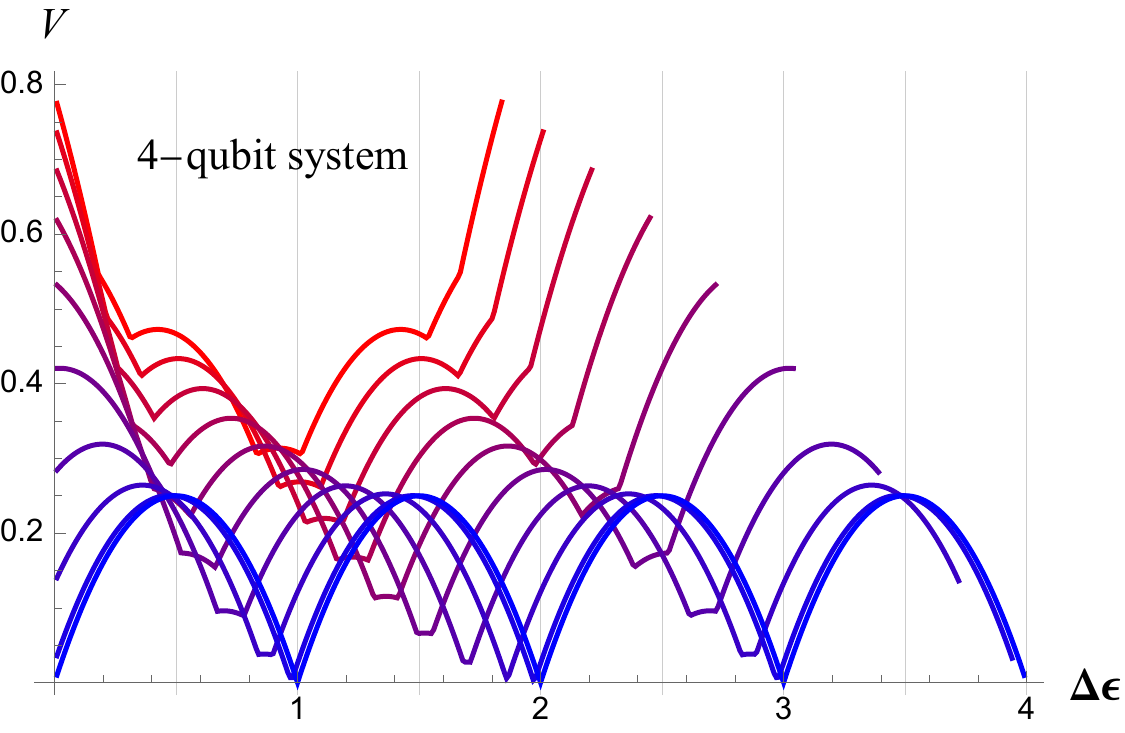}
\caption{\textbf{Optimal precision protocol}. The minimal energy variance $V$ (in units of $\omega^{2}$) obtained via the optimal-precision protocol is shown as a function of the energy input $\Delta\epsilon$ for (a) a 5-dimensional quantum system with equally spaced energy levels, and (b) a system of $4$ identical qubits for temperatures (in units of $\hbar \omega/k_{\textup{B}}$) from $T = 0.1$ (blue, bottom) to $T=1$ (red, top) in steps of $0.1$.}
\label{fig:qudit}
\end{figure*}

\noindent From the resulting probability distribution we obtain the average energy \begin{align}
    \tilde{\epsilon}_{\textup{I}} &=\,\sum\limits_{s=1}^{2^N}m(s)\tilde{p}(s)
    \,=\,\sum\limits_{m=0}^N\sum_{i_m=1}^{g(m)}m \, \tilde{p}_{m,i_m},
\end{align}
which is generally not equal to the desired target energy, $\tilde{\epsilon}_{\textup{I}}\neq\epsilon$, but may be smaller or larger than $\epsilon$ in any of the four cases (i)-(iv). Consequently, the ASD from the desired energy is generally different from the energy variance for the distribution arising from step I. Therefore, the energy of the system must be changed to reach the target $\epsilon$, which is the purpose of step II.\\

\noindent\textbf{Step II of the protocol:} Now, we want to adjust the average energy by using a sequence of unitary two-level rotations. Each of the transformations slightly alters the average energy to achieve $\epsilon$, but since the ASD was globally minimized (within the unitary orbit of the initial state) by the first step of the protocol, step 2 will increase the ASD. We hence select the transformations such that each of them increases the ASD only minimally per unit of energy change. Here we need to find the optimal sequence of these two-level rotations.

To do so, we first consider a two-level rotation between two arbitrary levels $m$ and $n$ with weights $\tilde{p}_m$ and $\tilde{p}_n$ and energies $E_m$ and $E_n$, respectively, and parameterize the rotation by an angle $\theta$. Starting from a diagonal density matrix (with respect to the energy eigenbasis), this transformation can be represented as the map
\begin{align}
    (\tilde{p}_m,\,\tilde{p}_n)\ \mapsto\ (\tilde{p}_m \cos^2{\!\theta}+\tilde{p}_n \sin^2\!\theta,\,\tilde{p}_n \cos^2{\!\theta}+\tilde{p}_m \sin^{2}\!\theta).
    \label{eq:two-level rotation map}
\end{align}
The associated energy change is given by
\begin{align}
    \Delta \tilde{\epsilon}_{\textup{II}}\,=\,\left(\tilde{p}_m-\tilde{p}_n\right)\,
    \frac{E_n-E_m}{\omega}\,\sin^{2}\!\theta.
\end{align}
Similarly, the change in the ASD is
\begin{align}
\label{eq:Ave. sq. two level}
    \frac{\Delta \tilde{V}}{\omega^{2}}&\,=\, \left(\tilde{p}_m-\tilde{p}_n\right)\left(\left(\tfrac{E_n}{\omega}-\epsilon\right)^2-\left(\tfrac{E_m}{\omega}-\epsilon\right)^2\right)\,\sin^{2}\!\theta\nonumber\\
    &\,=\, \left(\tfrac{E_m+E_n}{\omega}-2\epsilon\right)\,\Delta \tilde{\epsilon}_{\textup{II}}.
\end{align}
{From this expression we see that} we have to apply a two-level rotation between {levels} $n$ and $m${, chosen such that $(\tfrac{E_m+E_n}{\omega}-2\epsilon)$ is minimized} while bringing the average energy closer to $\epsilon${, in order to obtain the minimum ASD increase per unit energy.} To identify these pairs of levels, it is convenient to choose a relabelling relative to the value $k$ from Eq.~(\ref{eq:k}). That is, instead of $m$ and $n$, we introduce the variables $l\in \mathds{N}_0$ and $j\in \mathds{Z}$ such that $m=k-l$ and $n=k+l+j$. The average energy change associated to the two-level rotation in Eq.~(\ref{eq:two-level rotation map}) can then be written as
\begin{align}
    \Delta \tilde{\epsilon}_{\textup{II}}\,=\,\left(\tilde{p}_{m,i_m}-\tilde{p}_{n,i_n}\right)\, \left(2l+j\right)\,\sin^{2}\!\theta.
\end{align}
Using Eq.~(\ref{eq:Ave. sq. two level}), we can also obtain the change of $\tilde{V}$ per unit energy change, i.e.,
\begin{align}
\label{eq:Ave. sq. two level Nqubit}
   \frac{1}{\omega}\frac{ \Delta \tilde{V}}{\Delta \tilde{\epsilon}_{\textup{II}}}\,=\,\bigl(2\left(k-\epsilon\right)+j\bigr) .
\end{align}
We thus see that the variable $j$ determines a hierarchy of possible two-level rotations that increase $\tilde{V}$ the least, while ensuring that $\Delta \tilde{\epsilon}_{\textup{II}}$ and $\tfrac{ \Delta \tilde{V}}{\Delta \tilde{\epsilon}_{\textup{II}}}$ have the desired sign. That is, when $\tilde{\epsilon}_{\textup{I}}<\epsilon$, the average energy needs to be increased, suggesting that we have to select index pairs such that $\tilde{p}_{m,i_m}>\tilde{p}_{n,i_n}$ and $2l+j>0$ while minimizing $j$. In contrast, when $\tilde{\epsilon}_{\textup{I}}>\epsilon$, we should select levels with $\tilde{p}_{m,i_m}<\tilde{p}_{n,i_n}$ and $2l+j>0$, making the maximal values of $j$ desirable.
According to these rules, we can identify the optimal value $j_{\mathrm{opt}}$ of $j$ and the corresponding set of admissible values $l_{\mathrm{opt},i}$ of $l$ for any given probability distribution $\tilde{p}(s)$. Since the energy levels can be degenerate, for any fixed pair $(j_{\mathrm{opt}},l_{\mathrm{opt},i})$, corresponding to some variables $(m,n)$, one can then find two sets of labels, $\{(n,i_n)\}_{i_n=1}^{g(n)}$ and $\{(m,i_m)\}_{i_m=1}^{g(m)}$ with the desired properties. In contrast to the non-degenerate case discussed in Appendix~A.1.II of  Ref.~\cite{FriisHuber2018} this means that there are now $x_{\mathrm{max},i}:=\min\{g(k-l_{\mathrm{opt},i}),g(k+l_{\mathrm{opt},i}+j_{\mathrm{opt}})\}$ possible pairs of levels between which one may rotate for any fixed choice of $(j_{\mathrm{opt}},l_{\mathrm{opt},i})$, and we label these pairs by a subscript $x$, i.e., $(j_{\mathrm{opt}},l_{\mathrm{opt},i})_{x}$ with $x=1,\ldots,x_{\mathrm{max},i}$. For a given probability distribution $\tilde{p}(s)$ we can thus generate a set $P_{\mathrm{opt}}$ given by
\vspace*{-2mm}
\begin{align}
    P_{\mathrm{opt}}    =\,\bigcup\limits_{i}\Bigl\{(j_{\mathrm{opt}},l_{\mathrm{opt},i})_{x}|x=1,\ldots,x_{\mathrm{max},i}\Bigr\}.
\end{align}

\noindent For each pair of levels in $P_{\mathrm{opt}}$, we can then perform a two-level rotation. In principle, one has the freedom to adjust the angles of all possible rotations specified by the pairs in $P_{\mathrm{opt}}$ individually to approach the desired target energy. For instance, one may perform individual operations one after the other with maximal angles $\theta=\tfrac{\pi}{2}$ until one is close enough to the target energy so that the adjustment of a single rotation angle to a suitable value $0<\theta<\tfrac{\pi}{2}$ reaches the value $\epsilon$ exactly. Irrespective of the order or particular distributions of these angles, the resulting energy variance is always the same, as long as the target average energy is reached. However, we note that different choices of these angles may lead to different results as far as other figures of merit for the charging process are concerned. In particular, adjusting the angles individually can result in discontinuities of the work fluctuations associated with the protocol as a function of $\epsilon$ at the transition points between the cases (i)-(iv) above.

Here, we therefore choose a common rotation angle $\theta$ for all pairs of levels in $P_{\mathrm{opt}}$. If the target energy can be reached by a suitable choice of $\theta$, then one selects this value, performs the operations, updates the probability distribution and the protocol is concluded. If the chosen rotation does not reach the target energy, one carries out the rotations with $\theta=\tfrac{\pi}{2}$ for all pairs in $P_{\mathrm{opt}}$, updates the probability distribution and generates the corresponding new set $P'_{\mathrm{opt}}$. This procedure is continued until the target energy is reached.

Not only can we obtain the optimal-precision protocol in this way, we also observe that the final probability distributions change continuously at the transition points at the values $\epsilon\,=\,(n+1)\,\frac{1}{2}$ for $n\in \mathds{N}$. As described, this approach for obtaining the optimal-precision protocol is independent of the degeneracy of the energy levels. Therefore, it can be applied for any system whose Hamiltonian can be written in the form of Eq.~(\ref{eq:Hamiltonian Nqubit}), for instance, $N$ identical qudit systems with equally spaced Hamiltonians.

In panels (a) and (b) of Fig.~\ref{fig:qudit}, the results of the protocol are illustrated for a $5$-dimensional system ($d=5$, $N=1$) and a $4$-qubit system ($d=2$, $N=4$, same $\omega$), respectively, showing the minimal unitarily achievable energy variance $V$ (in units of $\omega^{2}$) as a function of the energy input $\Delta\epsilon$. In both case, the systems are initially in thermal states with respect to their local Hamiltonians with temperatures (in units $\hbar\omega/k_{\textup{B}}$) from $0.1$ to $1$ in steps of $0.1$. From Ref.~\cite{FriisHuber2018}, it can easily be seen that the fundamental variance limit for pure initial states (zero temperature, ground state) is
\begin{align}
    V\,=\, \omega^2
    \left(\Delta\epsilon-\lfloor\Delta\epsilon\rfloor\right)\,\left(\lceil \Delta\epsilon\rceil-\Delta\epsilon\right).
    \label{eq:min var pore}
\end{align}
{From this formula we conclude} that if the energy input is an integer multiple of $\omega$, the variance vanishes and its maximum value is an integer multiple of $\frac{\omega}{2}$. In both plots, we see that for temperatures close to zero (in our case for $T=0.1$), the minimum variance is well approximated by Eq.~(\ref{eq:min var pore}). When the initial temperature is raised, the exact periodic behaviour of the minimal variance for the harmonic oscillator discussed in Ref.~\cite{FriisHuber2018} disappears. Instead, one now observes that the finite system dimension leads to a symmetric behaviour with respect to the point where the input energy $\Delta\epsilon$ is exactly half-way between its minimal and maximal values as specified in~(\ref{eq:input energy range}). The positions of the local minima and local maxima of the minimal variance $V(\Delta\epsilon)$ for each fixed Hamiltonian depend on the initial temperature.

Here, it is interesting to observe two competing effects when comparing panels (a) and (b): For zero initial temperatures, the $d$-dimensional system and a $(d-1)$-qubit system appear as equivalently useful work-storage devices as far as the maximally achievable charge $\Delta\epsilon_{\mathrm{max}}$ and the minimal achievable variance $V(\Delta\epsilon)$ are concerned. However, for higher temperatures, certain trade-offs become evident. On the one hand, the degeneracy in the energy-level structure of the multi-qubit system means that $\Delta\epsilon_{\mathrm{max}}$ is decreased more strongly with increasing $T$ than for a single $d$-level system. On the other hand, the degeneracy also means that the global minimum, $\min_{\Delta\epsilon}V(\Delta\epsilon)$, remains at smaller values as compared with the $d$-level system.


\section{Fundamental work fluctuation limit for arbitrary temperature}\label{sec:fluctuation}

We now turn our attention to the work fluctuations during the charging process of a general finite-dimensional system. For {ease of presentation in this section we write the system Hamiltonian as} $H=\sum_{n=1}^d E_n\,\ketbra{n}{n}$, where $E_n$ are sorted in increasing order{. T}he work fluctuations $\Delta W$ are obtained from Eq.~(\ref{eq:def. fluct.}), and we again focus on two particular cases (cf. Sec.~\ref{sec:framework}): a qudit with equally-spaced Hamiltonian, and a syste{m} consisting of $N$ identical qubits, each with the same local Hamiltonian.\\

\begin{figure}[t!]
\label{fig:Initial and final distribution}
(a)\,\includegraphics[width=0.4\textwidth,trim={0cm 3.5mm 0cm 0mm}]{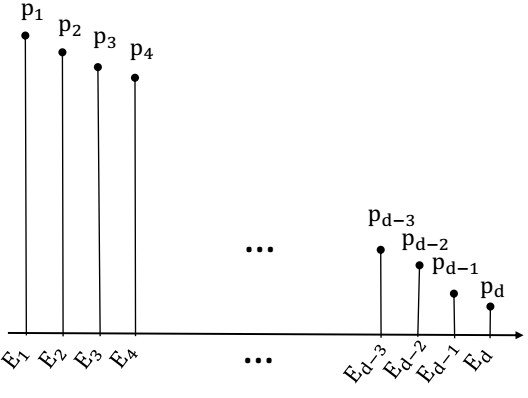}\\
(b)\,\includegraphics[width=0.4\textwidth,trim={0cm 3.5mm 0cm 0mm}]{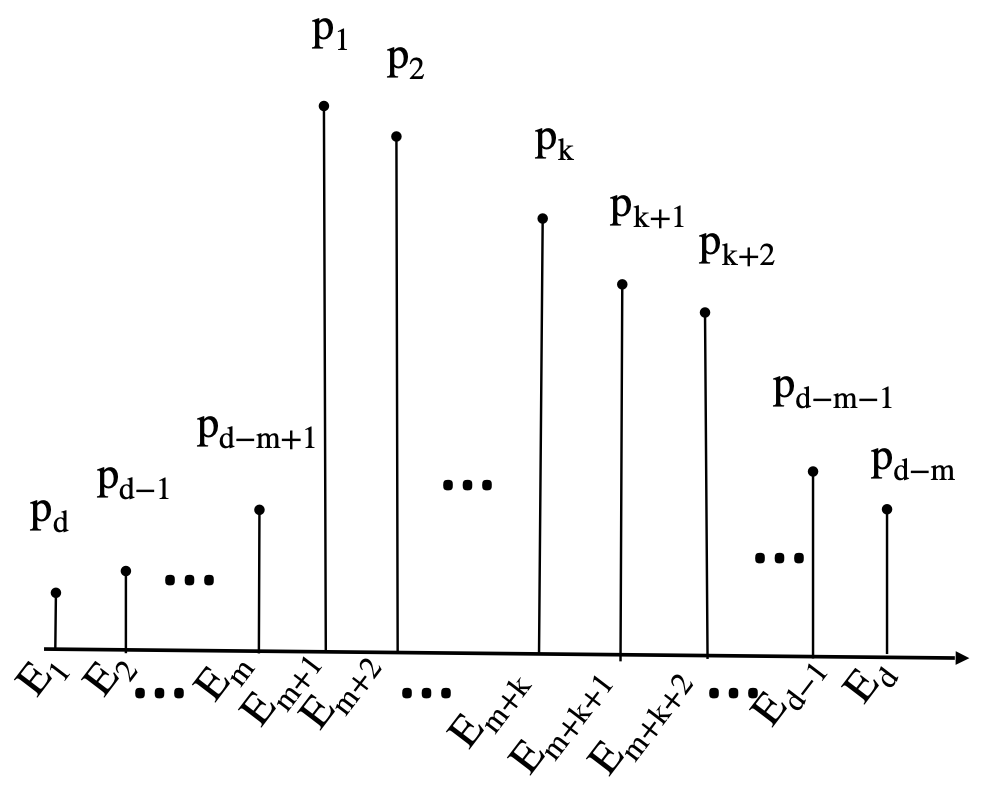}\\
\caption{\textbf{Ansatz for minimizing work fluctuations}.
To illustrate the working principle of the proposed protocol to suppress work fluctuations during the charging process as much as possible, the initial and final probability distributions corresponding to the diagonal of the battery-system density operator with respect to the energy eigenbasis are shown for increasing the energy of the system by $\Delta \tilde{\epsilon}({m},{k})$.}
\end{figure}

\noindent\textbf{Qudit protocol}.\ For this scenario, we first make an ansatz for increasing the system's energy by formulating a protocol that consists of a sequence of permutations of populations of pairs of energy levels, as illustrated in Fig.~\ref{fig:Initial and final distribution}. The sequence is specified by a free parameter, $m$. The parameter is a non-negative integer and the energy differences that can be achieved by all sequences labelled by $m$ hence form a discrete set (for fixed initial temperature). As a consequence, this ansatz cannot reach arbitrary final energies, and so further fine-tuning is required in subsequent modifications of the protocol.\\[-1mm]

\noindent\textbf{Qudit protocol{\textemdash}Phase One}.\ To increase the system's energy starting from an initial thermal state we start with a sequence of permutations that move the populations of the $m$ energy levels with {the} largest energies to the $m$ {levels with the lowest energy}. The starting point for this sequence is to exchange the smallest population $p_{d}$ of any energy level (which initially corresponds to the largest eigenvalue $E_{d}$ of the Hamiltonian) with the population of the adjacent energy level with lower energy $E_{d-1}$. Taking the new population of this, now second largest energy level, as a new starting point, we exchange it with the population of its adjacent lower-energy neighbour $E_{d-2}$ and repeat this procedure, step-by-step, until we reach the lowest energy level $E_{1}$. In the process, all other populations are shifted upwards by one energy level.

Before we proceed with the remaining $m-1$ levels, let us motivate this procedure by considering the limiting case where $E_{d}\rightarrow\infty$, i.e., a harmonic oscillator (since we assume equally spaced energy levels). In this limit, the result of the first sequence of pairwise permutations is a shift of the average energy by exactly one unit. At the same time, the original population $p_{d}$ of the only energy level experiencing a shift different from one unit vanishes in this limit, $\lim_{E_{d}\rightarrow\infty}p_{d}=0${. As a result} the work fluctuations associated to this {limiting-case} process vanish. For a finite-dimensional system with finite energy gaps we have $p_{d}>0$ and so the overall shift in average energy will be less than one, and the fluctuations will not vanish but will be proportional to the smallest population $p_{d}$. For a harmonic oscillator, the procedure can be repeated any number of times to raise the average energy by any non-negative integer {number} of units with vanishing work fluctuations~\cite{FriisHuber2018}.

For finite dimensions and energy gaps we can also repeat the procedure carried out above for $p_{d}$, now for the second-smallest population $p_{d-1}$ but stop when it has reached the second-smallest energy level $E_{2}$, and similarly for all of the remaining $m-2$ populations among the smallest $m$ populations. As a result, the smallest $m$ populations end up as the populations, in increasing order, of the smallest $m$ energy levels. It is clear that the resulting density matrix at the end of this step is diagonal with respect to the energy eigenbasis. The new probability weights with respect to this basis are given by
\begin{align}
\tilde{p}_n\,=\,
\begin{cases}
 p_{d-n+1}    &n \,\leq\, m  \\[0.5mm]
  p_{n+m}  & n\,\geq\,m
\end{cases}.
\end{align}
Here, $m$ remains as a free parameter that allows us to adjust the average energy of the final state. At this point, the total energy $\tilde{\epsilon}(m)$ of the system reads
\begin{equation}
\tilde{\epsilon}(m)\,\omega =\,\sum_{n=1}^{k}E_{n}\, p_{d-n+1}+\sum_{n=1}^{d-m}E_{n+m}\, p_{n},\label{eq:av. en fluct. qudit}
\end{equation}
where $m\in\{0,1,\hdots, d-1\}$. Note that if $m=0$, the first sum in Eq.~\eqref{eq:av. en fluct. qudit} does not contribute to the average energy.

To better understand the proposed protocol, let us now again make a comparison with the optimal protocol for a harmonic oscillator from~\cite{FriisHuber2018} by considering the simple case of a qudit system with equally spaced energy levels. In this case, by considering $E_n\,=\, (n-1)\,\omega$, one can rewrite the corresponding energy {increase} of the system for given $m$ in Eq.~(\ref{eq:av. en fluct. qudit}) as
\begin{align}
\tilde{\epsilon}(m)
&\,=\,\epsilon_0 +m+\sum_{n=d-m+1}^{d}\left(d-2n-m\right)p_{n}\,.
\end{align}
In general, we know that $\Delta \tilde{\epsilon}(m)\,=\,\tilde{\epsilon}(m)-\epsilon_0$ is a monotonically increasing function of $m$ that takes discrete values. Therefore we are not able to cover all possible energy increases $\Delta \epsilon$ using this approach. To remedy this, we first need to find a parameter $m$ for a given $\Delta \epsilon$ which minimizes $\Delta \tilde{\epsilon}_{\textup{I}}(m):=\Delta \epsilon-\Delta\tilde{\epsilon}(m)$ under the constraint that $\Delta \tilde{\epsilon}_{\textup{I}}(m)\geq0$, such that
\begin{align}
\Delta \tilde{\epsilon}_{\textup{I}}(\tilde{m})\,=\,\min_{m}\bigl\{
\Delta \tilde{\epsilon}_{\textup{I}}(m)
\,|\,\Delta \tilde{\epsilon}_{\textup{I}}(m)
\geqslant 0\bigr\}.
\label{En mk}
\end{align}
That is, we find the parameter for the protocol described above for which the energy that is reached is as close as possible but still smaller than the desired target energy.\\

\noindent\textbf{Qudit protocol{\textemdash}Phase Two}.\ During the second phase of the protocol we then compensate for the missing energy $\Delta \tilde{\epsilon}_{\textup{I}}(\tilde{m})$ by transforming the probability weights associated with the levels $n=\,\tilde{m}+1,\tilde{m}+2,...,d $ according to
\begin{align}
&\left(\tilde{p}_{\tilde{m}+1},\,\tilde{p}_{\tilde{m}+2},\hdots, \,\tilde{p}_{d-1},\,\tilde{p}_{d} \right)\,=\, \left(p_{1},\,p_{2},\hdots, \,p_{d-\tilde{m}-1},\,p_{d-\tilde{m}} \right)\nonumber\\[1mm]
&\ \mapsto~~\Bigl(p_{1}\cos^{2}\!\theta\, +p_{d-{\tilde{m}}}\sin^2 \theta,\,p_{2}\cos^{2}\!\theta\, +p_{1}\sin^2 \theta,\,\hdots\,,\nonumber\\
&\,p_{d-\tilde{m}-1}\cos^{2}\!\theta\, +p_{d-{\tilde{m}}-2}\sin^2 \theta,\,p_{d-\tilde{m}}\cos^{2}\!\theta\, +p_{d-{\tilde{m}}-1}\sin^2 \theta
 \Bigr).
\end{align}
These pairwise rotations (of the largest and smallest, second-largest and second smallest, etc.) of the probabilities in the considered subset about a common angle $\theta$ result in an energy change with respect to the first phase of the protocol given by
\begin{align}
\Delta \tilde{\epsilon}_{\textup{I}}(\tilde{m})\,=\, \left(\tilde{\epsilon}(\tilde{m}+1)-\tilde{\epsilon}(\tilde{m})\right)\,\sin^{2}\!\theta.
\label{En rotation}
\end{align}
Due to the minimization in Eq.~(\ref{En mk}) the rotation between these levels must be sufficient to reach the desired energy. Using Eqs.~(\ref{En mk}) and~(\ref{En rotation}), the required angle for the rotation to compensate for the rest of the energy is obtained from
\begin{align}
\theta_{\textup{I}}(\tilde{m})\,=\, \arcsin \sqrt{\frac{\Delta \tilde{\epsilon}_{\textup{I}}(\tilde{m})}{\left(\tilde{\epsilon}(\tilde{m}+1)-\tilde{\epsilon}(\tilde{m})\right)}}
\end{align}
\begin{align}
(\Delta W)^2\,&=\,(\Delta W)_{  \leqslant d-\tilde{m}-1}^2+(\Delta W)_{ d-\tilde{m}}^2+(\Delta W)_{  \geqslant d-\tilde{m}+1}^2\nonumber\\[1mm]
& =\,\sum_{n=1}^{d-\tilde{m}-1}p_n\, \big[\cos^{2}\!\theta\, \,(E_{n+\tilde{m}}-E_{n}-\omega\,\Delta\epsilon)^2 \nonumber\\[1mm]
& \ \ \ +\, \sin^{2}\!\theta\, \,(E_{n+\tilde{m}+1}-E_{n}-\omega\,\Delta\epsilon)^2    \big]\label{eq:fluc d dim}\\
&\ \ \ +\,p_{ d-\tilde{m}}\, \big[\cos^{2}\!\theta\, \,(E_{d}-E_{{ d-\tilde{m}}}-\omega\,\Delta\epsilon)^2 \nonumber\\[1mm]
&\ \ \ + \, \sin^{2}\!\theta\, \,(E_{\tilde{m}+1}-E_{{ d-\tilde{m}}}-\omega\,\Delta\epsilon)^2 \big] \nonumber\\[1mm]
&\ \ \ + \,
\sum_{n=d-\tilde{m}+1}^{d}p_n \, (E_{d-n+1}-E_{n}-\omega\,\Delta\epsilon)^2.
\nonumber
\end{align}

\begin{figure*}[t!]
(a) \includegraphics[width=0.45\textwidth,trim={0cm 0mm 0cm 0mm}]{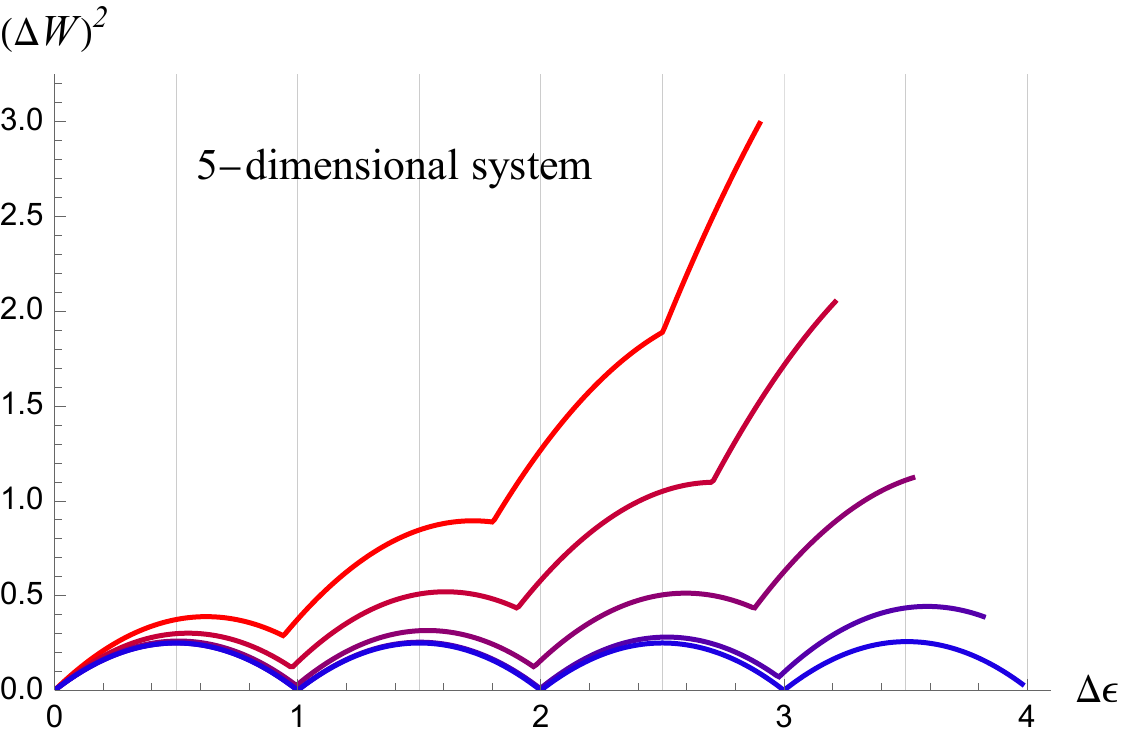}
(b) \includegraphics[width=0.45\textwidth,trim={0cm 0mm 0cm 0mm}]{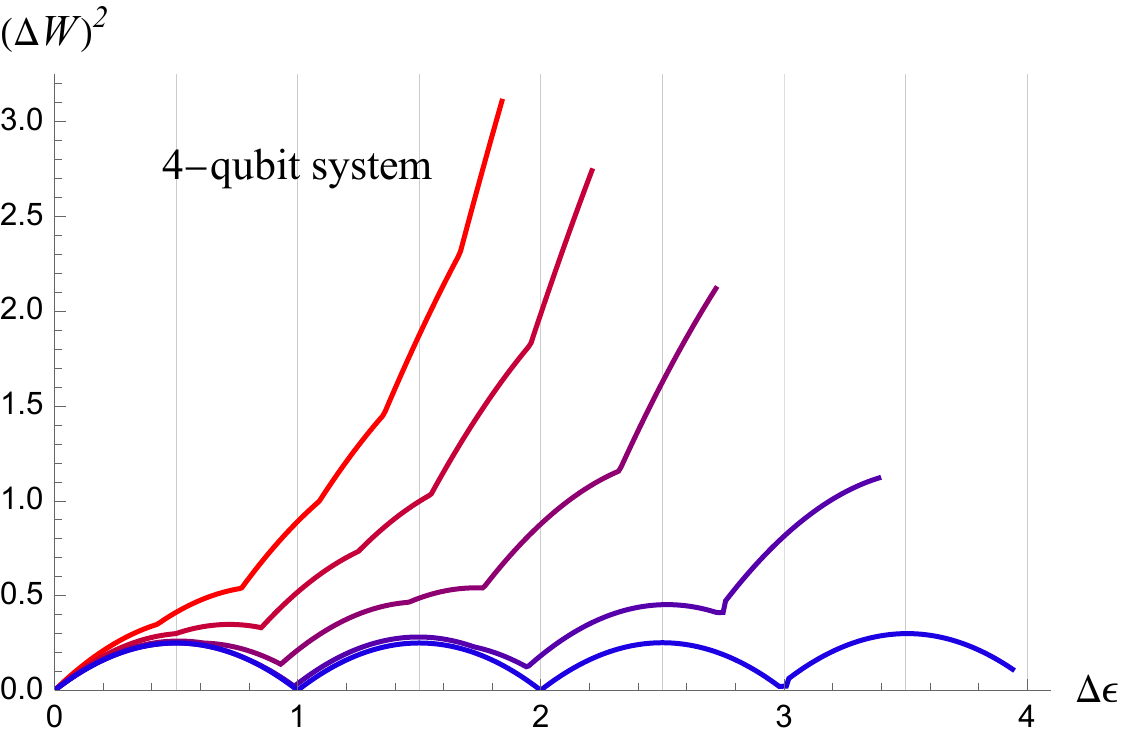}
\caption{\textbf{Proposed protocol for minimizing work fluctuation}. The curves show the work fluctuations $(\Delta W)^{2}$ (in units of $\omega^{2}$) obtained from the protocol proposed in Sec.~\ref{sec:fluctuation} as a function of the input energy $\Delta\epsilon$ (in units of $\omega$) for (a) a $5$-dimensional quantum system, and (b) for a $4$-qubit system,
for temperatures (in units of $\hbar \omega/k_{\textup{B}}$) from $T = 0.2$ (blue) to $T=1$ (red) in steps $0.2$. For the chosen parameters (input energies, temperatures, and system dimension/qubit number) results from a numerical optimization of the work fluctuations are included but cannot be distinguished by the naked eye from the curves resulting from our proposed protocol, supporting the conclusion that the latter is at least a close approximation of the true optimum.}
\label{fig:Opt. fluctuation}
\end{figure*}

\noindent\textbf{Performance of the protocol}.\ We now wish to examine how well, in particular, how close to the optimum, the proposed protocol performs. To do this, we once again consult the case of the harmonic oscillator{. H}ere this corresponds to the limit $d\to \infty$ of a $d$-dimensional system with equally spaced energy levels. For such a situation, the protocol described above reduces to the protocol from \cite{FriisHuber2018} which was shown to minimize the work fluctuations for fixed energy input: {T}here, $\tilde{m}$ is fully determined by $\lfloor\Delta \epsilon\rfloor$ which leads to a minimization of all terms except {for} the last sum in Eq.~(\ref{eq:fluc d dim}). One therefore only needs to investigate the {contribution from} last term in the work fluctuation on its own. The crucial step of {the protocol from}~\cite{FriisHuber2018} is then to realize that the probability weights in th{e mentioned} sum have all been shifted from some level with label $n$ to another level with label $m<n$, whose energy gap $E_m-E_n$ might diverge{. I}n particular, $(E_m-E_n-\omega\,\Delta\epsilon)^{2}$ can diverge, but the associated probabilities go to zero much faster (exponentially with $E_{m}-E_{n}$), $\lim_{n\to \infty}\, p_n=0$. So the last sum containing the probability weights $p_{n}$ for $n=d-\tilde{m}+1,\ldots,d$ in Eq.~(\ref{eq:fluc d dim}) vanishes in the case of the harmonic oscillator.

In the finite-dimensional case we consider here, however, all energy gaps and all weights $p_n$ remain finite and give a nonzero contribution to the last sum in the work fluctuation in Eq.~(\ref{eq:fluc d dim}). In this case, it is generally complicated to confirm the optimality of the constructed protocol. However, in the regime of small temperatures (with respect to the maximum energy gap) it can be argued that the last sum in Eq.~(\ref{eq:fluc d dim}) is negligible for sufficiently small input energy and the protocol thus (at least) approximates the true optimal protocol. Furthermore, it is obvious that the optimal fluctuation protocol for pure states of any qudit or $N$-qubit system consist of partial shifts of the ground-state population to levels $\floor{\Delta  \epsilon}$  and $\ceil{\Delta \epsilon}$ such that the corresponding energy is equal to $\Delta \epsilon$, which is compatible with the optimal precision protocol.

To check our approach quantitatively we have numerically calculated the work fluctuations arising from the proposed protocol for a qudit system with varying input energy and for different initial temperatures, and compared them with a brute-force numerical search for the corresponding optimal protocol, as illustrated in Fig.~\ref{fig:Opt. fluctuation}~(a) for $d=5$. For the parameters we have considered the numerical differences between our protocol and the optimum are non-zero but invisible to the naked eye. We therefore now continue with an analysis of the properties of our proposed near-optimal protocol.\\[-1.5mm]

In the regime where the input energy is small compared to the inverse temperature of the initial state or compared with the system dimension, one can observe almost periodic behaviour of the fluctuations as functions of the input energy, approximating the periodic behaviour of the harmonic-oscillator case \cite{FriisHuber2018}. In particular, if the system is infinite-dimensional or in a pure state, one may reach arbitrary input energies while keeping the fluctuations bounded from above by $\Delta W\leq \frac{\omega}{2}$, and for all input energies that are integer multiples of $\omega$ one can reach $\Delta W=0$.\\[-1.5mm]

However, in finite-dimensional systems with finite temperatures, the periodic behaviour and local minima and maxima gradually disappears with increasing input energy. The fluctuation becomes a monotonically increasing function of $\Delta\epsilon$ as one approaches the maximum of the energy that can be transferred to the system unitarily.\\

\noindent\textbf{Protocol for $N$ qubits}.\ So far, we have discussed fluctuations for a single-qudit system with equally spaced energy levels. But, as illustrated in Fig.~\ref{fig:Opt. fluctuation}~(b), one can also observe the same qualitative features when applying the proposed protocol to a system of $N$ non-interacting qubits.
To this end, we relabel the eigenvalues and eigenstates of the $N$-qubit Hamiltonian via a variable that we have already encountered in the description presented at the end of Sec.~\ref{sec:nqubit}, i.e. $H_{\textup{tot}}=\sum_{s=1}^{2^N}E_s \ketbra{s}{s}$, where $s\equiv s(m, i_m)= \sum_{i=0}^{m-1} g(i)+i_m-\delta_{m,0}$. With this, we can easily apply the protocol derived for qudit systems above.\\[-1.5mm]

In Fig.~\ref{fig:Opt. fluctuation} we showcase the performance of this protocol for a 5-dimensional system and for a 4-qubit system in terms of the work fluctuations as functions of the input energy for different temperatures. These plots illustrate that for small input energies one observes the discussed approximately periodic behaviour (as one encounters for the harmonic oscillator \cite{FriisHuber2018}) in finite-dimensional systems. It is clear that the periodic behavior is a result of transferring probability weights close to zero from high-energy levels to low-energy level with a negligible fluctuation cost. It tell us that the number of periodic cycles is given by the number of energy levels with negligible probability weights. However, in contrast to the harmonic-oscillator case, the work fluctuations observed in finite-dimensional systems strongly increase for larger energy inputs and any residual periodicity is gradually lost.


\section{Comparison of the protocols}
\label{sec:protocol comparison }

\begin{figure*}[ht!]
(a) \includegraphics[width=0.45\textwidth,trim={0cm 0mm 0cm 0mm}]{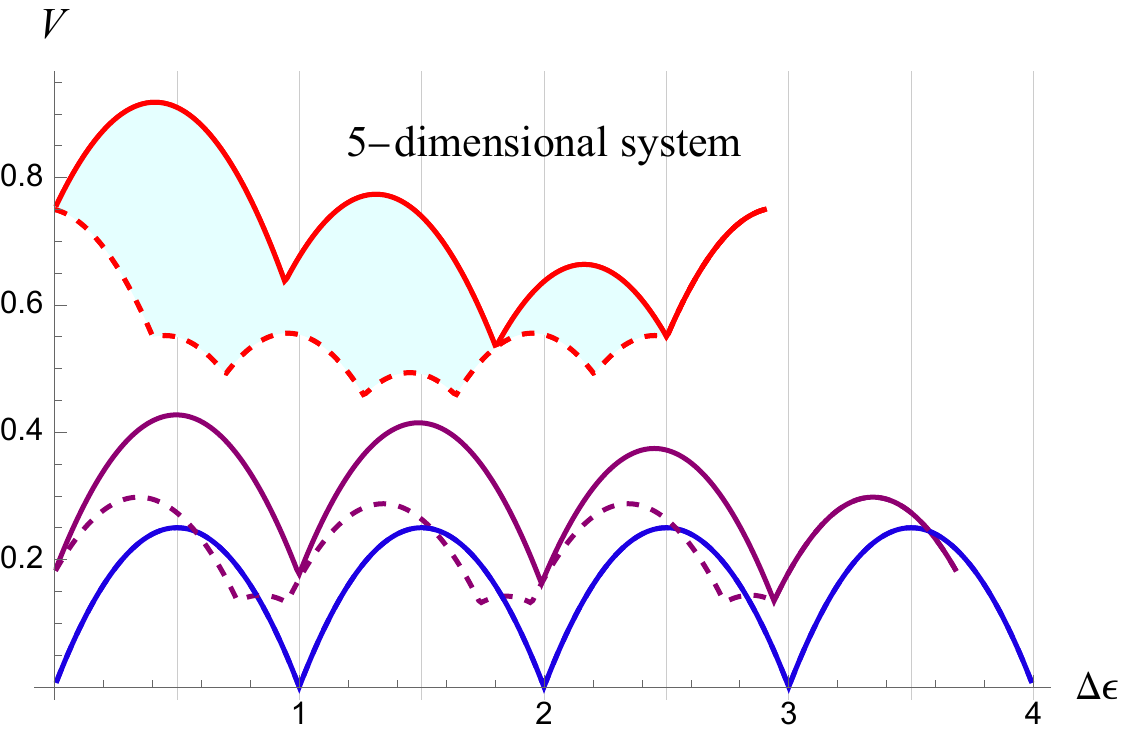}
(b) \includegraphics[width=0.45\textwidth,trim={0cm 0mm 0cm 0mm}]{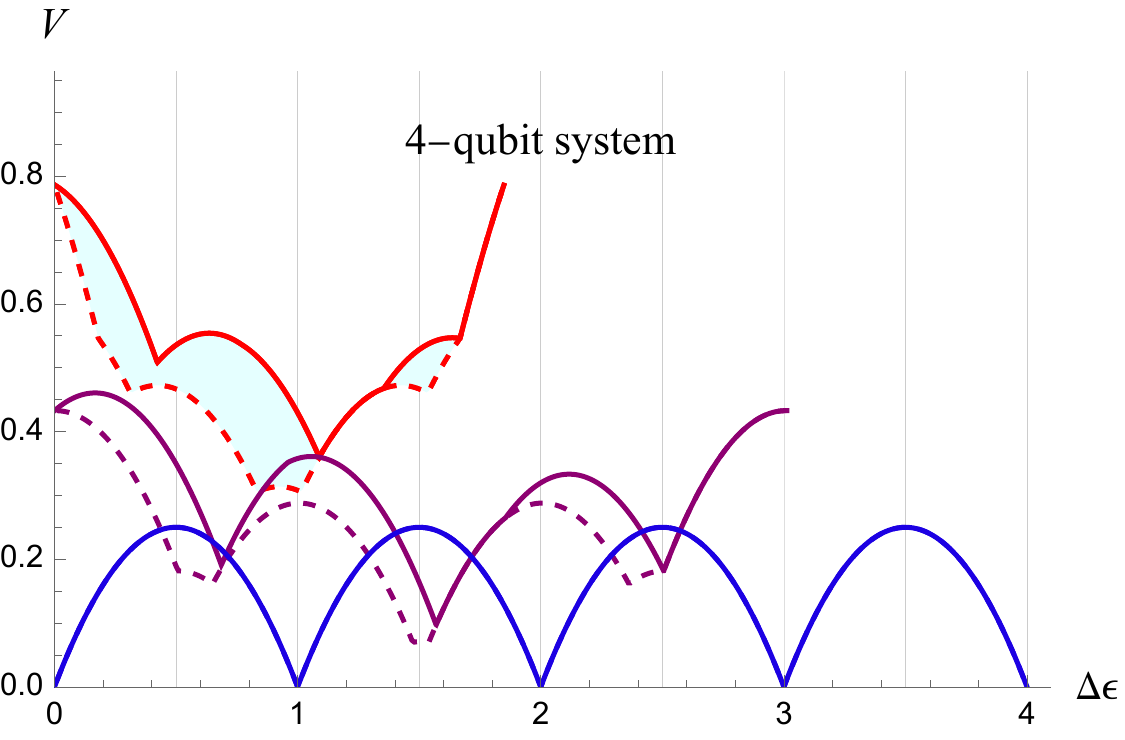}\\
(c) \includegraphics[width=0.45\textwidth,trim={0cm 0mm 0cm 0mm}]{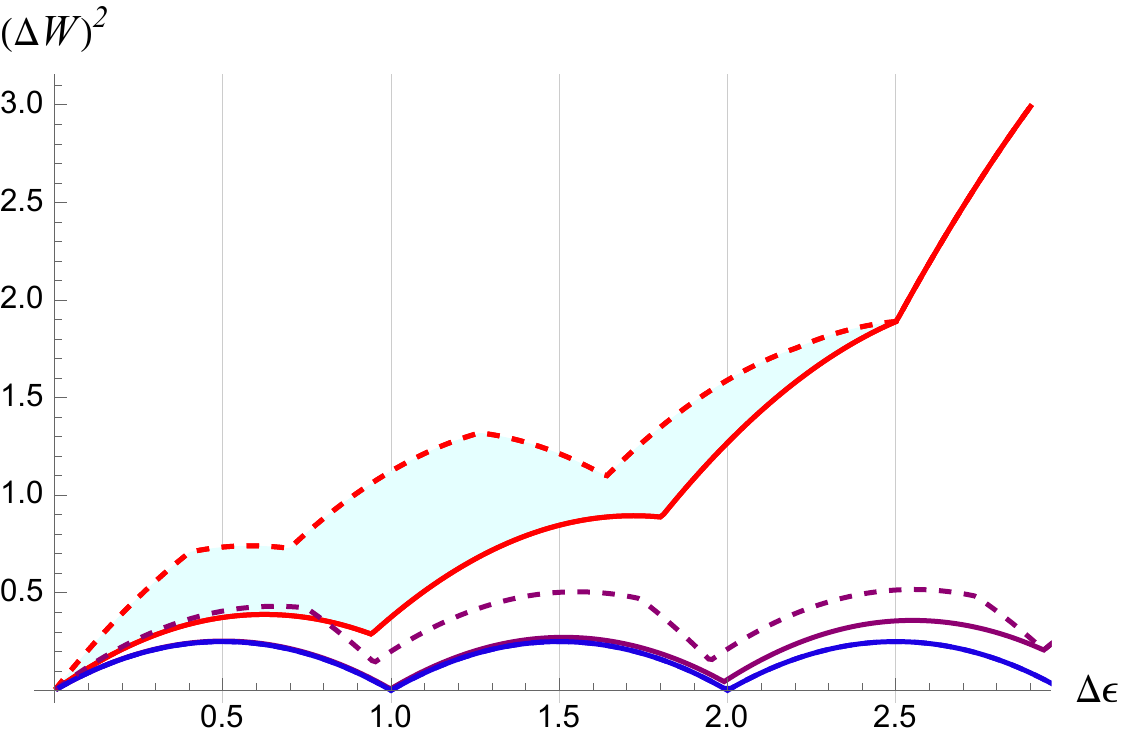}
(d) \includegraphics[width=0.45\textwidth,trim={0cm 0mm 0cm 0mm}]{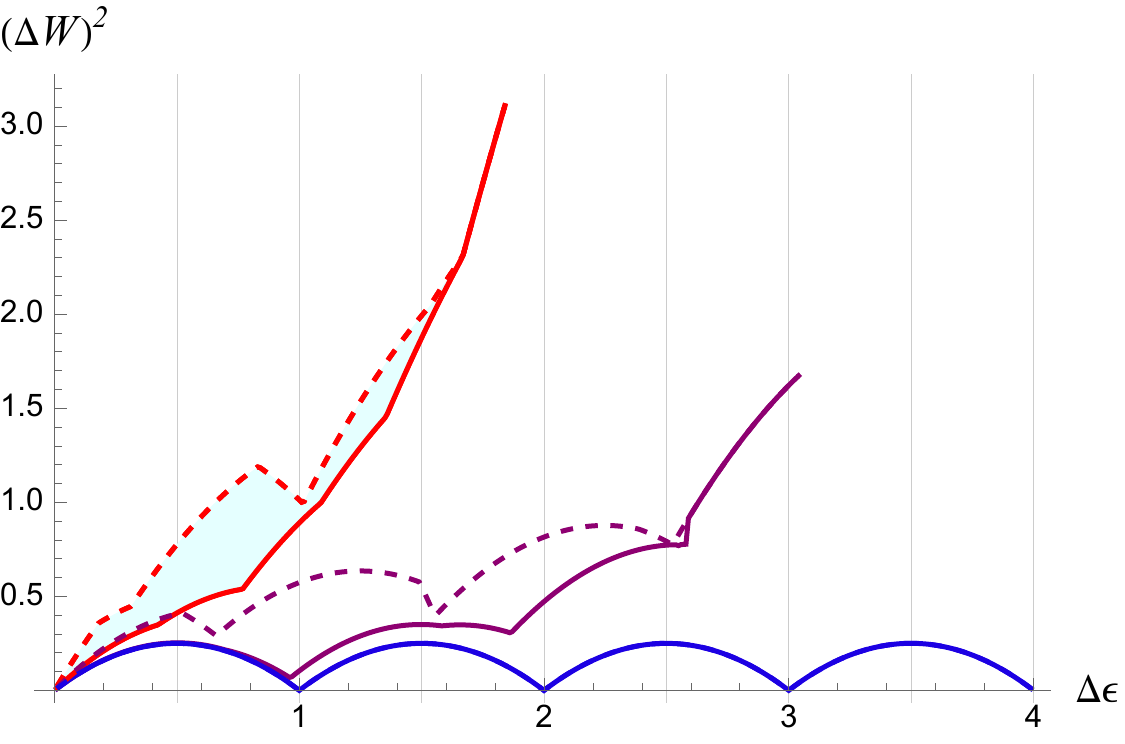}
\caption{\textbf{Comparison of precision and work fluctuations in both protocols}. We compare the protocol that minimizes the variance ({dashed} lines) and the {proposed} protocol aiming to minimize the work fluctuations (solid curves) by calculating both the variance [upper panels, (a) and (b)] and the work fluctuations [lower panels, (c) and (d)] in units of $\omega^{2}$, for both protocols, for both a $5$-dimensional [(a) and (c)] and for a $4$-qubit [(b) and (d)] system as functions of the energy input $\Delta \epsilon$ (in units of $\omega$)
for temperatures (in units of $\hbar \omega/k_{\textup{B}}$) for $T = 0.1$ (blue), $T=0.5$ (purple), and $T=1$ (red).}
\label{fig:protocol comparison}
\end{figure*}

After introducing the protocols that minimize the variance and work fluctuations in Secs.~\ref{sec:Precision} and~\ref{sec:fluctuation}, respectively, we now wish to investigate the trade-offs between the two quantities by checking how well the protocols designed for minimizing one of them perform in terms of the respective other quantity. For pure initial states, it is known~\cite{FriisHuber2018} that the protocols coincide, and so both the variance and the fluctuations can be minimized simultaneously. For finite temperatures, the protocols generally do not coincide. In Fig.~\ref{fig:protocol comparison} we therefore show the energy variance and the work fluctuations associated to the optimal-precision and the fluctuation protocol as functions of the invested energy $\Delta \epsilon$ for both systems of interest and for different initial temperatures. However, we note that no further optimization of either protocol is carried out here in order to further adapt it to the respective second figure of merit.

As expected, the protocols coincide for small temperatures (e.g., as can be seen from the blue curves on the very bottom of all panels in Fig.~\ref{fig:protocol comparison}), but the differences between the protocols become apparent for increasing temperatures. For the example shown in Fig.~\ref{fig:protocol comparison}, the differences between the variances obtained from the fluctuation protocol and from the optimal-precision protocol [panels (a) and (b)] are smaller on average than the differences in the work fluctuations from using the optimal-precision protocol as opposed to the fluctuations protocol [panels (c) and (d)] when taking into account the different units on the vertical axes of panels (a) and (b) with respect to (c) and (d).

For example, for the highest shown temperature ($T=1$, red curves), the ratio $d_{\mathrm{max}}\suptiny{0}{0}{V}/d_{\mathrm{max}}\suptiny{0}{0}{\Delta W}$ of the maximum difference $d_{\mathrm{max}}\suptiny{0}{0}{V}:=\max\limits_{\Delta\epsilon}(V\subtiny{0}{0}{\mathrm{opt.}\Delta W}-V\subtiny{0}{0}{\mathrm{opt.}V})$ between the function values in (a) and the difference $d_{\mathrm{max}}\suptiny{0}{0}{\Delta W}:=\max\limits_{\Delta\epsilon}\bigl((\Delta W\subtiny{0}{0}{\mathrm{opt.}V})^{2}-(\Delta W\subtiny{0}{0}{\mathrm{opt.}\Delta W})^{2}\bigr)$ between the function values in (c) is $d_{\mathrm{max}}\suptiny{0}{0}{V}/d_{\mathrm{max}}\suptiny{0}{0}{\Delta W}=\,0.48$ $<1$, and the ratio of the areas enclosed between the respective curves [shaded red areas in (a) and (c)] is $A\suptiny{0}{0}{V}/A\suptiny{0}{0}{\Delta W}=\,0.51\,<1$.

From the examples we have considered it thus appears that it is better on average to employ the optimal-precision protocol if one wishes to keep both the variance and fluctuations low with equal priority. At the same time, we observe that the differences between the protocols become less pronounced when the system under consideration offers more degeneracy in its energy levels: The ratios for the highest-temperature curves for the $4$-qubit system in (b) and (d) evaluate to $d_{\mathrm{max}}\suptiny{0}{0}{V}/d_{\mathrm{max}}\suptiny{0}{0}{\Delta W}=\,0.35$ and $A\suptiny{0}{0}{V}/A\suptiny{0}{0}{\Delta W}=\,0.34$, respectively.

However, for both figures of merit we see that the differences between the two protocols become less pronounced, and partially even vanish altogether, when the input energy reaches sufficiently large values. Therefore, if we want to almost fully charge the quantum battery, there is no apparent priority to use one of the mentioned protocols rather than the other. We attribute the latter convergence of the protocols to the fact that the finite system dimension severely limits the possible options for protocols to differ when the input energy is large.

In our discussion up to this point, we have considered unitary operations that act globally on the entire Hilbert space for both types of considered systems. However, in particular for the $N$-qubit case, one might argue that non-local operations, i.e., operations that act jointly on several (or all) qubits and may entangle them, might be considerably more difficult to implement. We therefore briefly want to examine the role of local operations for the task at hand in the next section.


\section{Local vs non-local operations}\label{sec:non-local}

Here we are interested in investigating the role of non-local operations in charging multipartite quantum batteries. In particular, we focus on the work fluctuations and the charging precision when unitarily transferring energy into a system comprising $N$ non-interacting qubits with equal energy gaps. To do so, we compare the generically non-local processes that arise from applying the protocols considered in Secs.~\ref{sec:Precision} and~\ref{sec:fluctuation} to $N$-qubits with two alternatives: first, with a specific local process that we refer to as \emph{symmetric local charging}, and second, with the numerically obtained optimal local processes for work fluctuations and the charging precision.

The motivation for this comparison are twofold: Apart from the observation that local operations might be easier to implement than non-local ones, the fact that local operations do not create any correlations between the qubits can help us to better understand the role of correlations in charging processes. Meanwhile, it is expected that, much like in the reverse process of (unitary) work extraction~\cite{PerarnauLlobetUzdin2019}, the increase of the average energy via local unitaries performs worse than the corresponding global protocols.\\[-2mm]

\begin{figure*}[ht!]
(a) \includegraphics[width=0.45\textwidth,trim={0cm 0mm 0cm 0mm}]{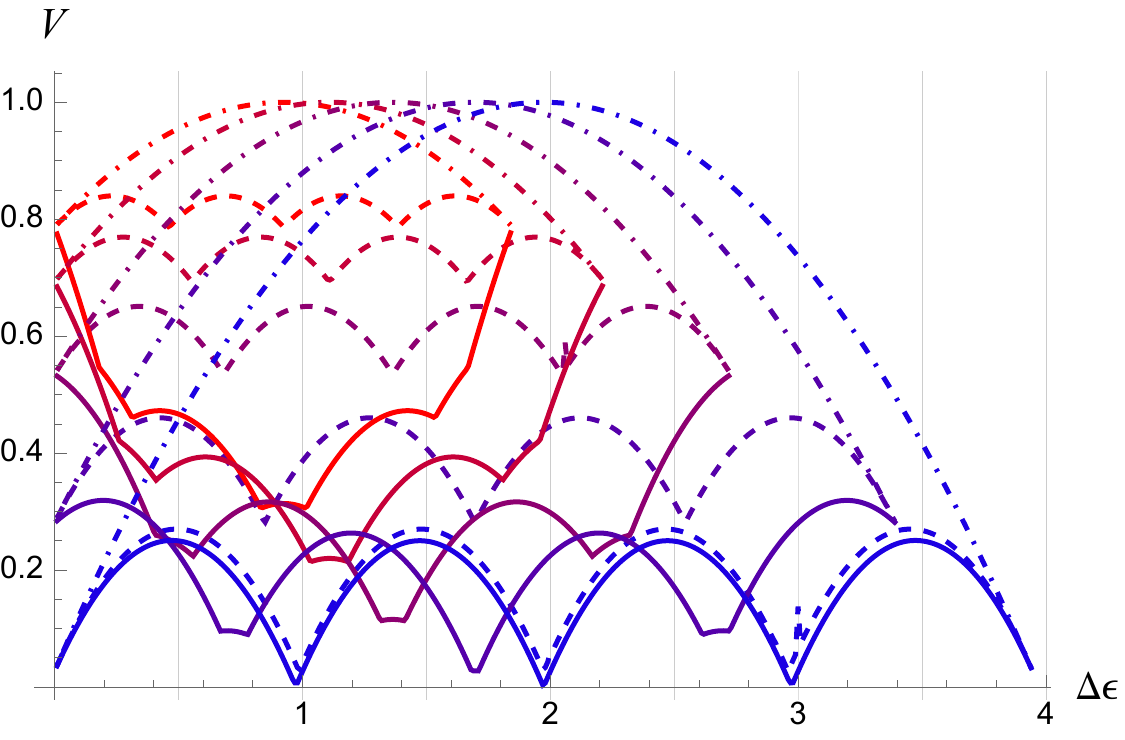}
(b) \includegraphics[width=0.45\textwidth,trim={0cm 0mm 0cm 0mm}]{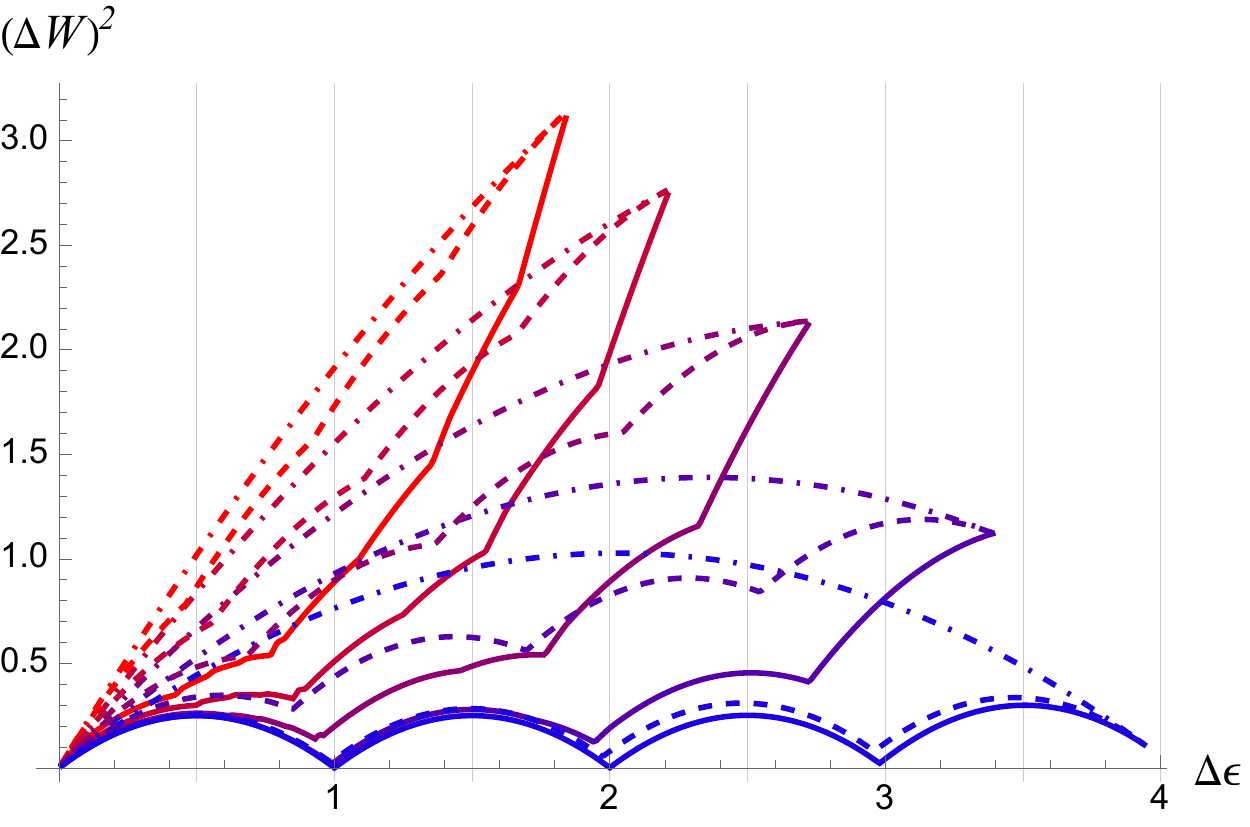}\\
\caption{\textbf{Local vs non-local operations}.
The variance $V$ and work fluctuations $(\Delta W)^{2}$ (both in units of $\omega^{2}$) obtained from the protocols (conjectured to) minimize the respective quantities are plotted as solid lines in (a) and (b), respectively, as functions of the energy input $\Delta\epsilon$ (in units of $\omega$) of a $4$-qubit system for temperatures (in units of $\hbar \omega/k_{\textup{B}}$) from $T = 0.2$ (blue) to $T=1$ (red) in steps $0.2$. The dot-dashed and dashed lines represent the respective quantities obtained from the local charging processes, and are the analytically worst-case and numerically best-case local processes for both of the quantities, respectively.}
\label{fig:Nqubit loc vs nonloc}
\end{figure*}

\noindent{\textbf{Local charging processes}.}\ To study the charging process in terms of local operations, we consider the $N$-qubit system as a many-body quantum battery whose energy increase is brought about via local unitary operations $U_{\mathrm{loc}}$, i.e., $U_{\mathrm{loc}}\,=\,\bigotimes_{i=1}^N\, U_i$, where $U_i$ acts unitarily on the Hilbert space of the $i$th qubit. If we assume that the initial state of the total system is an uncorrelated thermal state at inverse temperature $\beta$, then there also is no correlation in the final state. We then have
\begin{align}
    \varrho_{\textup{tot}}&= U_{\mathrm{loc}}\, \tau(\beta)^{\otimes N}U^{\dagger}_{\mathrm{loc}}=\bigotimes_{i=1}^N\, \varrho_i,
\end{align}
where $\varrho_i= U_i\, \tau(\beta)\, U^{\dagger}_i$. Due to the identical local Hamiltonians, the work deposited in the single-qubit batteries through these operations can be obtained as
\begin{align}
    \Delta \epsilon_{\mathrm{loc}}=\sum_{i=1}^N \,\tr[\varrho_i\, H]\,-\, N \tr[\tau(\beta)H].
\end{align}
In this case, since there are neither classical nor quantum correlations present, one can easily show that the variance with respect to the local Hamiltonians is given by the sum of the local variances,
\begin{align}
\label{eq:local variance}
   V(\varrho_{\textup{tot}})=  \sum_{i=1}^N \,\tr[\varrho_i\, H^2]\,-\,\sum_{i=1}^N \,{\tr[\varrho_i\, H]}^2.
\end{align}
Similarly, it is straightforward to show in this case that the work fluctuation of the total system can be written as a sum of the local work fluctuations,
\begin{align}
\label{eq: loc fluct}
(\Delta W_{\textup{tot}})^2& \,=\, \sum_{i=1}^N (\Delta W_{i})^2=V(\varrho_i)+V(\tau(\beta))\\
&\ \ -\,2\big( \tr[U_i^{\dagger}H\,U_iH\tau(\beta)]-\tr[H\tau(\beta)]\,\tr[H\varrho_i]\big).
\nonumber
\end{align}

\noindent{\textbf{Symmetric local charging}.}\ Now we specialize our discussion to a specific type of local charging process that we call \emph{symmetric local charging process} (SLCP). In such a process, for a given total energy increase $\Delta \epsilon$, the energy of each of the $N$ qubits increases by a fixed amount $\frac{\Delta \epsilon }{N}$ via some fixed local unitary, i.e., $U_i=U$ for all $i$. For general systems, e.g., harmonic oscillators, such a{n} SLCP can result in variances and fluctuations that can be larger or smaller than those of other local charging processes at the same energy input, see, e.g., \cite[Fig.~4]{FriisHuber2018}. Here, however, we show that SLCPs for $N$ qubits generate the maximal
amount of work fluctuations and variance for a given energy increase among all local unitary charging processes. We thus find that
the worst-case local scenarios for both quantities of interest are realized by the same protocol, see Appendix~\ref{app:worst local protocol fluc}.

However, this worst-case local process needs to be compared to two different optimal ({local/}global) protocols. As optimal global procedures we consider the protocols introduced in the previous sections of this article. For the optimal local strategies, we numerically determine the minimum variances and work fluctuations. As shown in Fig.~\ref{fig:Nqubit loc vs nonloc} (a) and (b) for a $4$-qubit system, SLCPs generally do not coincide with either the global or local optimal protocols, even when the initial state is an energy eigenstate (the ground state, in the scenario we consider). The exceptions we observe are only the trivial cases of zero charge and maximal charge. In the considered temperature range, both the variance and work fluctuations obtained from SLCPs therefore maximize the difference to the respective quantities from both the optimal local and the optimal global processes.

For low initial temperatures, there does not appear to be a discernible difference between applying the optimal local or global operations, with perfect matching for the ground state. However, for increasing temperatures we observe that the gap between the optimal local strategy and the optimal global strategy increases, while the gap between the optimal local process and the SLCP decreases.

Regarding the correlation in the final state, we note that there can be many global unitaries generating the same probability distribution (on the diagonal of the density operator) for a given initial state. In particular, this distribution can arise from correlated or uncorrelated states, and which is the case depends both on the unitary and on the initial state. In addition, degeneracy in the energy basis results in a set of different optimal distributions for a given energy input. All of these factor suggest that relation between the specific correlations of the final state and achieved figure of merit are both difficult to determine and likely of no practical concern. Nevertheless it is clear that the advantage of the non-local operations comes from the fact that the set of  probability distributions in the energy basis that is are reachable via such unitary operations is larger than the corresponding set for local operations.

Finally, let us remark that we do not know if the local worst case also represents the overall worst case among all (also global unitary) processes. For the case of harmonic oscillators this is trivially the case because the variances and fluctuations can diverge for both the local and global processes due to the infinite Hilbert-space dimensions of each individual oscillators. At the same time we do not know the optimal local strategies for $N$ qubits for either precision or fluctuations. Determining such strategies would require an optimization over all ways of splitting the energy contributions among the considered qubits. We leave both of these questions as open problems for future work.


\vspace*{-4mm}
\section{Conclusion}\label{sec:discussion}
\vspace*{-2mm}

In this article we have investigated the process of battery charging, i.e., depositing work, for finite-dimensional quantum systems. Starting from thermal states with no extractable work, we have considered processes that raise the average energy unitarily so that all deposited energy can be extracted again in principle, and we have focused our discussion on two figures of merit that characterise such a process: charging precision and work fluctuations. We have further centered our investigation on two exemplary systems of interest: $d$-dimensional quantum systems with equally spaced Hamiltonians, and systems of $N$ non-interacting qubits with identical energy gaps.

For these systems we designed protocols with the purpose of optimizing the charging precision, that is, minimizing the variance of the final average energy, and minimizing the work fluctuations during the charging process, respectively. While we show that our protocol for minimizing the precision is indeed optimal, a similar proof of optimality of the competing protocol for the work fluctuations remains elusive. Yet, all evidence we have gathered seems to suggest that the protocol is indeed optimal, in particular, it reduces to the optimal protocol known for harmonic oscillators~\cite{FriisHuber2018} in the limit $d\rightarrow\infty$. However, except for the notable case of initial zero temperature or the special case where the dimension of the qudit diverges, the two protocols generally differ and so optimizing with respect to one figure of merit comes at the expense of suboptimal performance in the other. We have therefore compared the performance of the two protocols with respect to both figures of merit.
Based on the evidence we have currently available, we see no fundamental reason to generally pick one of these protocols over the other. Although there are some isolated parameter regimes (see Fig.~\ref{fig:protocol comparison}) where the two protocols give the same performance for one of the two quantities (but not the other), in general one of the protocols outperforms the other. Consequently, there is a trade-off between these quantities: Optimizing performance with respect to precision means sub-optimal fluctuations and vice versa. The choice of optimal protocol therefore strongly depends on the weighting one assigns to the two figures of merit, the specific system (dimension, composition, Hamiltonian) under consideration, and on the choice of input energy.

Another question that comes into play in the $N$-qubit scenario concerns the potential added complication of requiring non-local operations acting jointly on all qubits in order to achieve optimal performance. To illustrate this problem we have compared our protocols to a simple local charging scenario that requires only identical unitary operations to be performed individually on all $N$ qubits. We show that such an approach results in the worst possible performance, which highlights that access and control over global unitary transformations is another resource that needs to be considered in this context (see, e.g., the discussion in~\cite{TarantoBakhshinezhadEtAl2023}). We have also numerically compared both protocols with the optimal local protocol. We numerically show the advantage of non-local operations in the charging process when the system is initially in a state far from the ground state.

Meanwhile, the energetic correspondence between the two systems considered, $N$ qubits versus a single qudit of dimension $N+1$ with matching energy-level spacing, provides an opportunity to examine the role of the internal structure and the system dimension, $2^N$ versus $N+1$, in determining advantageous charging strategies. We show that for given initial energy and energy input, accessing higher Hilbert-space dimensions ($N$ qubits) allows us to achieve smaller fluctuations and higher precision as compared to lower Hilbert-space dimensions ($N+1$-dimensional qudit).

For future work we envisage an even more all-encompassing approach towards studying the trade-offs between the identified resources in achieving optimal or near-optimal performance in terms of the figures of merit considered here, but also beyond, taking into account such aspects as the charging speed and stabilisation of the charge.


\begin{acknowledgments}
P. B. is supported by grant number FQXi Grant Number: FQXi-IAF19-07 from the Foundational Questions Institute Fund, a donor advised fund of Silicon Valley Community Foundation. P. B. also acknowledges funding from the Austrian Science Fund (FWF) through the project P~31339-N27, from the European Research Council (Consolidator grant ’Cocoquest’ 101043705), as well as support by the Ministry of Science, 1920 Research, and Technology of Iran (through funding for graduate research visits) and Sharif University of Technology’s Office of Vice President for Research and Technology through Grant QA960512.
B. R. J. is supported by the Austrian Science Fund (FWF) project: Y879-N27 (START).
F. C. B. acknowledges support by Irish Research Council under grant number IRCLA/2022/3922 and the European Union's Horizon 2020 research and innovation programme under the Marie Sk{\l}odowska-Curie grant agreement No. 801110. This project reflects only the authors' view, the EU Agency is not responsible for any use that may be made of the information it contains.
N.F. acknowledges support from the Austrian Science Fund (FWF) through the project P~31339-N27 and through the project P 36478-N funded by the European Union - NextGenerationEU, as well as from the Austrian Federal Ministry of Education, Science and Research via the Austrian Research Promotion Agency (FFG) through the flagship project HPQC (FO999897481) funded by the European Union – NextGenerationEU.
This publication was made possible through the support of Grant 62423 from the John Templeton Foundation. The opinions expressed in this publication are those of the author(s) and do not necessarily reflect the views of the John Templeton Foundation.
\end{acknowledgments}


\bibliographystyle{apsrev4-1fixed_with_article_titles_full_names_new}
\bibliography{Master_Bib_File}

\begin{thebibliography}{65}%
\makeatletter
\providecommand \@ifxundefined [1]{%
 \@ifx{#1\undefined}
}%
\providecommand \@ifnum [1]{%
 \ifnum #1\expandafter \@firstoftwo
 \else \expandafter \@secondoftwo
 \fi
}%
\providecommand \@ifx [1]{%
 \ifx #1\expandafter \@firstoftwo
 \else \expandafter \@secondoftwo
 \fi
}%
\providecommand \natexlab [1]{#1}%
\providecommand \enquote  [1]{#1}%
\providecommand \bibnamefont  [1]{#1}%
\providecommand \bibfnamefont [1]{#1}%
\providecommand \citenamefont [1]{#1}%
\providecommand \href@noop [0]{\@secondoftwo}%
\providecommand \href [0]{\begingroup \@sanitize@url \@href}%
\providecommand \@href[1]{\@@startlink{#1}\@@href}%
\providecommand \@@href[1]{\endgroup#1\@@endlink}%
\providecommand \@sanitize@url [0]{\catcode `\\12\catcode `\$12\catcode
  `\&12\catcode `\#12\catcode `\^12\catcode `\_12\catcode `\%12\relax}%
\providecommand \@@startlink[1]{}%
\providecommand \@@endlink[0]{}%
\providecommand \url  [0]{\begingroup\@sanitize@url \@url }%
\providecommand \@url [1]{\endgroup\@href {#1}{\urlprefix }}%
\providecommand \urlprefix  [0]{URL }%
\providecommand \Eprint [0]{\href }%
\providecommand \doibase [0]{https://doi.org/}%
\providecommand \selectlanguage [0]{\@gobble}%
\providecommand \bibinfo  [0]{\@secondoftwo}%
\providecommand \bibfield  [0]{\@secondoftwo}%
\providecommand \translation [1]{[#1]}%
\providecommand \BibitemOpen [0]{}%
\providecommand \bibitemStop [0]{}%
\providecommand \bibitemNoStop [0]{.\EOS\space}%
\providecommand \EOS [0]{\spacefactor3000\relax}%
\providecommand \BibitemShut  [1]{\csname bibitem#1\endcsname}%
\let\auto@bib@innerbib\@empty
\bibitem [{\citenamefont {Dowling}\ and\ \citenamefont
  {Milburn}(2003)}]{DowlingMilburn2003}%
  \BibitemOpen
  \bibfield  {author} {\bibinfo {author} {\bibfnamefont {Jonathan~P.}\
  \bibnamefont {Dowling}}\ and\ \bibinfo {author} {\bibfnamefont {Gerard~J.}\
  \bibnamefont {Milburn}},\ }\emph {\enquote {\bibinfo {title} {Quantum
  technology: the second quantum revolution},}\ }\href
  {https://doi.org/10.1098/rsta.2003.1227} {\bibfield  {journal} {\bibinfo
  {journal} {Phil. Trans. R. Soc. A}\ }\textbf {\bibinfo {volume} {361}},\
  \bibinfo {pages} {1655} (\bibinfo {year} {2003})},\ \Eprint
  {http://arxiv.org/abs/quant-ph/0206091} {arXiv:quant-ph/0206091}\BibitemShut
  {NoStop}%
\bibitem [{\citenamefont {Preskill}(2018)}]{Preskill2018}%
  \BibitemOpen
  \bibfield  {author} {\bibinfo {author} {\bibfnamefont {John}\ \bibnamefont
  {Preskill}},\ }\emph {\enquote {\bibinfo {title} {{Quantum Computing in the
  NISQ era and beyond}},}\ }\href {https://doi.org/10.22331/q-2018-08-06-79}
  {\bibfield  {journal} {\bibinfo  {journal} {Quantum}\ }\textbf {\bibinfo
  {volume} {2}},\ \bibinfo {pages} {79} (\bibinfo {year} {2018})},\ \Eprint
  {http://arxiv.org/abs/1801.00862} {arXiv:1801.00862}\BibitemShut {NoStop}%
\bibitem [{\citenamefont {Goold}\ \emph {et~al.}(2016)\citenamefont {Goold},
  \citenamefont {Huber}, \citenamefont {Riera}, \citenamefont {del Rio},\ and\
  \citenamefont {Skrzypczyk}}]{GooldHuberRieraDelRioSkrzypczyk2016}%
  \BibitemOpen
  \bibfield  {author} {\bibinfo {author} {\bibfnamefont {John}\ \bibnamefont
  {Goold}}, \bibinfo {author} {\bibfnamefont {Marcus}\ \bibnamefont {Huber}},
  \bibinfo {author} {\bibfnamefont {Arnau}\ \bibnamefont {Riera}}, \bibinfo
  {author} {\bibfnamefont {L{\'i}dia}\ \bibnamefont {del Rio}}, \ and\ \bibinfo
  {author} {\bibfnamefont {Paul}\ \bibnamefont {Skrzypczyk}},\ }\emph {\enquote
  {\bibinfo {title} {The role of quantum information in thermodynamics
  \textemdash\ a topical review},}\ }\href
  {https://doi.org/10.1088/1751-8113/49/14/143001} {\bibfield  {journal}
  {\bibinfo  {journal} {J. Phys. A: Math. Theor.}\ }\textbf {\bibinfo {volume}
  {49}},\ \bibinfo {pages} {143001} (\bibinfo {year} {2016})},\ \Eprint
  {http://arxiv.org/abs/1505.07835} {arXiv:1505.07835}\BibitemShut {NoStop}%
\bibitem [{\citenamefont {Vinjanampathy}\ and\ \citenamefont
  {Anders}(2016)}]{VinjanampathyAnders2016}%
  \BibitemOpen
  \bibfield  {author} {\bibinfo {author} {\bibfnamefont {Sai}\ \bibnamefont
  {Vinjanampathy}}\ and\ \bibinfo {author} {\bibfnamefont {Janet}\ \bibnamefont
  {Anders}},\ }\emph {\enquote {\bibinfo {title} {Quantum thermodynamics},}\
  }\href {https://doi.org/10.1080/00107514.2016.1201896} {\bibfield  {journal}
  {\bibinfo  {journal} {Contemp. Phys.}\ }\textbf {\bibinfo {volume} {57}},\
  \bibinfo {pages} {545} (\bibinfo {year} {2016})},\ \Eprint
  {http://arxiv.org/abs/1508.06099} {arXiv:1508.06099}\BibitemShut {NoStop}%
\bibitem [{\citenamefont {Nernst}(1906)}]{Nernst1906}%
  \BibitemOpen
  \bibfield  {author} {\bibinfo {author} {\bibfnamefont {Walther}\ \bibnamefont
  {Nernst}},\ }\emph {\enquote {\bibinfo {title} {{{\"{U}}ber die Beziehung
  zwischen W{\"{a}}rmeentwicklung und maximaler Arbeit bei kondensierten
  Systemen.}}}\ }in\ \href {https://archive.org/details/mobot31753002089495}
  {\emph {\bibinfo {booktitle} {Sitzungsberichte der K{\"{o}}nliglich
  Preussischen Akademie der Wissenschaften}}}\ (\bibinfo {address} {Berlin},\
  \bibinfo {year} {1906})\ pp.\ \bibinfo {pages} {933--940}\BibitemShut
  {NoStop}%
\bibitem [{\citenamefont {Freitas}\ \emph {et~al.}(2018)\citenamefont
  {Freitas}, \citenamefont {Gallego}, \citenamefont {Masanes},\ and\
  \citenamefont {Paz}}]{FreitasGallegoMasanesPaz2018}%
  \BibitemOpen
  \bibfield  {author} {\bibinfo {author} {\bibfnamefont {Nahuel}\ \bibnamefont
  {Freitas}}, \bibinfo {author} {\bibfnamefont {Rodrigo}\ \bibnamefont
  {Gallego}}, \bibinfo {author} {\bibfnamefont {Llu{\'{i}}s}\ \bibnamefont
  {Masanes}}, \ and\ \bibinfo {author} {\bibfnamefont {Juan~Pablo}\
  \bibnamefont {Paz}},\ }\emph {\enquote {\bibinfo {title} {{Cooling to
  Absolute Zero: The Unattainability Principle}},}\ }in\ \href
  {https://doi.org/10.1007/978-3-319-99046-0_25} {\emph {\bibinfo {booktitle}
  {{Thermodynamics in the Quantum Regime}}}},\ \bibinfo {editor} {edited by\
  \bibinfo {editor} {\bibfnamefont {Felix}\ \bibnamefont {Binder}}, \bibinfo
  {editor} {\bibfnamefont {Luis~A}\ \bibnamefont {Correa}}, \bibinfo {editor}
  {\bibfnamefont {Christian}\ \bibnamefont {Gogolin}}, \bibinfo {editor}
  {\bibfnamefont {Janet}\ \bibnamefont {Anders}}, \ and\ \bibinfo {editor}
  {\bibfnamefont {Gerardo}\ \bibnamefont {Adesso}}}\ (\bibinfo  {publisher}
  {Springer},\ \bibinfo {address} {Cham},\ \bibinfo {year} {2018})\
  Chap.~\bibinfo {chapter} {25}, pp.\ \bibinfo {pages} {597--622},\ \Eprint
  {http://arxiv.org/abs/1911.06377} {arXiv:1911.06377}\BibitemShut {NoStop}%
\bibitem [{\citenamefont {Taranto}\ \emph {et~al.}(2023)\citenamefont
  {Taranto}, \citenamefont {Bakhshinezhad}, \citenamefont {Bluhm},
  \citenamefont {Silva}, \citenamefont {Friis}, \citenamefont {Lock},
  \citenamefont {Vitagliano}, \citenamefont {Binder}, \citenamefont {Debarba},
  \citenamefont {Schwarzhans}, \citenamefont {Clivaz},\ and\ \citenamefont
  {Huber}}]{TarantoBakhshinezhadEtAl2023}%
  \BibitemOpen
  \bibfield  {author} {\bibinfo {author} {\bibfnamefont {Philip}\ \bibnamefont
  {Taranto}}, \bibinfo {author} {\bibfnamefont {Faraj}\ \bibnamefont
  {Bakhshinezhad}}, \bibinfo {author} {\bibfnamefont {Andreas}\ \bibnamefont
  {Bluhm}}, \bibinfo {author} {\bibfnamefont {Ralph}\ \bibnamefont {Silva}},
  \bibinfo {author} {\bibfnamefont {Nicolai}\ \bibnamefont {Friis}}, \bibinfo
  {author} {\bibfnamefont {Maximilian P.~E.}\ \bibnamefont {Lock}}, \bibinfo
  {author} {\bibfnamefont {Giuseppe}\ \bibnamefont {Vitagliano}}, \bibinfo
  {author} {\bibfnamefont {Felix~C.}\ \bibnamefont {Binder}}, \bibinfo {author}
  {\bibfnamefont {Tiago}\ \bibnamefont {Debarba}}, \bibinfo {author}
  {\bibfnamefont {Emanuel}\ \bibnamefont {Schwarzhans}}, \bibinfo {author}
  {\bibfnamefont {Fabien}\ \bibnamefont {Clivaz}}, \ and\ \bibinfo {author}
  {\bibfnamefont {Marcus}\ \bibnamefont {Huber}},\ }\emph {\enquote {\bibinfo
  {title} {{Landauer Versus Nernst: What is the True Cost of Cooling a Quantum
  System?}}}\ }\href {https://doi.org/10.1103/PRXQuantum.4.010332} {\bibfield
  {journal} {\bibinfo  {journal} {PRX Quantum}\ }\textbf {\bibinfo {volume}
  {4}},\ \bibinfo {pages} {010332} (\bibinfo {year} {2023})},\ \Eprint
  {http://arxiv.org/abs/2106.05151} {arXiv:2106.05151}\BibitemShut {NoStop}%
\bibitem [{\citenamefont {Clivaz}\ \emph
  {et~al.}(2019{\natexlab{a}})\citenamefont {Clivaz}, \citenamefont {Silva},
  \citenamefont {Haack}, \citenamefont {Bohr~Brask}, \citenamefont {Brunner},\
  and\ \citenamefont {Huber}}]{ClivazSilvaHaackBohrBraskBrunnerHuber2019a}%
  \BibitemOpen
  \bibfield  {author} {\bibinfo {author} {\bibfnamefont {Fabien}\ \bibnamefont
  {Clivaz}}, \bibinfo {author} {\bibfnamefont {Ralph}\ \bibnamefont {Silva}},
  \bibinfo {author} {\bibfnamefont {G{\'e}raldine}\ \bibnamefont {Haack}},
  \bibinfo {author} {\bibfnamefont {Jonatan}\ \bibnamefont {Bohr~Brask}},
  \bibinfo {author} {\bibfnamefont {Nicolas}\ \bibnamefont {Brunner}}, \ and\
  \bibinfo {author} {\bibfnamefont {Marcus}\ \bibnamefont {Huber}},\ }\emph
  {\enquote {\bibinfo {title} {Unifying paradigms of quantum refrigeration:
  fundamental limits of cooling and associated work costs},}\ }\href
  {https://doi.org/10.1103/PhysRevE.100.042130} {\bibfield  {journal} {\bibinfo
   {journal} {Phys. Rev. E}\ }\textbf {\bibinfo {volume} {100}},\ \bibinfo
  {pages} {042130} (\bibinfo {year} {2019}{\natexlab{a}})},\ \Eprint
  {http://arxiv.org/abs/1710.11624} {arXiv:1710.11624}\BibitemShut {NoStop}%
\bibitem [{\citenamefont {Clivaz}\ \emph
  {et~al.}(2019{\natexlab{b}})\citenamefont {Clivaz}, \citenamefont {Silva},
  \citenamefont {Haack}, \citenamefont {Bohr~Brask}, \citenamefont {Brunner},\
  and\ \citenamefont {Huber}}]{ClivazSilvaHaackBohrBraskBrunnerHuber2019b}%
  \BibitemOpen
  \bibfield  {author} {\bibinfo {author} {\bibfnamefont {Fabien}\ \bibnamefont
  {Clivaz}}, \bibinfo {author} {\bibfnamefont {Ralph}\ \bibnamefont {Silva}},
  \bibinfo {author} {\bibfnamefont {G{\'e}raldine}\ \bibnamefont {Haack}},
  \bibinfo {author} {\bibfnamefont {Jonatan}\ \bibnamefont {Bohr~Brask}},
  \bibinfo {author} {\bibfnamefont {Nicolas}\ \bibnamefont {Brunner}}, \ and\
  \bibinfo {author} {\bibfnamefont {Marcus}\ \bibnamefont {Huber}},\ }\emph
  {\enquote {\bibinfo {title} {{Unifying Paradigms of Quantum Refrigeration: A
  Universal and Attainable Bound on Cooling}},}\ }\href
  {https://doi.org/10.1103/PhysRevLett.123.170605} {\bibfield  {journal}
  {\bibinfo  {journal} {Phys. Rev. Lett.}\ }\textbf {\bibinfo {volume} {123}},\
  \bibinfo {pages} {170605} (\bibinfo {year} {2019}{\natexlab{b}})},\ \Eprint
  {http://arxiv.org/abs/1903.04970} {arXiv:1903.04970}\BibitemShut {NoStop}%
\bibitem [{\citenamefont {Scovil}\ and\ \citenamefont
  {Schulz-DuBois}(1959)}]{ScovilSchulzDuBois1959}%
  \BibitemOpen
  \bibfield  {author} {\bibinfo {author} {\bibfnamefont {Henry E.~D.}\
  \bibnamefont {Scovil}}\ and\ \bibinfo {author} {\bibfnamefont {Erich~O.}\
  \bibnamefont {Schulz-DuBois}},\ }\emph {\enquote {\bibinfo {title}
  {{Three-Level Masers as Heat Engines}},}\ }\href
  {https://doi.org/10.1103/PhysRevLett.2.262} {\bibfield  {journal} {\bibinfo
  {journal} {Phys. Rev. Lett.}\ }\textbf {\bibinfo {volume} {2}},\ \bibinfo
  {pages} {262} (\bibinfo {year} {1959})}\BibitemShut {NoStop}%
\bibitem [{\citenamefont {Kosloff}\ and\ \citenamefont
  {Levy}(2014)}]{KosloffLevy2014}%
  \BibitemOpen
  \bibfield  {author} {\bibinfo {author} {\bibfnamefont {Ronnie}\ \bibnamefont
  {Kosloff}}\ and\ \bibinfo {author} {\bibfnamefont {Amikam}\ \bibnamefont
  {Levy}},\ }\emph {\enquote {\bibinfo {title} {{Quantum Heat Engines and
  Refrigerators: Continuous Devices}},}\ }\href
  {https://doi.org/10.1146/annurev-physchem-040513-103724} {\bibfield
  {journal} {\bibinfo  {journal} {Annu. Rev. Phys. Chem.}\ }\textbf {\bibinfo
  {volume} {65}},\ \bibinfo {pages} {365} (\bibinfo {year} {2014})},\ \Eprint
  {http://arxiv.org/abs/1310.0683} {arXiv:1310.0683}\BibitemShut {NoStop}%
\bibitem [{\citenamefont {Uzdin}\ \emph {et~al.}(2015)\citenamefont {Uzdin},
  \citenamefont {Levy},\ and\ \citenamefont {Kosloff}}]{UzdinLevyKosloff2015}%
  \BibitemOpen
  \bibfield  {author} {\bibinfo {author} {\bibfnamefont {Raam}\ \bibnamefont
  {Uzdin}}, \bibinfo {author} {\bibfnamefont {Amikam}\ \bibnamefont {Levy}}, \
  and\ \bibinfo {author} {\bibfnamefont {Ronnie}\ \bibnamefont {Kosloff}},\
  }\emph {\enquote {\bibinfo {title} {{Equivalence of Quantum Heat Machines,
  and Quantum-Thermodynamic Signatures}},}\ }\href
  {https://doi.org/10.1103/PhysRevX.5.031044} {\bibfield  {journal} {\bibinfo
  {journal} {Phys. Rev. X}\ }\textbf {\bibinfo {volume} {5}},\ \bibinfo {pages}
  {031044} (\bibinfo {year} {2015})},\ \Eprint
  {http://arxiv.org/abs/1502.06592} {arXiv:1502.06592}\BibitemShut {NoStop}%
\bibitem [{\citenamefont {Levy}\ and\ \citenamefont
  {Gelbwaser-Klimovsky}(2019)}]{LevyGelbwaserKlimovsky2019}%
  \BibitemOpen
  \bibfield  {author} {\bibinfo {author} {\bibfnamefont {Amikam}\ \bibnamefont
  {Levy}}\ and\ \bibinfo {author} {\bibfnamefont {David}\ \bibnamefont
  {Gelbwaser-Klimovsky}},\ }\emph {\enquote {\bibinfo {title} {Quantum features
  and signatures of quantum thermal machines},}\ }in\ \href
  {https://doi.org/10.1007/978-3-319-99046-0_4} {\emph {\bibinfo {booktitle}
  {Thermodynamics in the Quantum Regime}}},\ \bibinfo {editor} {edited by\
  \bibinfo {editor} {\bibfnamefont {Felix}\ \bibnamefont {Binder}}, \bibinfo
  {editor} {\bibfnamefont {Luis~A.}\ \bibnamefont {Correa}}, \bibinfo {editor}
  {\bibfnamefont {Christian}\ \bibnamefont {Gogolin}}, \bibinfo {editor}
  {\bibfnamefont {Janet}\ \bibnamefont {Anders}}, \ and\ \bibinfo {editor}
  {\bibfnamefont {Gerardo}\ \bibnamefont {Adesso}}}\ (\bibinfo  {publisher}
  {Springer},\ \bibinfo {address} {Cham},\ \bibinfo {year} {2019})\
  Chap.~\bibinfo {chapter} {4}, pp.\ \bibinfo {pages} {87--126},\ \Eprint
  {http://arxiv.org/abs/1803.05586} {arXiv:1803.05586}\BibitemShut {NoStop}%
\bibitem [{\citenamefont {Mitchison}(2019)}]{Mitchison2019}%
  \BibitemOpen
  \bibfield  {author} {\bibinfo {author} {\bibfnamefont {Mark~T.}\ \bibnamefont
  {Mitchison}},\ }\emph {\enquote {\bibinfo {title} {Quantum thermal absorption
  machines: refrigerators, engines and clocks},}\ }\href
  {https://doi.org/10.1080/00107514.2019.1631555} {\bibfield  {journal}
  {\bibinfo  {journal} {Contemp. Phys.}\ }\textbf {\bibinfo {volume} {60}},\
  \bibinfo {pages} {164} (\bibinfo {year} {2019})},\ \Eprint
  {http://arxiv.org/abs/1902.02672} {arXiv:1902.02672}\BibitemShut {NoStop}%
\bibitem [{\citenamefont {Woods}\ \emph {et~al.}(2019)\citenamefont {Woods},
  \citenamefont {Ng},\ and\ \citenamefont {Wehner}}]{WoodsNgWehner2019}%
  \BibitemOpen
  \bibfield  {author} {\bibinfo {author} {\bibfnamefont {Mischa~P.}\
  \bibnamefont {Woods}}, \bibinfo {author} {\bibfnamefont {Nelly Huei~Ying}\
  \bibnamefont {Ng}}, \ and\ \bibinfo {author} {\bibfnamefont {Stephanie}\
  \bibnamefont {Wehner}},\ }\emph {\enquote {\bibinfo {title} {The maximum
  efficiency of nano heat engines depends on more than temperature},}\ }\href
  {https://doi.org/10.22331/q-2019-08-19-177} {\bibfield  {journal} {\bibinfo
  {journal} {{Quantum}}\ }\textbf {\bibinfo {volume} {3}},\ \bibinfo {pages}
  {177} (\bibinfo {year} {2019})},\ \Eprint {http://arxiv.org/abs/1506.02322}
  {arXiv:1506.02322}\BibitemShut {NoStop}%
\bibitem [{\citenamefont {Alicki}\ and\ \citenamefont
  {Fannes}(2013)}]{AlickiFannes2013}%
  \BibitemOpen
  \bibfield  {author} {\bibinfo {author} {\bibfnamefont {Robert}\ \bibnamefont
  {Alicki}}\ and\ \bibinfo {author} {\bibfnamefont {Mark}\ \bibnamefont
  {Fannes}},\ }\emph {\enquote {\bibinfo {title} {Entanglement boost for
  extractable work from ensembles of quantum batteries},}\ }\href
  {https://doi.org/10.1103/PhysRevE.87.042123} {\bibfield  {journal} {\bibinfo
  {journal} {Phys. Rev. E}\ }\textbf {\bibinfo {volume} {87}},\ \bibinfo
  {pages} {042123} (\bibinfo {year} {2013})},\ \Eprint
  {http://arxiv.org/abs/1211.1209} {arXiv:1211.1209}\BibitemShut {NoStop}%
\bibitem [{\citenamefont {Campaioli}\ \emph {et~al.}(2019)\citenamefont
  {Campaioli}, \citenamefont {Pollock},\ and\ \citenamefont
  {Vinjanampathy}}]{CampaioliPollockVinjanampathy2019}%
  \BibitemOpen
  \bibfield  {author} {\bibinfo {author} {\bibfnamefont {Francesco}\
  \bibnamefont {Campaioli}}, \bibinfo {author} {\bibfnamefont {Felix~A.}\
  \bibnamefont {Pollock}}, \ and\ \bibinfo {author} {\bibfnamefont {Sai}\
  \bibnamefont {Vinjanampathy}},\ }\emph {\enquote {\bibinfo {title} {{Quantum
  Batteries}},}\ }in\ \href {https://doi.org/10.1007/978-3-319-99046-0_8}
  {\emph {\bibinfo {booktitle} {Thermodynamics in the Quantum Regime}}},\
  \bibinfo {editor} {edited by\ \bibinfo {editor} {\bibfnamefont {Felix}\
  \bibnamefont {Binder}}, \bibinfo {editor} {\bibfnamefont {Luis~A.}\
  \bibnamefont {Correa}}, \bibinfo {editor} {\bibfnamefont {Christian}\
  \bibnamefont {Gogolin}}, \bibinfo {editor} {\bibfnamefont {Janet}\
  \bibnamefont {Anders}}, \ and\ \bibinfo {editor} {\bibfnamefont {Gerardo}\
  \bibnamefont {Adesso}}}\ (\bibinfo  {publisher} {Springer},\ \bibinfo
  {address} {Cham},\ \bibinfo {year} {2019})\ Chap.~\bibinfo {chapter} {8},
  pp.\ \bibinfo {pages} {207--225},\ \Eprint {http://arxiv.org/abs/1805.05507}
  {arXiv:1805.05507}\BibitemShut {NoStop}%
\bibitem [{\citenamefont {Campaioli}\ \emph {et~al.}(2023)\citenamefont
  {Campaioli}, \citenamefont {Gherardini}, \citenamefont {Quach}, \citenamefont
  {Polini},\ and\ \citenamefont
  {Andolina}}]{CampaioliGherardiniQuachPoliniAndolina}%
  \BibitemOpen
  \bibfield  {author} {\bibinfo {author} {\bibfnamefont {Francesco}\
  \bibnamefont {Campaioli}}, \bibinfo {author} {\bibfnamefont {Stefano}\
  \bibnamefont {Gherardini}}, \bibinfo {author} {\bibfnamefont {James~Q.}\
  \bibnamefont {Quach}}, \bibinfo {author} {\bibfnamefont {Marco}\ \bibnamefont
  {Polini}}, \ and\ \bibinfo {author} {\bibfnamefont {Gian~Marcello}\
  \bibnamefont {Andolina}},\ }\href@noop {} {\emph {\enquote {\bibinfo {title}
  {{Colloquium: Quantum Batteries}},}\ }}\Eprint
  {http://arxiv.org/abs/2308.02277} {arXiv:2308.02277} [quant-ph] (\bibinfo
  {year} {2023})\BibitemShut {NoStop}%
\bibitem [{\citenamefont {Binder}\ \emph {et~al.}(2015)\citenamefont {Binder},
  \citenamefont {Vinjanampathy}, \citenamefont {Modi},\ and\ \citenamefont
  {Goold}}]{BinderVinjanampathyModiGoold2015}%
  \BibitemOpen
  \bibfield  {author} {\bibinfo {author} {\bibfnamefont {Felix~C.}\
  \bibnamefont {Binder}}, \bibinfo {author} {\bibfnamefont {Sai}\ \bibnamefont
  {Vinjanampathy}}, \bibinfo {author} {\bibfnamefont {Kavan}\ \bibnamefont
  {Modi}}, \ and\ \bibinfo {author} {\bibfnamefont {John}\ \bibnamefont
  {Goold}},\ }\emph {\enquote {\bibinfo {title} {{Quantacell: Powerful charging
  of quantum batteries}},}\ }\href
  {https://doi.org/10.1088/1367-2630/17/7/075015} {\bibfield  {journal}
  {\bibinfo  {journal} {New J. Phys.}\ }\textbf {\bibinfo {volume} {17}},\
  \bibinfo {pages} {075015} (\bibinfo {year} {2015})},\ \Eprint
  {http://arxiv.org/abs/1503.07005} {arXiv:1503.07005}\BibitemShut {NoStop}%
\bibitem [{\citenamefont {Campaioli}\ \emph {et~al.}(2017)\citenamefont
  {Campaioli}, \citenamefont {Pollock}, \citenamefont {Binder}, \citenamefont
  {C{\'e}leri}, \citenamefont {Goold}, \citenamefont {Vinjanampathy},\ and\
  \citenamefont
  {Modi}}]{CampaioliPollockBinderCeleriGooldVinjanampathyModi2017}%
  \BibitemOpen
  \bibfield  {author} {\bibinfo {author} {\bibfnamefont {Francesco}\
  \bibnamefont {Campaioli}}, \bibinfo {author} {\bibfnamefont {Felix~A.}\
  \bibnamefont {Pollock}}, \bibinfo {author} {\bibfnamefont {Felix~C.}\
  \bibnamefont {Binder}}, \bibinfo {author} {\bibfnamefont {Lucas~C.}\
  \bibnamefont {C{\'e}leri}}, \bibinfo {author} {\bibfnamefont {John}\
  \bibnamefont {Goold}}, \bibinfo {author} {\bibfnamefont {Sai}\ \bibnamefont
  {Vinjanampathy}}, \ and\ \bibinfo {author} {\bibfnamefont {Kavan}\
  \bibnamefont {Modi}},\ }\emph {\enquote {\bibinfo {title} {{Enhancing the
  charging power of quantum batteries}},}\ }\href
  {https://doi.org/10.1103/PhysRevLett.118.150601} {\bibfield  {journal}
  {\bibinfo  {journal} {Phys. Rev. Lett.}\ }\textbf {\bibinfo {volume} {118}},\
  \bibinfo {pages} {150601} (\bibinfo {year} {2017})},\ \Eprint
  {http://arxiv.org/abs/1612.04991} {arXiv:1612.04991}\BibitemShut {NoStop}%
\bibitem [{\citenamefont {Gyhm}\ \emph {et~al.}(2022)\citenamefont {Gyhm},
  \citenamefont {{\v{S}}afr{\'{a}}nek},\ and\ \citenamefont {Rosa}}]{Gyhm2022}%
  \BibitemOpen
  \bibfield  {author} {\bibinfo {author} {\bibfnamefont {Ju~Yeon}\ \bibnamefont
  {Gyhm}}, \bibinfo {author} {\bibfnamefont {Dominik}\ \bibnamefont
  {{\v{S}}afr{\'{a}}nek}}, \ and\ \bibinfo {author} {\bibfnamefont {Dario}\
  \bibnamefont {Rosa}},\ }\emph {\enquote {\bibinfo {title} {{Quantum Charging
  Advantage Cannot Be Extensive without Global Operations}},}\ }\href
  {https://doi.org/10.1103/PhysRevLett.128.140501} {\bibfield  {journal}
  {\bibinfo  {journal} {Phys. Rev. Lett.}\ }\textbf {\bibinfo {volume} {128}},\
  \bibinfo {pages} {140501} (\bibinfo {year} {2022})},\ \Eprint
  {http://arxiv.org/abs/2108.02491} {arXiv:2108.02491}\BibitemShut {NoStop}%
\bibitem [{\citenamefont {Le}\ \emph {et~al.}(2018)\citenamefont {Le},
  \citenamefont {Levinsen}, \citenamefont {Modi}, \citenamefont {Parish},\ and\
  \citenamefont {Pollock}}]{LeLevinsenModiParishPollock2018}%
  \BibitemOpen
  \bibfield  {author} {\bibinfo {author} {\bibfnamefont {Thao~P.}\ \bibnamefont
  {Le}}, \bibinfo {author} {\bibfnamefont {Jesper}\ \bibnamefont {Levinsen}},
  \bibinfo {author} {\bibfnamefont {Kavan}\ \bibnamefont {Modi}}, \bibinfo
  {author} {\bibfnamefont {Meera}\ \bibnamefont {Parish}}, \ and\ \bibinfo
  {author} {\bibfnamefont {Felix~A.}\ \bibnamefont {Pollock}},\ }\emph
  {\enquote {\bibinfo {title} {{Spin-chain model of a many-body quantum
  battery}},}\ }\href {https://doi.org/10.1103/PhysRevA.97.022106} {\bibfield
  {journal} {\bibinfo  {journal} {Phys. Rev. A}\ }\textbf {\bibinfo {volume}
  {97}},\ \bibinfo {pages} {022106} (\bibinfo {year} {2018})},\ \Eprint
  {http://arxiv.org/abs/1712.03559} {arXiv:1712.03559}\BibitemShut {NoStop}%
\bibitem [{\citenamefont {Ferraro}\ \emph {et~al.}(2018)\citenamefont
  {Ferraro}, \citenamefont {Campisi}, \citenamefont {Andolina}, \citenamefont
  {Pellegrini},\ and\ \citenamefont
  {Polini}}]{FerraroCampisiAndolinaPellegriniPolini2018}%
  \BibitemOpen
  \bibfield  {author} {\bibinfo {author} {\bibfnamefont {Dario}\ \bibnamefont
  {Ferraro}}, \bibinfo {author} {\bibfnamefont {Michele}\ \bibnamefont
  {Campisi}}, \bibinfo {author} {\bibfnamefont {Gian~Marcello}\ \bibnamefont
  {Andolina}}, \bibinfo {author} {\bibfnamefont {Vittorio}\ \bibnamefont
  {Pellegrini}}, \ and\ \bibinfo {author} {\bibfnamefont {Marco}\ \bibnamefont
  {Polini}},\ }\emph {\enquote {\bibinfo {title} {{High-Power Collective
  Charging of a Solid-State Quantum Battery}},}\ }\href
  {https://doi.org/10.1103/PhysRevLett.120.117702} {\bibfield  {journal}
  {\bibinfo  {journal} {Phys. Rev. Lett.}\ }\textbf {\bibinfo {volume} {120}},\
  \bibinfo {pages} {117702} (\bibinfo {year} {2018})},\ \Eprint
  {http://arxiv.org/abs/1707.04930} {arXiv:1707.04930}\BibitemShut {NoStop}%
\bibitem [{\citenamefont {Andolina}\ \emph {et~al.}(2018)\citenamefont
  {Andolina}, \citenamefont {Farina}, \citenamefont {Mari}, \citenamefont
  {Pellegrini}, \citenamefont {Giovannetti},\ and\ \citenamefont
  {Polini}}]{AndolinaFarinaMariPellegriniGiovannettiPolini2018}%
  \BibitemOpen
  \bibfield  {author} {\bibinfo {author} {\bibfnamefont {Gian~Marcello}\
  \bibnamefont {Andolina}}, \bibinfo {author} {\bibfnamefont {Donato}\
  \bibnamefont {Farina}}, \bibinfo {author} {\bibfnamefont {Andrea}\
  \bibnamefont {Mari}}, \bibinfo {author} {\bibfnamefont {Vittorio}\
  \bibnamefont {Pellegrini}}, \bibinfo {author} {\bibfnamefont {Vittorio}\
  \bibnamefont {Giovannetti}}, \ and\ \bibinfo {author} {\bibfnamefont {Marco}\
  \bibnamefont {Polini}},\ }\emph {\enquote {\bibinfo {title} {Charger-mediated
  energy transfer in exactly-solvable models for quantum batteries},}\ }\href
  {https://doi.org/10.1103/PhysRevB.98.205423} {\bibfield  {journal} {\bibinfo
  {journal} {Phys. Rev. B}\ }\textbf {\bibinfo {volume} {98}},\ \bibinfo
  {pages} {205423} (\bibinfo {year} {2018})},\ \Eprint
  {http://arxiv.org/abs/1807.04031} {arXiv:1807.04031}\BibitemShut {NoStop}%
\bibitem [{\citenamefont {Andolina}\ \emph
  {et~al.}(2019{\natexlab{a}})\citenamefont {Andolina}, \citenamefont {Keck},
  \citenamefont {Mari}, \citenamefont {Campisi}, \citenamefont {Giovannetti},\
  and\ \citenamefont {Polini}}]{AndolinaEtAl2019}%
  \BibitemOpen
  \bibfield  {author} {\bibinfo {author} {\bibfnamefont {Gian~Marcello}\
  \bibnamefont {Andolina}}, \bibinfo {author} {\bibfnamefont {Maximilian}\
  \bibnamefont {Keck}}, \bibinfo {author} {\bibfnamefont {Andrea}\ \bibnamefont
  {Mari}}, \bibinfo {author} {\bibfnamefont {Michele}\ \bibnamefont {Campisi}},
  \bibinfo {author} {\bibfnamefont {Vittorio}\ \bibnamefont {Giovannetti}}, \
  and\ \bibinfo {author} {\bibfnamefont {Marco}\ \bibnamefont {Polini}},\
  }\emph {\enquote {\bibinfo {title} {{Extractable Work, the Role of
  Correlations, and Asymptotic Freedom in Quantum Batteries}},}\ }\href
  {https://doi.org/10.1103/PhysRevLett.122.047702} {\bibfield  {journal}
  {\bibinfo  {journal} {Phys. Rev. Lett.}\ }\textbf {\bibinfo {volume} {122}},\
  \bibinfo {pages} {047702} (\bibinfo {year} {2019}{\natexlab{a}})},\ \Eprint
  {http://arxiv.org/abs/1807.08656} {arXiv:1807.08656}\BibitemShut {NoStop}%
\bibitem [{\citenamefont {Andolina}\ \emph
  {et~al.}(2019{\natexlab{b}})\citenamefont {Andolina}, \citenamefont {Keck},
  \citenamefont {Mari}, \citenamefont {Giovannetti},\ and\ \citenamefont
  {Polini}}]{AndolinaKeckMariGiovannettiPolini2019}%
  \BibitemOpen
  \bibfield  {author} {\bibinfo {author} {\bibfnamefont {Gian~Marcello}\
  \bibnamefont {Andolina}}, \bibinfo {author} {\bibfnamefont {Maximilian}\
  \bibnamefont {Keck}}, \bibinfo {author} {\bibfnamefont {Andrea}\ \bibnamefont
  {Mari}}, \bibinfo {author} {\bibfnamefont {Vittorio}\ \bibnamefont
  {Giovannetti}}, \ and\ \bibinfo {author} {\bibfnamefont {Marco}\ \bibnamefont
  {Polini}},\ }\emph {\enquote {\bibinfo {title} {Quantum versus classical
  many-body batteries},}\ }\href {https://doi.org/10.1103/PhysRevB.99.205437}
  {\bibfield  {journal} {\bibinfo  {journal} {Phys. Rev. B}\ }\textbf {\bibinfo
  {volume} {99}},\ \bibinfo {pages} {205437} (\bibinfo {year}
  {2019}{\natexlab{b}})},\ \Eprint {http://arxiv.org/abs/1812.04669}
  {arXiv:1812.04669}\BibitemShut {NoStop}%
\bibitem [{\citenamefont {Farina}\ \emph {et~al.}(2019)\citenamefont {Farina},
  \citenamefont {Andolina}, \citenamefont {Mari}, \citenamefont {Polini},\ and\
  \citenamefont {Giovannetti}}]{FarinaAndolinaMariPoliniGiovannetti2019}%
  \BibitemOpen
  \bibfield  {author} {\bibinfo {author} {\bibfnamefont {Donato}\ \bibnamefont
  {Farina}}, \bibinfo {author} {\bibfnamefont {Gian~Marcello}\ \bibnamefont
  {Andolina}}, \bibinfo {author} {\bibfnamefont {Andrea}\ \bibnamefont {Mari}},
  \bibinfo {author} {\bibfnamefont {Marco}\ \bibnamefont {Polini}}, \ and\
  \bibinfo {author} {\bibfnamefont {Vittorio}\ \bibnamefont {Giovannetti}},\
  }\emph {\enquote {\bibinfo {title} {Charger-mediated energy transfer for
  quantum batteries: an open system approach},}\ }\href
  {https://doi.org/10.1103/PhysRevB.99.035421} {\bibfield  {journal} {\bibinfo
  {journal} {Phys. Rev. B}\ }\textbf {\bibinfo {volume} {99}},\ \bibinfo
  {pages} {035421} (\bibinfo {year} {2019})},\ \Eprint
  {http://arxiv.org/abs/1810.10890} {arXiv:1810.10890}\BibitemShut {NoStop}%
\bibitem [{\citenamefont {Rossini}\ \emph {et~al.}(2019)\citenamefont
  {Rossini}, \citenamefont {Andolina},\ and\ \citenamefont
  {Polini}}]{RossiniAndolinaPolini2019}%
  \BibitemOpen
  \bibfield  {author} {\bibinfo {author} {\bibfnamefont {Davide}\ \bibnamefont
  {Rossini}}, \bibinfo {author} {\bibfnamefont {Gian~Marcello}\ \bibnamefont
  {Andolina}}, \ and\ \bibinfo {author} {\bibfnamefont {Marco}\ \bibnamefont
  {Polini}},\ }\emph {\enquote {\bibinfo {title} {Many-body localized quantum
  batteries},}\ }\href {https://doi.org/10.1103/PhysRevB.100.115142} {\bibfield
   {journal} {\bibinfo  {journal} {Phys. Rev. B}\ }\textbf {\bibinfo {volume}
  {100}},\ \bibinfo {pages} {115142} (\bibinfo {year} {2019})},\ \Eprint
  {http://arxiv.org/abs/1906.00644} {arXiv:1906.00644}\BibitemShut {NoStop}%
\bibitem [{\citenamefont {Crescente}\ \emph
  {et~al.}(2020{\natexlab{a}})\citenamefont {Crescente}, \citenamefont
  {Carrega}, \citenamefont {Sassetti},\ and\ \citenamefont
  {Ferraro}}]{CrescenteCarregaSassettiFerraro2020a}%
  \BibitemOpen
  \bibfield  {author} {\bibinfo {author} {\bibfnamefont {Alba}\ \bibnamefont
  {Crescente}}, \bibinfo {author} {\bibfnamefont {Matteo}\ \bibnamefont
  {Carrega}}, \bibinfo {author} {\bibfnamefont {Maura}\ \bibnamefont
  {Sassetti}}, \ and\ \bibinfo {author} {\bibfnamefont {Dario}\ \bibnamefont
  {Ferraro}},\ }\emph {\enquote {\bibinfo {title} {{Ultrafast charging in a
  two-photon Dicke quantum battery}},}\ }\href
  {https://doi.org/10.1103/PhysRevB.102.245407} {\bibfield  {journal} {\bibinfo
   {journal} {Phys. Rev. B}\ }\textbf {\bibinfo {volume} {102}},\ \bibinfo
  {pages} {245407} (\bibinfo {year} {2020}{\natexlab{a}})},\ \Eprint
  {http://arxiv.org/abs/2009.09791} {arXiv:2009.09791}\BibitemShut {NoStop}%
\bibitem [{\citenamefont {Rossini}\ \emph {et~al.}(2020)\citenamefont
  {Rossini}, \citenamefont {Andolina}, \citenamefont {Rosa}, \citenamefont
  {Carrega},\ and\ \citenamefont
  {Polini}}]{RossiniAndolinaRosaCarregaPolini2020}%
  \BibitemOpen
  \bibfield  {author} {\bibinfo {author} {\bibfnamefont {Davide}\ \bibnamefont
  {Rossini}}, \bibinfo {author} {\bibfnamefont {Gian~Marcello}\ \bibnamefont
  {Andolina}}, \bibinfo {author} {\bibfnamefont {Dario}\ \bibnamefont {Rosa}},
  \bibinfo {author} {\bibfnamefont {Matteo}\ \bibnamefont {Carrega}}, \ and\
  \bibinfo {author} {\bibfnamefont {Marco}\ \bibnamefont {Polini}},\ }\emph
  {\enquote {\bibinfo {title} {{Quantum Advantage in the Charging Process of
  Sachdev-Ye-Kitaev Batteries}},}\ }\href
  {https://doi.org/10.1103/PhysRevLett.125.236402} {\bibfield  {journal}
  {\bibinfo  {journal} {Phys. Rev. Lett.}\ }\textbf {\bibinfo {volume} {125}},\
  \bibinfo {pages} {236402} (\bibinfo {year} {2020})},\ \Eprint
  {http://arxiv.org/abs/1912.07234} {arXiv:1912.07234}\BibitemShut {NoStop}%
\bibitem [{\citenamefont {Centrone}\ \emph {et~al.}(2023)\citenamefont
  {Centrone}, \citenamefont {Mancino},\ and\ \citenamefont
  {Paternostro}}]{CentroneMancinoPaternostro2021}%
  \BibitemOpen
  \bibfield  {author} {\bibinfo {author} {\bibfnamefont {Federico}\
  \bibnamefont {Centrone}}, \bibinfo {author} {\bibfnamefont {Luca}\
  \bibnamefont {Mancino}}, \ and\ \bibinfo {author} {\bibfnamefont {Mauro}\
  \bibnamefont {Paternostro}},\ }\emph {\enquote {\bibinfo {title} {Charging
  batteries with quantum squeezing},}\ }\href
  {https://doi.org/10.1103/PhysRevA.108.052213} {\bibfield  {journal} {\bibinfo
   {journal} {Phys. Rev. A}\ }\textbf {\bibinfo {volume} {108}},\ \bibinfo
  {pages} {052213} (\bibinfo {year} {2023})},\ \Eprint
  {http://arxiv.org/abs/2106.07899} {arXiv:2106.07899}\BibitemShut {NoStop}%
\bibitem [{\citenamefont {Seah}\ \emph {et~al.}(2021)\citenamefont {Seah},
  \citenamefont {Perarnau-Llobet}, \citenamefont {Haack}, \citenamefont
  {Brunner},\ and\ \citenamefont
  {Nimmrichter}}]{SeahPerarnauHaackBrunnerNimmrichter2021}%
  \BibitemOpen
  \bibfield  {author} {\bibinfo {author} {\bibfnamefont {Stella}\ \bibnamefont
  {Seah}}, \bibinfo {author} {\bibfnamefont {Mart{\'i}}\ \bibnamefont
  {Perarnau-Llobet}}, \bibinfo {author} {\bibfnamefont {G{\'e}raldine}\
  \bibnamefont {Haack}}, \bibinfo {author} {\bibfnamefont {Nicolas}\
  \bibnamefont {Brunner}}, \ and\ \bibinfo {author} {\bibfnamefont {Stefan}\
  \bibnamefont {Nimmrichter}},\ }\emph {\enquote {\bibinfo {title} {{Quantum
  Speed-Up in Collisional Battery Charging}},}\ }\href
  {https://doi.org/10.1103/PhysRevLett.127.100601} {\bibfield  {journal}
  {\bibinfo  {journal} {Phys. Rev. Lett.}\ }\textbf {\bibinfo {volume} {127}},\
  \bibinfo {pages} {100601} (\bibinfo {year} {2021})},\ \Eprint
  {http://arxiv.org/abs/2105.01863} {arXiv:2105.01863} [quant-ph]\BibitemShut
  {NoStop}%
\bibitem [{\citenamefont {Shaghaghi}\ \emph {et~al.}(2022)\citenamefont
  {Shaghaghi}, \citenamefont {Singh}, \citenamefont {Benenti},\ and\
  \citenamefont {Rosa}}]{Shaghaghi2022}%
  \BibitemOpen
  \bibfield  {author} {\bibinfo {author} {\bibfnamefont {Vahid}\ \bibnamefont
  {Shaghaghi}}, \bibinfo {author} {\bibfnamefont {Varinder}\ \bibnamefont
  {Singh}}, \bibinfo {author} {\bibfnamefont {Giuliano}\ \bibnamefont
  {Benenti}}, \ and\ \bibinfo {author} {\bibfnamefont {Dario}\ \bibnamefont
  {Rosa}},\ }\emph {\enquote {\bibinfo {title} {{Micromasers as quantum
  batteries}},}\ }\href {https://doi.org/10.1088/2058-9565/ac8829} {\bibfield
  {journal} {\bibinfo  {journal} {Quantum Sci. Technol.}\ }\textbf {\bibinfo
  {volume} {7}},\ \bibinfo {pages} {04LT01} (\bibinfo {year} {2022})},\ \Eprint
  {http://arxiv.org/abs/2204.09995} {arXiv:2204.09995}\BibitemShut {NoStop}%
\bibitem [{\citenamefont {Salvia}\ \emph {et~al.}(2023)\citenamefont {Salvia},
  \citenamefont {Perarnau-Llobet}, \citenamefont {Haack}, \citenamefont
  {Brunner},\ and\ \citenamefont
  {Nimmrichter}}]{SalviaPerarnauHaackBrunnerNimmrichter2022}%
  \BibitemOpen
  \bibfield  {author} {\bibinfo {author} {\bibfnamefont {Raffaele}\
  \bibnamefont {Salvia}}, \bibinfo {author} {\bibfnamefont {Mart{\'i}}\
  \bibnamefont {Perarnau-Llobet}}, \bibinfo {author} {\bibfnamefont
  {G{\'e}raldine}\ \bibnamefont {Haack}}, \bibinfo {author} {\bibfnamefont
  {Nicolas}\ \bibnamefont {Brunner}}, \ and\ \bibinfo {author} {\bibfnamefont
  {Stefan}\ \bibnamefont {Nimmrichter}},\ }\emph {\enquote {\bibinfo {title}
  {{Quantum advantage in charging cavity and spin batteries by repeated
  interactions}},}\ }\href {https://doi.org/10.1103/PhysRevResearch.5.013155}
  {\bibfield  {journal} {\bibinfo  {journal} {Phys. Rev. Res.}\ }\textbf
  {\bibinfo {volume} {5}},\ \bibinfo {pages} {013155} (\bibinfo {year}
  {2023})},\ \Eprint {http://arxiv.org/abs/2205.00026}
  {arXiv:2205.00026}\BibitemShut {NoStop}%
\bibitem [{\citenamefont {Rosa}\ \emph {et~al.}(2020)\citenamefont {Rosa},
  \citenamefont {Rossini}, \citenamefont {Andolina}, \citenamefont {Polini},\
  and\ \citenamefont {Carrega}}]{RosaRossiniAndolinaPoliniCarrega2020}%
  \BibitemOpen
  \bibfield  {author} {\bibinfo {author} {\bibfnamefont {Dario}\ \bibnamefont
  {Rosa}}, \bibinfo {author} {\bibfnamefont {Davide}\ \bibnamefont {Rossini}},
  \bibinfo {author} {\bibfnamefont {Gian~Marcello}\ \bibnamefont {Andolina}},
  \bibinfo {author} {\bibfnamefont {Marco}\ \bibnamefont {Polini}}, \ and\
  \bibinfo {author} {\bibfnamefont {Matteo}\ \bibnamefont {Carrega}},\ }\emph
  {\enquote {\bibinfo {title} {{Ultra-stable charging of fast-scrambling SYK
  quantum batteries}},}\ }\href {https://doi.org/10.1007/JHEP11(2020)067}
  {\bibfield  {journal} {\bibinfo  {journal} {J. High Energ. Phys.}\ }\textbf
  {\bibinfo {volume} {2020}},\ \bibinfo {pages} {67} (\bibinfo {year}
  {2020})},\ \Eprint {http://arxiv.org/abs/1912.07247}
  {arXiv:1912.07247}\BibitemShut {NoStop}%
\bibitem [{\citenamefont {Gherardini}\ \emph {et~al.}(2020)\citenamefont
  {Gherardini}, \citenamefont {Campaioli}, \citenamefont {Caruso},\ and\
  \citenamefont {Binder}}]{GherardiniCampaioliFilippoBinder2020}%
  \BibitemOpen
  \bibfield  {author} {\bibinfo {author} {\bibfnamefont {Stefano}\ \bibnamefont
  {Gherardini}}, \bibinfo {author} {\bibfnamefont {Francesco}\ \bibnamefont
  {Campaioli}}, \bibinfo {author} {\bibfnamefont {Filippo}\ \bibnamefont
  {Caruso}}, \ and\ \bibinfo {author} {\bibfnamefont {Felix~C.}\ \bibnamefont
  {Binder}},\ }\emph {\enquote {\bibinfo {title} {Stabilizing open quantum
  batteries by sequential measurements},}\ }\href
  {https://doi.org/10.1103/PhysRevResearch.2.013095} {\bibfield  {journal}
  {\bibinfo  {journal} {Phys. Rev. Research}\ }\textbf {\bibinfo {volume}
  {2}},\ \bibinfo {pages} {013095} (\bibinfo {year} {2020})},\ \Eprint
  {http://arxiv.org/abs/1910.02458} {arXiv:1910.02458}\BibitemShut {NoStop}%
\bibitem [{\citenamefont {Hovhannisyan}\ \emph {et~al.}(2020)\citenamefont
  {Hovhannisyan}, \citenamefont {Barra},\ and\ \citenamefont
  {Imparato}}]{HovhannisyanBarraImparato2020}%
  \BibitemOpen
  \bibfield  {author} {\bibinfo {author} {\bibfnamefont {Karen~V.}\
  \bibnamefont {Hovhannisyan}}, \bibinfo {author} {\bibfnamefont {Felipe}\
  \bibnamefont {Barra}}, \ and\ \bibinfo {author} {\bibfnamefont {Alberto}\
  \bibnamefont {Imparato}},\ }\emph {\enquote {\bibinfo {title} {{Charging
  assisted by thermalization}},}\ }\href
  {https://doi.org/10.1103/PhysRevResearch.2.033413} {\bibfield  {journal}
  {\bibinfo  {journal} {Phys. Rev. Research}\ }\textbf {\bibinfo {volume}
  {2}},\ \bibinfo {pages} {033413} (\bibinfo {year} {2020})},\ \Eprint
  {http://arxiv.org/abs/2001.07696} {arXiv:2001.07696}\BibitemShut {NoStop}%
\bibitem [{\citenamefont {Mitchison}\ \emph {et~al.}(2021)\citenamefont
  {Mitchison}, \citenamefont {Goold},\ and\ \citenamefont
  {Prior}}]{MitchisonGooldPrior2021}%
  \BibitemOpen
  \bibfield  {author} {\bibinfo {author} {\bibfnamefont {Mark~T.}\ \bibnamefont
  {Mitchison}}, \bibinfo {author} {\bibfnamefont {John}\ \bibnamefont {Goold}},
  \ and\ \bibinfo {author} {\bibfnamefont {Javier}\ \bibnamefont {Prior}},\
  }\emph {\enquote {\bibinfo {title} {{Charging a quantum battery with linear
  feedback control}},}\ }\href {https://doi.org/10.22331/q-2021-07-13-500}
  {\bibfield  {journal} {\bibinfo  {journal} {Quantum}\ }\textbf {\bibinfo
  {volume} {5}},\ \bibinfo {pages} {500} (\bibinfo {year} {2021})},\ \Eprint
  {http://arxiv.org/abs/2012.00350} {arXiv:2012.00350}\BibitemShut {NoStop}%
\bibitem [{\citenamefont {Caravelli}\ \emph {et~al.}(2021)\citenamefont
  {Caravelli}, \citenamefont {Yan}, \citenamefont {Garc{\'i}a-Pintos},\ and\
  \citenamefont {Hamma}}]{CaravelliYanGarciaPintosHamma2021}%
  \BibitemOpen
  \bibfield  {author} {\bibinfo {author} {\bibfnamefont {Francesco}\
  \bibnamefont {Caravelli}}, \bibinfo {author} {\bibfnamefont {Bin}\
  \bibnamefont {Yan}}, \bibinfo {author} {\bibfnamefont {Luis~Pedro}\
  \bibnamefont {Garc{\'i}a-Pintos}}, \ and\ \bibinfo {author} {\bibfnamefont
  {Alioscia}\ \bibnamefont {Hamma}},\ }\emph {\enquote {\bibinfo {title}
  {{Energy storage and coherence in closed and open quantum batteries}},}\
  }\href {https://doi.org/10.22331/q-2021-07-15-505} {\bibfield  {journal}
  {\bibinfo  {journal} {Qantum}\ }\textbf {\bibinfo {volume} {5}},\ \bibinfo
  {pages} {505} (\bibinfo {year} {2021})},\ \Eprint
  {http://arxiv.org/abs/2012.15026} {arXiv:2012.15026}\BibitemShut {NoStop}%
\bibitem [{\citenamefont {Barra}(2019)}]{Barra2019}%
  \BibitemOpen
  \bibfield  {author} {\bibinfo {author} {\bibfnamefont {Felipe}\ \bibnamefont
  {Barra}},\ }\emph {\enquote {\bibinfo {title} {{Dissipative Charging of a
  Quantum Battery}},}\ }\href {https://doi.org/10.1103/PhysRevLett.122.210601}
  {\bibfield  {journal} {\bibinfo  {journal} {Phys. Rev. Lett.}\ }\textbf
  {\bibinfo {volume} {122}},\ \bibinfo {pages} {210601} (\bibinfo {year}
  {2019})},\ \Eprint {http://arxiv.org/abs/1902.00422}
  {arXiv:1902.00422}\BibitemShut {NoStop}%
\bibitem [{\citenamefont {Alicki}(2019)}]{Alicki2019}%
  \BibitemOpen
  \bibfield  {author} {\bibinfo {author} {\bibfnamefont {Robert}\ \bibnamefont
  {Alicki}},\ }\emph {\enquote {\bibinfo {title} {{A quantum open system model
  of molecular battery charged by excitons}},}\ }\href
  {https://doi.org/10.1063/1.5096772} {\bibfield  {journal} {\bibinfo
  {journal} {J. Chem. Phys.}\ }\textbf {\bibinfo {volume} {150}},\ \bibinfo
  {pages} {214110} (\bibinfo {year} {2019})},\ \Eprint
  {http://arxiv.org/abs/1903.12140} {arXiv:1903.12140}\BibitemShut {NoStop}%
\bibitem [{\citenamefont {Garcia-Pintos}\ \emph {et~al.}(2020)\citenamefont
  {Garcia-Pintos}, \citenamefont {Hamma},\ and\ \citenamefont {del
  Campo}}]{GarciaPintosHammaDelCampo2020}%
  \BibitemOpen
  \bibfield  {author} {\bibinfo {author} {\bibfnamefont {Luis~Pedro}\
  \bibnamefont {Garcia-Pintos}}, \bibinfo {author} {\bibfnamefont {Alioscia}\
  \bibnamefont {Hamma}}, \ and\ \bibinfo {author} {\bibfnamefont {Adolfo}\
  \bibnamefont {del Campo}},\ }\emph {\enquote {\bibinfo {title} {{Fluctuations
  in Extractable Work Bound the Charging Power of Quantum Batteries}},}\ }\href
  {https://doi.org/10.1103/PhysRevLett.125.040601} {\bibfield  {journal}
  {\bibinfo  {journal} {Phys. Rev. Lett.}\ }\textbf {\bibinfo {volume} {125}},\
  \bibinfo {pages} {040601} (\bibinfo {year} {2020})},\ \Eprint
  {http://arxiv.org/abs/1909.03558} {arXiv:1909.03558}\BibitemShut {NoStop}%
\bibitem [{\citenamefont {Cusumano}\ and\ \citenamefont
  {Rudnicki}(2021)}]{CusumanoRudnicki2021}%
  \BibitemOpen
  \bibfield  {author} {\bibinfo {author} {\bibfnamefont {Stefano}\ \bibnamefont
  {Cusumano}}\ and\ \bibinfo {author} {\bibfnamefont {{\L}ukasz}\ \bibnamefont
  {Rudnicki}},\ }\emph {\enquote {\bibinfo {title} {{Comment on ``Fluctuations
  in Extractable Work Bound the Charging Power of Quantum Batteries''}},}\
  }\href {https://doi.org/10.1103/PhysRevLett.127.028901} {\bibfield  {journal}
  {\bibinfo  {journal} {Phys. Rev. Lett.}\ }\textbf {\bibinfo {volume} {127}},\
  \bibinfo {pages} {028901} (\bibinfo {year} {2021})},\ \Eprint
  {http://arxiv.org/abs/2102.05627} {arXiv:2102.05627}\BibitemShut {NoStop}%
\bibitem [{\citenamefont {Wang}(2021)}]{Wang2021}%
  \BibitemOpen
  \bibfield  {author} {\bibinfo {author} {\bibfnamefont {Shang-Yung}\
  \bibnamefont {Wang}},\ }\href@noop {} {\emph {\enquote {\bibinfo {title}
  {{Comment on ``Fluctuations in Extractable Work Bound the Charging Power of
  Quantum Batteries''}},}\ }}\Eprint {http://arxiv.org/abs/2102.04921}
  {arXiv:2102.04921} [quant-ph] (\bibinfo {year} {2021})\BibitemShut {NoStop}%
\bibitem [{\citenamefont {Imai}\ \emph {et~al.}(2023)\citenamefont {Imai},
  \citenamefont {G\"uhne},\ and\ \citenamefont
  {Nimmrichter}}]{ImaiGuehneNimmrichter}%
  \BibitemOpen
  \bibfield  {author} {\bibinfo {author} {\bibfnamefont {Satoya}\ \bibnamefont
  {Imai}}, \bibinfo {author} {\bibfnamefont {Otfried}\ \bibnamefont {G\"uhne}},
  \ and\ \bibinfo {author} {\bibfnamefont {Stefan}\ \bibnamefont
  {Nimmrichter}},\ }\emph {\enquote {\bibinfo {title} {Work fluctuations and
  entanglement in quantum batteries},}\ }\href
  {https://doi.org/10.1103/PhysRevA.107.022215} {\bibfield  {journal} {\bibinfo
   {journal} {Phys. Rev. A}\ }\textbf {\bibinfo {volume} {107}},\ \bibinfo
  {pages} {022215} (\bibinfo {year} {2023})},\ \Eprint
  {http://arxiv.org/abs/2205.08447} {arXiv:2205.08447}\BibitemShut {NoStop}%
\bibitem [{\citenamefont {Friis}\ and\ \citenamefont
  {Huber}(2018)}]{FriisHuber2018}%
  \BibitemOpen
  \bibfield  {author} {\bibinfo {author} {\bibfnamefont {Nicolai}\ \bibnamefont
  {Friis}}\ and\ \bibinfo {author} {\bibfnamefont {Marcus}\ \bibnamefont
  {Huber}},\ }\emph {\enquote {\bibinfo {title} {Precision and {W}ork
  {F}luctuations in {G}aussian {B}attery {C}harging},}\ }\href
  {https://doi.org/10.22331/q-2018-04-23-61} {\bibfield  {journal} {\bibinfo
  {journal} {{Quantum}}\ }\textbf {\bibinfo {volume} {2}},\ \bibinfo {pages}
  {61} (\bibinfo {year} {2018})},\ \Eprint {http://arxiv.org/abs/1708.00749}
  {arXiv:1708.00749}\BibitemShut {NoStop}%
\bibitem [{\citenamefont {Crescente}\ \emph
  {et~al.}(2020{\natexlab{b}})\citenamefont {Crescente}, \citenamefont
  {Carrega}, \citenamefont {Sassetti},\ and\ \citenamefont
  {Ferraro}}]{CrescenteCarregaSassettiFerraro2020b}%
  \BibitemOpen
  \bibfield  {author} {\bibinfo {author} {\bibfnamefont {Alba}\ \bibnamefont
  {Crescente}}, \bibinfo {author} {\bibfnamefont {Matteo}\ \bibnamefont
  {Carrega}}, \bibinfo {author} {\bibfnamefont {Maura}\ \bibnamefont
  {Sassetti}}, \ and\ \bibinfo {author} {\bibfnamefont {Dario}\ \bibnamefont
  {Ferraro}},\ }\emph {\enquote {\bibinfo {title} {{Charging and energy
  fluctuations of a driven quantum battery}},}\ }\href
  {https://doi.org/10.1088/1367-2630/ab91fc} {\bibfield  {journal} {\bibinfo
  {journal} {New J. Phys.}\ }\textbf {\bibinfo {volume} {22}},\ \bibinfo
  {pages} {063057} (\bibinfo {year} {2020}{\natexlab{b}})},\ \Eprint
  {http://arxiv.org/abs/2005.05068} {arXiv:2005.05068}\BibitemShut {NoStop}%
\bibitem [{\citenamefont {Julia-Farre}\ \emph {et~al.}(2020)\citenamefont
  {Julia-Farre}, \citenamefont {Salamon}, \citenamefont {Riera}, \citenamefont
  {Bera},\ and\ \citenamefont
  {Lewenstein}}]{JuliaFarreSalamonRieraBeraLewenstein2020}%
  \BibitemOpen
  \bibfield  {author} {\bibinfo {author} {\bibfnamefont {Sergi}\ \bibnamefont
  {Julia-Farre}}, \bibinfo {author} {\bibfnamefont {Tymoteusz}\ \bibnamefont
  {Salamon}}, \bibinfo {author} {\bibfnamefont {Arnau}\ \bibnamefont {Riera}},
  \bibinfo {author} {\bibfnamefont {Manabendra~N.}\ \bibnamefont {Bera}}, \
  and\ \bibinfo {author} {\bibfnamefont {Maciej}\ \bibnamefont {Lewenstein}},\
  }\emph {\enquote {\bibinfo {title} {Bounds on the capacity and power of
  quantum batteries},}\ }\href
  {https://doi.org/10.1103/PhysRevResearch.2.023113} {\bibfield  {journal}
  {\bibinfo  {journal} {Phys. Rev. Research}\ }\textbf {\bibinfo {volume}
  {2}},\ \bibinfo {pages} {023113} (\bibinfo {year} {2020})},\ \Eprint
  {http://arxiv.org/abs/1811.04005} {arXiv:1811.04005}\BibitemShut {NoStop}%
\bibitem [{\citenamefont {Caravelli}\ \emph {et~al.}(2020)\citenamefont
  {Caravelli}, \citenamefont {Coulter-De~Wit}, \citenamefont {Garcia-Pintos},\
  and\ \citenamefont {Hamma}}]{CaravelliCoulterDeWitGarciaPintosHamma2020}%
  \BibitemOpen
  \bibfield  {author} {\bibinfo {author} {\bibfnamefont {Francesco}\
  \bibnamefont {Caravelli}}, \bibinfo {author} {\bibfnamefont {Ghislaine}\
  \bibnamefont {Coulter-De~Wit}}, \bibinfo {author} {\bibfnamefont
  {Luis~Pedro}\ \bibnamefont {Garcia-Pintos}}, \ and\ \bibinfo {author}
  {\bibfnamefont {Alioscia}\ \bibnamefont {Hamma}},\ }\emph {\enquote {\bibinfo
  {title} {{Random Quantum Batteries}},}\ }\href
  {https://doi.org/10.1103/PhysRevResearch.2.023095} {\bibfield  {journal}
  {\bibinfo  {journal} {Phys. Rev. Research}\ }\textbf {\bibinfo {volume}
  {2}},\ \bibinfo {pages} {023095} (\bibinfo {year} {2020})},\ \Eprint
  {http://arxiv.org/abs/1908.08064} {arXiv:1908.08064}\BibitemShut {NoStop}%
\bibitem [{\citenamefont {Ferraro}\ \emph {et~al.}(2019)\citenamefont
  {Ferraro}, \citenamefont {Andolina}, \citenamefont {Campisi}, \citenamefont
  {Pellegrini},\ and\ \citenamefont
  {Polini}}]{FerraroAndolinaCampisiPellegriniPolini2019}%
  \BibitemOpen
  \bibfield  {author} {\bibinfo {author} {\bibfnamefont {Dario}\ \bibnamefont
  {Ferraro}}, \bibinfo {author} {\bibfnamefont {Gian~Marcello}\ \bibnamefont
  {Andolina}}, \bibinfo {author} {\bibfnamefont {Michele}\ \bibnamefont
  {Campisi}}, \bibinfo {author} {\bibfnamefont {Vittorio}\ \bibnamefont
  {Pellegrini}}, \ and\ \bibinfo {author} {\bibfnamefont {Marco}\ \bibnamefont
  {Polini}},\ }\emph {\enquote {\bibinfo {title} {Quantum supercapacitors},}\
  }\href {https://doi.org/10.1103/PhysRevB.100.075433} {\bibfield  {journal}
  {\bibinfo  {journal} {Phys. Rev. B}\ }\textbf {\bibinfo {volume} {100}},\
  \bibinfo {pages} {075433} (\bibinfo {year} {2019})},\ \Eprint
  {http://arxiv.org/abs/1902.06474} {arXiv:1902.06474}\BibitemShut {NoStop}%
\bibitem [{\citenamefont {Jablonski}(2019)}]{JablonskiMSc2019}%
  \BibitemOpen
  \bibfield  {author} {\bibinfo {author} {\bibfnamefont {Beniamin~Radomir}\
  \bibnamefont {Jablonski}},\ }\emph {\bibinfo {title} {Charging precision for
  finite-dimensional quantum batteries}},\ \href
  {https://doi.org/10.25365/thesis.56195} {Master's thesis},\ \bibinfo
  {school} {University of Vienna} (\bibinfo {year} {2019})\BibitemShut
  {NoStop}%
\bibitem [{\citenamefont {Pusz}\ and\ \citenamefont
  {Woronowicz}(1978)}]{PuszWoronowicz1978}%
  \BibitemOpen
  \bibfield  {author} {\bibinfo {author} {\bibfnamefont {Wies{\l}aw}\
  \bibnamefont {Pusz}}\ and\ \bibinfo {author} {\bibfnamefont
  {Stanis{\l}aw~L.}\ \bibnamefont {Woronowicz}},\ }\emph {\enquote {\bibinfo
  {title} {{Passive states and KMS states for general quantum systems}},}\
  }\href {https://doi.org/10.1007/BF01614224} {\bibfield  {journal} {\bibinfo
  {journal} {Comm. Math. Phys.}\ }\textbf {\bibinfo {volume} {58}},\ \bibinfo
  {pages} {273} (\bibinfo {year} {1978})},\ \bibinfo {note}
  {\href{https://projecteuclid.org/euclid.cmp/1103901491}{https://projecteuclid.org/euclid.cmp/1103901491}}\BibitemShut
  {NoStop}%
\bibitem [{\citenamefont {Brown}\ \emph {et~al.}(2016)\citenamefont {Brown},
  \citenamefont {Friis},\ and\ \citenamefont {Huber}}]{BrownFriisHuber2016}%
  \BibitemOpen
  \bibfield  {author} {\bibinfo {author} {\bibfnamefont {Eric~G.}\ \bibnamefont
  {Brown}}, \bibinfo {author} {\bibfnamefont {Nicolai}\ \bibnamefont {Friis}},
  \ and\ \bibinfo {author} {\bibfnamefont {Marcus}\ \bibnamefont {Huber}},\
  }\emph {\enquote {\bibinfo {title} {{Passivity and practical work extraction
  using Gaussian operations}},}\ }\href
  {https://doi.org/10.1088/1367-2630/18/11/113028} {\bibfield  {journal}
  {\bibinfo  {journal} {New J. Phys.}\ }\textbf {\bibinfo {volume} {18}},\
  \bibinfo {pages} {113028} (\bibinfo {year} {2016})},\ \Eprint
  {http://arxiv.org/abs/1608.04977} {arXiv:1608.04977}\BibitemShut {NoStop}%
\bibitem [{\citenamefont {Reeb}\ and\ \citenamefont
  {Wolf}(2014)}]{ReebWolf2014}%
  \BibitemOpen
  \bibfield  {author} {\bibinfo {author} {\bibfnamefont {David}\ \bibnamefont
  {Reeb}}\ and\ \bibinfo {author} {\bibfnamefont {Michael~M.}\ \bibnamefont
  {Wolf}},\ }\emph {\enquote {\bibinfo {title} {{An improved Landauer Principle
  with finite-size corrections}},}\ }\href
  {https://doi.org/10.1088/1367-2630/16/10/103011} {\bibfield  {journal}
  {\bibinfo  {journal} {New J. Phys.}\ }\textbf {\bibinfo {volume} {16}},\
  \bibinfo {pages} {103011} (\bibinfo {year} {2014})},\ \Eprint
  {http://arxiv.org/abs/1306.4352} {arXiv:1306.4352}\BibitemShut {NoStop}%
\bibitem [{\citenamefont {Alicki}\ \emph {et~al.}(2004)\citenamefont {Alicki},
  \citenamefont {Horodecki}, \citenamefont {Horodecki},\ and\ \citenamefont
  {Horodecki}}]{AlickiHorodeckiMPR2004}%
  \BibitemOpen
  \bibfield  {author} {\bibinfo {author} {\bibfnamefont {Robert}\ \bibnamefont
  {Alicki}}, \bibinfo {author} {\bibfnamefont {Micha{\l}}\ \bibnamefont
  {Horodecki}}, \bibinfo {author} {\bibfnamefont {Pawe{\l}}\ \bibnamefont
  {Horodecki}}, \ and\ \bibinfo {author} {\bibfnamefont {Ryszard}\ \bibnamefont
  {Horodecki}},\ }\emph {\enquote {\bibinfo {title} {{Thermodynamics of Quantum
  Information Systems -- Hamiltonian Description}},}\ }\href
  {https://doi.org/10.1023/B:OPSY.0000047566.72717.71} {\bibfield  {journal}
  {\bibinfo  {journal} {Open Syst. Inf. Dyn.}\ }\textbf {\bibinfo {volume}
  {11}},\ \bibinfo {pages} {205} (\bibinfo {year} {2004})},\ \Eprint
  {http://arxiv.org/abs/quant-ph/0402012} {arXiv:quant-ph/0402012}\BibitemShut
  {NoStop}%
\bibitem [{\citenamefont {Brunner}\ \emph {et~al.}(2012)\citenamefont
  {Brunner}, \citenamefont {Linden}, \citenamefont {Popescu},\ and\
  \citenamefont {Skrzypczyk}}]{BrunnerLindenPopescuSkrzypczyk2012}%
  \BibitemOpen
  \bibfield  {author} {\bibinfo {author} {\bibfnamefont {Nicolas}\ \bibnamefont
  {Brunner}}, \bibinfo {author} {\bibfnamefont {Noah}\ \bibnamefont {Linden}},
  \bibinfo {author} {\bibfnamefont {Sandu}\ \bibnamefont {Popescu}}, \ and\
  \bibinfo {author} {\bibfnamefont {Paul}\ \bibnamefont {Skrzypczyk}},\ }\emph
  {\enquote {\bibinfo {title} {{Virtual qubits, virtual temperatures, and the
  foundations of thermodynamics}},}\ }\href
  {https://doi.org/10.1103/PhysRevE.85.051117} {\bibfield  {journal} {\bibinfo
  {journal} {Phys. Rev. E}\ }\textbf {\bibinfo {volume} {85}},\ \bibinfo
  {pages} {051117} (\bibinfo {year} {2012})},\ \Eprint
  {http://arxiv.org/abs/1106.2138} {arXiv:1106.2138}\BibitemShut {NoStop}%
\bibitem [{\citenamefont {{\AA}berg}(2013)}]{Aberg2013}%
  \BibitemOpen
  \bibfield  {author} {\bibinfo {author} {\bibfnamefont {Johan}\ \bibnamefont
  {{\AA}berg}},\ }\emph {\enquote {\bibinfo {title} {Truly work-like work
  extraction via a single-shot analysis},}\ }\href
  {https://doi.org/10.1038/ncomms2712} {\bibfield  {journal} {\bibinfo
  {journal} {Nat. Commun.}\ }\textbf {\bibinfo {volume} {4}},\ \bibinfo {pages}
  {1925} (\bibinfo {year} {2013})},\ \Eprint {http://arxiv.org/abs/1110.6121}
  {arXiv:1110.6121}\BibitemShut {NoStop}%
\bibitem [{\citenamefont {Gallego}\ \emph {et~al.}(2016)\citenamefont
  {Gallego}, \citenamefont {Eisert},\ and\ \citenamefont
  {Wilming}}]{GallegoEisertWilming2016}%
  \BibitemOpen
  \bibfield  {author} {\bibinfo {author} {\bibfnamefont {Rodrigo}\ \bibnamefont
  {Gallego}}, \bibinfo {author} {\bibfnamefont {Jens}\ \bibnamefont {Eisert}},
  \ and\ \bibinfo {author} {\bibfnamefont {Henrik}\ \bibnamefont {Wilming}},\
  }\emph {\enquote {\bibinfo {title} {{Thermodynamic work from operational
  principles}},}\ }\href {https://doi.org/10.1088/1367-2630/18/10/103017}
  {\bibfield  {journal} {\bibinfo  {journal} {New J. Phys.}\ }\textbf {\bibinfo
  {volume} {18}},\ \bibinfo {pages} {103017} (\bibinfo {year} {2016})},\
  \Eprint {http://arxiv.org/abs/1504.05056} {arXiv:1504.05056}\BibitemShut
  {NoStop}%
\bibitem [{\citenamefont {Niedenzu}\ \emph {et~al.}(2019)\citenamefont
  {Niedenzu}, \citenamefont {Huber},\ and\ \citenamefont
  {Boukobza}}]{NiedenzuHuberBoukobza2019}%
  \BibitemOpen
  \bibfield  {author} {\bibinfo {author} {\bibfnamefont {Wolfgang}\
  \bibnamefont {Niedenzu}}, \bibinfo {author} {\bibfnamefont {Marcus}\
  \bibnamefont {Huber}}, \ and\ \bibinfo {author} {\bibfnamefont {Erez}\
  \bibnamefont {Boukobza}},\ }\emph {\enquote {\bibinfo {title} {{Concepts of
  work in autonomous quantum heat engines}},}\ }\href
  {https://doi.org/10.22331/q-2019-10-14-195} {\bibfield  {journal} {\bibinfo
  {journal} {Quantum}\ }\textbf {\bibinfo {volume} {3}},\ \bibinfo {pages}
  {195} (\bibinfo {year} {2019})},\ \Eprint {http://arxiv.org/abs/1907.01353}
  {arXiv:1907.01353}\BibitemShut {NoStop}%
\bibitem [{\citenamefont {Beyer}\ \emph {et~al.}(2020)\citenamefont {Beyer},
  \citenamefont {Luoma},\ and\ \citenamefont {Strunz}}]{BeyerLuomaStrunz2020}%
  \BibitemOpen
  \bibfield  {author} {\bibinfo {author} {\bibfnamefont {Konstantin}\
  \bibnamefont {Beyer}}, \bibinfo {author} {\bibfnamefont {Kimmo}\ \bibnamefont
  {Luoma}}, \ and\ \bibinfo {author} {\bibfnamefont {Walter~T.}\ \bibnamefont
  {Strunz}},\ }\emph {\enquote {\bibinfo {title} {{Work as an external quantum
  observable and an operational quantum work fluctuation theorem}},}\ }\href
  {https://doi.org/10.1103/PhysRevResearch.2.033508} {\bibfield  {journal}
  {\bibinfo  {journal} {Phys. Rev. Research}\ }\textbf {\bibinfo {volume}
  {2}},\ \bibinfo {pages} {033508} (\bibinfo {year} {2020})},\ \Eprint
  {http://arxiv.org/abs/2003.06437} {arXiv:2003.06437}\BibitemShut {NoStop}%
\bibitem [{\citenamefont {Talkner}\ \emph {et~al.}(2007)\citenamefont
  {Talkner}, \citenamefont {Lutz},\ and\ \citenamefont
  {H{\"a}nggi}}]{TalknerLutzHaenggi2007}%
  \BibitemOpen
  \bibfield  {author} {\bibinfo {author} {\bibfnamefont {Peter}\ \bibnamefont
  {Talkner}}, \bibinfo {author} {\bibfnamefont {Eric}\ \bibnamefont {Lutz}}, \
  and\ \bibinfo {author} {\bibfnamefont {Peter}\ \bibnamefont {H{\"a}nggi}},\
  }\emph {\enquote {\bibinfo {title} {{Fluctuation theorems: Work is not an
  observable}},}\ }\href {https://doi.org/10.1103/PhysRevE.75.050102}
  {\bibfield  {journal} {\bibinfo  {journal} {Phys. Rev. E}\ }\textbf {\bibinfo
  {volume} {75}},\ \bibinfo {pages} {050102(R)} (\bibinfo {year} {2007})},\
  \Eprint {http://arxiv.org/abs/cond-mat/0703189}
  {arXiv:cond-mat/0703189}\BibitemShut {NoStop}%
\bibitem [{\citenamefont {Guryanova}\ \emph {et~al.}(2020)\citenamefont
  {Guryanova}, \citenamefont {Friis},\ and\ \citenamefont
  {Huber}}]{GuryanovaFriisHuber2020}%
  \BibitemOpen
  \bibfield  {author} {\bibinfo {author} {\bibfnamefont {Yelena}\ \bibnamefont
  {Guryanova}}, \bibinfo {author} {\bibfnamefont {Nicolai}\ \bibnamefont
  {Friis}}, \ and\ \bibinfo {author} {\bibfnamefont {Marcus}\ \bibnamefont
  {Huber}},\ }\emph {\enquote {\bibinfo {title} {{Ideal Projective Measurements
  Have Infinite Resource Costs}},}\ }\href
  {https://doi.org/10.22331/q-2020-01-13-222} {\bibfield  {journal} {\bibinfo
  {journal} {Quantum}\ }\textbf {\bibinfo {volume} {4}},\ \bibinfo {pages}
  {222} (\bibinfo {year} {2020})},\ \Eprint {http://arxiv.org/abs/1805.11899}
  {arXiv:1805.11899}\BibitemShut {NoStop}%
\bibitem [{\citenamefont {{Debarba}}\ \emph {et~al.}(2019)\citenamefont
  {{Debarba}}, \citenamefont {{Manzano}}, \citenamefont {{Guryanova}},
  \citenamefont {{Huber}},\ and\ \citenamefont {{Friis}}}]{DebarbaEtAl2019}%
  \BibitemOpen
  \bibfield  {author} {\bibinfo {author} {\bibfnamefont {Tiago}\ \bibnamefont
  {{Debarba}}}, \bibinfo {author} {\bibfnamefont {Gonzalo}\ \bibnamefont
  {{Manzano}}}, \bibinfo {author} {\bibfnamefont {Yelena}\ \bibnamefont
  {{Guryanova}}}, \bibinfo {author} {\bibfnamefont {Marcus}\ \bibnamefont
  {{Huber}}}, \ and\ \bibinfo {author} {\bibfnamefont {Nicolai}\ \bibnamefont
  {{Friis}}},\ }\emph {\enquote {\bibinfo {title} {{Work estimation and work
  fluctuations in the presence of non-ideal measurements}},}\ }\href
  {https://doi.org/10.1088/1367-2630/ab4d9d} {\bibfield  {journal} {\bibinfo
  {journal} {New J. Phys.}\ }\textbf {\bibinfo {volume} {21}},\ \bibinfo
  {pages} {113002} (\bibinfo {year} {2019})},\ \Eprint
  {http://arxiv.org/abs/1902.08568} {arXiv:1902.08568}\BibitemShut {NoStop}%
\bibitem [{\citenamefont {Spengler}\ \emph {et~al.}(2010)\citenamefont
  {Spengler}, \citenamefont {Huber},\ and\ \citenamefont
  {~}}]{SpenglerHuberHiesmayr2010}%
  \BibitemOpen
  \bibfield  {author} {\bibinfo {author} {\bibfnamefont {Christoph}\
  \bibnamefont {Spengler}}, \bibinfo {author} {\bibfnamefont {Marcus}\
  \bibnamefont {Huber}}, \ and\ \bibinfo {author} {\bibfnamefont {Beatrix~C.}\
  \bibnamefont {~}},\ }\emph {\enquote {\bibinfo {title} {{A composite
  parameterization of unitary groups, density matrices and subspaces}},}\
  }\href {https://doi.org/10.1088/1751-8113/43/38/385306} {\bibfield  {journal}
  {\bibinfo  {journal} {J. Phys. A: Math. Theor.}\ }\textbf {\bibinfo {volume}
  {43}},\ \bibinfo {pages} {385306} (\bibinfo {year} {2010})},\ \Eprint
  {http://arxiv.org/abs/1004.5252} {arXiv:1004.5252}\BibitemShut {NoStop}%
\bibitem [{\citenamefont {Perarnau-Llobet}\ and\ \citenamefont
  {Uzdin}(2019)}]{PerarnauLlobetUzdin2019}%
  \BibitemOpen
  \bibfield  {author} {\bibinfo {author} {\bibfnamefont {Mart{\'i}}\
  \bibnamefont {Perarnau-Llobet}}\ and\ \bibinfo {author} {\bibfnamefont
  {Raam}\ \bibnamefont {Uzdin}},\ }\emph {\enquote {\bibinfo {title}
  {{Collective operations can extremely reduce work fluctuations}},}\ }\href
  {https://doi.org/10.1088/1367-2630/ab36a9} {\bibfield  {journal} {\bibinfo
  {journal} {New J. Phys.}\ }\textbf {\bibinfo {volume} {21}},\ \bibinfo
  {pages} {083023} (\bibinfo {year} {2019})},\ \Eprint
  {http://arxiv.org/abs/1810.02237} {arXiv:1810.02237}\BibitemShut {NoStop}%
\end{thebibliography}%


\hypertarget{sec:appendix}
\appendix
\section*{Appendix}
\renewcommand{\thesubsubsection}{A.\arabic{subsection}.\Roman{subsubsection}}
\renewcommand{\thesubsection}{A.\arabic{subsection}}
\renewcommand{\thesection}{A}
%


\section{Maximal energy variance and work fluctuations via local unitary operations}
\label{app:worst local protocol fluc}

\subsection{Maximal variance for local operations}
\label{app:worst local protocol Variance}
In this appendix, we are concerned with finding the optimal or worst process for the variance and the work fluctuation when we are restricting ourselves to apply only local unitary operations to increase the energy of $N$-qubit systems that are initially in a thermal state with inverse temperature $\beta$. To do this, we particularly focus on the Lagrange-multiplier method to find the local minima and maxima of a function subject to one or more conditions that should be exactly satisfied. In charging processes, one can consider the work fluctuations or the charging precision as cost functions subject to the constraint of fixed energy increase.

We first focus on the energy variance when charging multi-qubit batteries by local unitary operations. If we start from an (uncorrelated) thermal state, no correlation can be created via local operations and the variance of the final state with respect to the local Hamiltonian can be written as the sum of the local variances [cf.~Eq.~(\ref{eq:local variance})]. Thus, we can describe the optimization problem at hand in terms of local variables describing the local unitary operations. In general, we may characterize any $2\times 2$ unitary operation by two real variables, $\theta$ and~$\phi$, as
\begin{align}
U(\theta,\phi)=\begin{bmatrix}
\,\cos{\theta}&& -e^{-i\phi}\,\sin{\theta}\,\\
 \,e^{i\phi}\,\sin{\theta}\,&&\cos{\theta}\,
\end{bmatrix}.
\label{eq: 2dim unitary}
\end{align}
In the present case, since off-diagonal elements of the density matrix do not play any role in the calculation of our quantities of interest and the variable $\phi$ only appears in these elements, we choose to set $\phi$ to zero and optimize the process with respect to $\theta$. Thus, the energy of any one qubit can be increased by a continuous rotation with an angle $\theta$. This rotation corresponds to a mapping of the diagonal elements given by
\begin{align}
    \begin{pmatrix}\tilde{p}_0\\\tilde{p}_1\end{pmatrix}\mapsto \, \begin{pmatrix}p_0\, \cos^{2}\!\theta\,+p_1\,\sin^{2}\!\theta\\\,p_0\, \sin^{2}\!\theta+p_1\,\cos^{2}\!\theta\,\end{pmatrix},
    \label{eq:single qu map}
\end{align}
where the $p_i$s are the energy populations of the initial thermal state. Employing this map, it is straightforward to calculate the energy variance of a single qubit:
\begin{align}
\label{eq: single qubit var}
    V(\tilde{\varrho})=\,\omega^2 \tilde{p}_1(1-\tilde{p}_1),
\end{align}
where the corresponding average energy is $\tilde{\epsilon}=\, \omega\,\tilde{p}_1$. For the sake of simplicity, one may choose the variable $\tilde{p}_1$ instead of $\theta$, following the simple relationship in Eq.~(\ref{eq:single qu map}).

As already noted, if the $N$ identical thermal qubits evolve via local unitary operations, i.e., $\tilde{\varrho}_{\textup{tot}}=\bigotimes_{i=1}^N \,U_i \tau (\beta)U_i^{\dagger}$, the total energy variance of the final state is the sum of the local variances. Making use of Eqs.~(\ref{eq:local variance}) and~(\ref{eq: single qubit var}) we thus have
\begin{align}
\label{eq: multi qubit var}
    V(\tilde{\varrho}_{\textup{tot}})=\sum_{i=1}^N V(\tilde{\varrho}_i)=\sum_{i=1}^N \,\omega^2 \tilde{p}_{1_i}(1-\tilde{p}_{1_i}),
\end{align}
where $p_{1_i}= \bra{1_i}\,U_i \tau (\beta)U_i^{\dagger}\,\ket{1_i}$, and $\ket{1_i}$ is the second eigenstate of the Hamiltonian of the $i$th qubit. In a similar way, we may define the total average energy as $\tilde{\epsilon}_{\textup{tot}}= \sum_{i=1}^N \,\omega\, \tilde{p}_{1_i}$. We are now ready to optimize the variance subject to a given value of the average energy $\tilde{\epsilon}_{\textup{tot}}=c$, where $c$ is a constant. We can solve this problem by using the Lagrange multiplier $\lambda_V$ as follows:
\begin{align}
  \mathcal{L}_V&=V(\tilde{\varrho}_{\textup{tot}})- \lambda_V (\tilde{\epsilon}_{\textup{tot}}-c)\nonumber\\
  &=\sum_{i=1}^N \,\omega^2 \tilde{p}_{1_i}(1-\tilde{p}_{1_i})-\lambda_V (\sum_{i=1}^N \,\omega \tilde{p}_{1_i}-c),
\end{align}
where $\mathcal{L}_V$ is a Lagrange function. In order to find the stationary point of the function subject to the energy constraint, we need to solve the following $N$ equations
\begin{align}
\frac{\partial \mathcal{L}_V}{\partial \tilde{p}_{1_i}}= \omega^2(1-\tilde{p}_{1_i})-\lambda_V  \,\omega \tilde{p}_{1_i}=0 ~~~~~\forall \, i \in \, \{1,\,2,\,\hdots,\,N\}.
\end{align}
It is obvious that our problem reduces to  $N$ independent linear equations, each of them related to the state transformation of one of the qubits. Due to the symmetry of the equations, we can show that the stationary point is characterized by the same value for all variables,
\begin{align}
    \tilde{p}_{1_i}=\frac{\omega}{\lambda_V -\omega}~~~~\forall \, i,
\end{align}
where $\lambda_V$ is determined by the energy constraint. Since the total system is initially uncorrelated and thermal, with inverse temperature $\beta$, the obtained result already implies that all qubits should be symmetrically transformed to the state that has the required energy, i.e., $U_{\mathrm{loc}}=U(\theta, 0)^{\otimes N}$. This transformation then raises the energy of each qubit by $\Delta \epsilon_{\textup{tot}}/N$ if we  want to invest the target energy $\Delta \epsilon_{\textup{tot}}$.
So far, we have shown that, for fixed energy input, the SLCP represents an extremal point among the local unitary charging processes. We now proceed to show that this extremal point indeed corresponds to the maximal variance achievable by a local unitary process subject to the energy constraint, and that SLCPs thus describe the worst-case local scenario for minimizing the variance. To do so, we consider a local unitary transformation leading us to the same final energy but a smaller variance compared to the SLCP.

Let us first assume that the energy of the $N-$qubit system is increased via an SLCP to reach a total energy $\tilde{\epsilon}_{\textup{tot}}$ by transforming ${p}_{1_i}\to \tilde{p}_{1_i}$  $\forall i$. For such a process, the energy variance of the final state is given by $V_{\textup{SLCP}}=N \,\omega^2 \tilde{p}_{1}(1-\tilde{p}_{1})$. In the next step, we use an asymmetric charging process in which the final populations of the excited states of two of the qubits (w.l.o.g., the first two qubits, labelled $1$ and $2$) are chosen to be $\tilde{p}_{1_1}=\tilde{p}_{1}+\delta$ and $\tilde{p}_{1_2}=\tilde{p}_{1}-\delta$, respectively, where $\delta$ is chosen to be a small but nonzero, positive real number. The populations of the remaining qubits are transformed in the same way as in the original SLCP, which leads us to the same total energy, thus satisfying the energy constraint. In such a process, the energy variance is
\begin{align}
\label{eq: multi qubit var assym delta}
    V_{\delta}&=\,\omega^2 \bigl((\tilde{p}_{1}+\delta)(1-\tilde{p}_{1}-\delta)+(\tilde{p}_{1}-\delta)(1-\tilde{p}_{1}+\delta)\bigr)\nonumber\\[1mm]
    &\ \ +\,(N-2) \,\omega^2\, \tilde{p}_{1}(1-\tilde{p}_{1})\nonumber\\[1mm]
    &=\,N \,\omega^2\, \tilde{p}_{1}(1-\tilde{p}_{1})-2\,\omega^2 \delta^2\,\leq \,V_{\textup{SLCP}}\,.
\end{align}
We thus see that the extreme point represented by the SLCP describes the maximal energy variance among all local unitary charging processes.

With this knowledge and using Eq.~(\ref {eq:Ave. sq. two level}), one can easily obtain the energy variance as a function of the average energy $\tilde{\epsilon}_{\textup{tot}}$ for such a process,
\begin{align}
    V(\tilde{\epsilon}_{\textup{tot}})= \tilde{\epsilon}_{\textup{tot}} (\omega-\frac{\tilde{\epsilon}_{\textup{tot}}}{N})
\end{align}
which is the maximum amount of energy variance that can be created through local unitary transformations when increasing the energy of the multipartite system from $\epsilon_0$ to $\tilde{\epsilon}_{\textup{tot}}$. Employing the Lagrange-multiplier method, we therefore obtain the worst-case scenario for minimizing the variance when charging an $N$-qubit system via local operations.


\subsection{Maximal work fluctuations for local operations}

In a similar way we would like to explore the stationary points of the work-fluctuation function constrained by a fixed energy input via local unitary operations. From Ref.~\cite{FriisHuber2018}, we already know that the fluctuations do not only depend on the initial and final states but also on the dynamics of the system. In our case, due to the commutation of the initial thermal state with the Hamiltonian, the fluctuations can be rewritten in a simplified operational form as
\begin{align}
\label{eq: operational fluct}
(\Delta W_{\textup{tot}})^2& =  V(\tilde{\varrho}_{\textup{tot}})+V(\tau(\beta)^{\otimes N})-2\big( \tr[\tilde{H}_{\textup{tot}}H_{\textup{tot}}\tau(\beta )^{\otimes N}]\nonumber\\[2mm]
&\ \ -\tr[H_{\textup{tot}}\tau(\beta )^{\otimes N}]\,\tr[H_{\textup{tot}}\tilde{\varrho}_{\textup{tot}}]\big),
\end{align}
where  $\tilde{H}_{\textup{tot}}=U_{\mathrm{loc}}^{\dagger}H_{\textup{tot}}\,U_{\mathrm{loc}}=\sum_{i=1}^N \tilde{H}_i$. Using the fact that all the operations contributing to the fluctuations are local operations, the initial and final variances are given by the sum of the respective local variances. However, we still need to express the last term in Eq.~(\ref{eq: operational fluct}) in terms of local quantities, i.e.,
\begin{align}
 & \tr[\tilde{H}_{\textup{tot}}H_{\textup{tot}}\tau(\beta )^{\otimes N}]=\sum_{i,j=1}^N \tr[\tilde{H}_{i}H_{j}\tau(\beta )^{\otimes N}]\nonumber\\
 &= \sum_{i\neq j}^N \tr[\tilde{H}_{i}\tau(\beta )]\,\tr[H_{j}\tau(\beta )]+\sum_{i=1}^N \tr[\tilde{H}_{i}H_i\tau(\beta )]\nonumber\\
 &= \sum_{i\neq j}^N \tr[H_i\tilde{\varrho}_i]\,\tr[H_{j}\tau(\beta )]+\sum_{i=1}^N \tr[\tilde{H}_{i}H_i\tau(\beta )],
 \label{eq:last fluct term}
\end{align}
where have made use of the cyclicity of the trace (i.e., $\tr[XY]=\tr[YX]$) in the last line. For the second term, we have
\begin{align}
   &\tr[H\tau(\beta)]\,\tr[H_{\textup{tot}}\tilde{\varrho}_{\textup{tot}}]=
   \label{eq:loc energy}\\
  &  \ \ =\sum_{i\neq j}^N \tr[H_i\tilde{\varrho}_i]\,\tr[H_{j}\tau(\beta )]+\sum_{i=1}^N \tr[H_i\tilde{\varrho}_i]\,\tr[H_{i}\tau(\beta )].
  \nonumber
\end{align}
Combining Eqs.~(\ref{eq: operational fluct}),~(\ref{eq:last fluct term}), and~(\ref{eq:loc energy}), the work fluctuations of the process may be described as a function of local quantities,
\begin{align}
\label{eq: loc operational fluct}
(\Delta W_{\textup{tot}})^2& =\sum_{i=1}^N \big[  V(\tilde{\varrho}_{i})+V(\tau(\beta))\\
&\ \ -2\big( \tr[\tilde{H}_{i}H_{i}\tau(\beta )]-\tr[H\tau(\beta)]\,\tr[H_{i}\tilde{\varrho}_{i}]\big)\big].\nonumber
\end{align}
In this case, one can obtain $(\Delta W_{\textup{tot}})^2$ as a function of $(\{\tilde{p}_{1_i}\}_{i},p_1,p_0)$ in similar way to the variance function by using Eqs.~(\ref{eq: 2dim unitary}) and~(\ref{eq:single qu map}),
\begin{align}
\label{eq:tot fluct as p}
(\Delta W_{\textup{tot}})^2& =\sum_{i=1}^N \big[  \,\omega^2 \tilde{p}_{1_i}(1-\tilde{p}_{1_i})+\,\omega^2 {p}_{1}(1-{p}_{1})\nonumber\\
&\ \ -2\big(\omega^2 p_1\,\tfrac{\tilde{p}_{1_i}-p_0}{p_1-p_0} -\omega^2 p_1 \,\tilde{p}_{1_i} ]\big)\big].
\end{align}
In this way Eq.~(\ref{eq:tot fluct as p}) provides an opportunity to find stationary points of the fluctuation function under the constraint of fixed energy input for such processes. To proceed, we can write the corresponding Lagrange function as
\begin{align}
  \mathcal{L}_F&=(\Delta W_{\textup{tot}})^2- \lambda_F (\tilde{\epsilon}_{\textup{tot}}-c)\label{eq:lagrange fun fluc}\\
  &=\sum_{i=1}^N \big[  \,\omega^2 \tilde{p}_{1_i}(1-\tilde{p}_{1_i})+\,\omega^2 {p}_{1}(1-{p}_{1})\nonumber\\
&\ \ -2\big(\omega^2 p_1\,\tfrac{\tilde{p}_{1_i}-p_0}{p_1-p_0} -\omega^2 p_1 \,\tilde{p}_{1_i} \big)-\lambda_F \bigl(\,\omega^2 \tilde{p}_{1_i}-\tfrac{c}{N}\bigr)\big].\nonumber
\end{align}
To determine the stationary points which are related to the best-case or worst-case scenario, we calculate the first partial derivatives of $\mathcal{L}_F$ with respect to the variables $\tilde{p}_{1_i}$,
\begin{align}
\frac{\partial \mathcal{L}_F}{\partial \tilde{p}_{1_i}}= \omega^2(1-\tilde{p}_{1_i}- \tfrac{p_1}{p_1-p_0})-\lambda_F  \,\omega\,\tilde{p}_{1_i} \ \ \forall \, i \in \, \{1,\,2,\,\hdots,\,N\}.
\end{align}
The condition of vanishing first derivative at the extremal points yields
\begin{align}
    \tilde{p}_{1_i}=\frac{\omega\,p_0}{(\lambda_F -\omega)(p_0-p_1)}~~~~\forall \, i,
\end{align}
which tells us that to reach the extremal point, the energy of the system should be increased via symmetric local unitary operations. By comparing the symmetric and asymmetric local transformations described in Appendix~\ref{app:worst local protocol Variance}, one then arrives at the same conclusion as before, that is, the obtained extremal point represents the local unitary process achieving maximal work fluctuations at fixed energy input. Thus, if we want to locally raise the energy of the $N$-qubit system by the amount $\Delta \epsilon_{\textup{tot}}$, the largest fluctuations are obtained when each of the qubits reaches a state with equal energy $\tilde{\epsilon}=\epsilon_0+\Delta \epsilon_{\textup{tot}}/N$. We thus see that for a given energy $\tilde{\epsilon}$, the obtained unique extremal point is given by $\tilde{p}_{1_i}=\tilde{\epsilon}/\omega$ which results in the maximal possible work fluctuations for locally charging an $N$-qubit system. Consequently, we have obtained the local-unitary charging protocol with the largest fluctuations and hence with the worst performance. Moreover, the local unitary worst-case scenarios of work fluctuation and variance are compatible and match for any initial temperature in such a process.

\end{document}